\documentclass[%
reprint,
 amsmath,amssymb,
 aps,
	]{revtex4-2}

\makeatletter
\newcounter{savesection}
\newcounter{apdxsection}
\renewcommand\appendix{\par
	\setcounter{savesection}{\value{section}}%
	\setcounter{section}{\value{apdxsection}}%
	\setcounter{subsection}{0}%
	\gdef\thesection{\@Alph\c@section}}
\newcommand\unappendix{\par
	\setcounter{apdxsection}{\value{section}}%
	\setcounter{section}{\value{savesection}}%
	\setcounter{subsection}{0}%
	\gdef\thesection{\@arabic\c@section}}
\makeatother

\usepackage{graphicx}
\usepackage{dcolumn}
\usepackage{bm}
   \usepackage{xcolor}
   \definecolor{best_acc}{rgb}{0,0.5,0} 
   \definecolor{refcol}{rgb}{0,0,1} 
   \definecolor{mygray}{gray}{0.85} 
   \definecolor{mycite}{rgb}{0,0,1} 
   \definecolor{edit1}{rgb}{1,0,0}
   \definecolor{edit2}{rgb}{0.62,.12,.941}
   \definecolor{edit1}{rgb}{0,0,0}
   \definecolor{edit2}{rgb}{0,0,0}
   \definecolor{myurl}{rgb}{0,.45,0.85} 
   \usepackage[colorlinks=blue]{hyperref}
   \hypersetup{
   	colorlinks=false,  
   	pdfborder={0 0 0}, 
   }
   \hypersetup{
   	colorlinks=true,
   	linkcolor= magenta,
   	citecolor= mycite,
   	filecolor=magenta,      
   	urlcolor=magenta,
   }
\usepackage{lipsum}  

\usepackage{lineno}

\begin{document}
	\title{Wavenumber Scattering and Inter-band Targeted Energy Transfer in Phononic Lattices with Local Vibro-Impact Nonlinearities}
	\author{Joshua R.~Tempelman}
	\author{Alexander F.~Vakakis}
	\author{Kathryn H.~Matlack}
	\affiliation{%
		Department of Mechanical Science and Engineering, University of Illinois at Urbana-Champaign, 1206 W Green St, Urbana, IL 61801
	}%

\begin{abstract}
 We propose a method for manipulating wave propagation in phononic lattices by employing local vibro-impact (VI) nonlinearities to \textit{scatter} energy across the underling linear band structure of the lattice, and \textit{transfer} energy from lower to higher optical bands.
 Inspired by recent developments in the field of nonlinear targeted energy transfer (TET) using \textit{non-resonant} energy exchanges, we achieve this using spatially localized VI forces that redistribute energy across the linear spectrum of the lattice in a non-resonant fashion.
 First, a 1-dimensional (1D), 2-band phononic lattice with embedded  VI unit cells is computationally studied to demonstrate that energy is scattered in the wavenumber domain, and this nonlinear scattering mechanism depends on the energy of the propagating wave. 
 Next, a 4-band lattice is studied with a similar technique to demonstrate the concept of inter-band targeted energy transfer (IBTET) and to establish analogous scaling relations with respect to energy. 
To interpret the results of IBTET, we study the nonlinear normal modes (NNMs) of a reduced order model (ROM) of the VI unit cell in the 4-band lattice, using the method of numerical continuation.
 Interestingly, the slope of the frequency-energy branches of the ROM
  corresponding to the 1:1 resonance NNM  matches remarkably well with the dependence of IBTET to input energy in the 4-band lattice. 
 In both phononic lattices, it is shown that there exists a maximum energy transfer at moderate input energies, followed by a power law decay of relative energy transfer either to the wavenumber domain or between bands on input energy; this power law dependence is additionally validated by the ROM.
  Moreover, relations between the dynamics of the VI lattice and the NNMs of the underlying Hamiltonian system provide physical interpretations for the relative energy transfers.
 Hence, we present a predictive framework to computationally explore non-resonant energy transfers across the linear band structure  of phononic lattices with local strong non-smooth nonlinearities and provide a comprehensive physics-based interpretation of these energy transfers based on the nonlinear dynamics of the lower-dimensional ROM.
\end{abstract}

\maketitle

\section{Introduction}
Periodicity has been leveraged to control acoustic and elastic energy propagation in linear time-invariant (LTI) phononic metamaterials~\cite{Cummer2016,Surjadi2019,Hussein2014}. 
Such systems are typically designed on a unit cell level whereby the application of the Bloch theorem allows one to \textit{engineer} a linear band structure which can  enable or augment specified wave phenomena with diverse applications such as lensing~\cite{Tol2019}, energy harvesting~\cite{Chen2014,Qi2016,Carrara2013}, vibration isolation~\cite{Reichl2017,Matlack2016,Mueller2019}, wave steering~\cite{Sirota2021}, mechanical logic circuits~\cite{Xia2018}, mechanical signal processing~\cite{ZangenehNejad2019}, and topological insulation~\cite{Pal2017,Chen2018,Tempelman2022a}. 

For LTI phononic systems, a propagating wave remains stationary on a prescribed subset of its band structure, and is invariant to amplitude (or energy) as the dynamics are completely described by the superposition principle~\cite{Hussein2014}. 
However, it is often desirable to predictively tune wave propagation in phononic materials such that the propagating wave shifts to a different subset of its band structure. 
To this end, one must either  manipulate the underlying band structure altogether by utilizing external forces or nonlinearity~\cite{Hussein2014,Patil2021}, or find methods to modify the distribution of (or, equivalently, passively manage) energy across a fixed underlying band structure. 

Whereas band manipulation has been achieved by introducing e.g., electromagnetic, magnetic, mechanical, or thermal fields~\cite{Vasseur2011,Allein2016,Pierce2020,Cheng2011,Yao2011,Zhao2022}, nonlinear mechanisms offer the key advantage of being passive and tunable (self-adaptive) to energy, frequency and wavenumber content~\cite{Patil2021,Denz2010}.
For instance, the effective dispersion relations of granular chains with Hertzian contact laws are tunable  by locally linearizing about various pre-compression states~\cite{Li2012,Chaunsali2016,Chaunsali2017}. 
Moreover, passive nonlinear mechanisms posses intrinsic frequency-amplitude dependencies, and the corresponding shifts to the band structures can be described by perturbations of the underlying linearized band structure~\cite{Narisetti2010} for low energy or by the nonlinear normal modes (NNMs) at high energy~\cite{Tempelman2021,Jayaprakash2010,Mojahed2019,Mojahed2019a}.
Aside from band structure manipulation, distributed nonlinearity in periodic chains has enabled exotic wave behavior in lattices with no properly defined band structure such as stegetons~\cite{Patil2022}, solitons~\cite{Zaera2018}, and breathers~\cite{Gendelman2008,Gendelman2013}. 

Herein, we aim to develop mechanisms to manipulate propagating energy in phononic metamaterials using \textit{localized} nonlinearities to \textit{transfer} energy across the underlying linear band structure.
In the absence of external actions, the transfer of energy across an underlying linear spectrum requires a nonlinear mechanism which has the capability to transfer energy form one modal subspace to another.
Such a mechanism is fundamental to achieving targeted energy transfer (TET), a concept which has been rigorously studied by the nonlinear dynamics community from the point of view of nonlinear modal dynamics~\cite{Vakakis2022}. 
TET is most commonly achieved by employing localized nonlinear energy sinks (NESs) which alter the global dynamics of a primary linear structure to which they are attached, with typical applications in vibration mitigation~\cite{Wang2015,Vakakis2000,Vakakis2001,Taleshi2016,Starosvetsky2007,Starosvetsky2010,Qiu2017,Huang2019,Georgiades2007,Gendelman2000,Gendelman2001,Gendelman2007,Darabi2016,Dai2017,Bergman2008}.
The TET phenomenon relies on resonance capture of the NES to a resonance manifold, and thus traditional TET is intrinsically suited for systems with smooth nonlinearities and periodic excitations~\cite{Vakakis2022}. However, theoretical and numerical support has recently been extended to systems with non-stationary dynamics~\cite{Manevitch2015} and systems with non-smooth nonlinearities such as idealized vibro-impact (VI) laws~\cite{Nucera2007,Gendelman2012,Li2016}.  

The use of nonlinear attachments in acoustic wave guides (either bulk or periodic) have demonstrated unprecedented properties in acoustical systems~\cite{Gendelman2018}.
For instance, a small mass supported by an essential (non-linearizable) stiffness nonlinearity in parallel to a viscous damper attached to a periodic array of oscillators has been shown to host a rich variety of nonlinear dynamics when interacting with traveling waves~\cite{Rothos2009}, and are even capable of arresting incoming pulses~\cite{Vakakis2014}.
Moreover, with the incorporation of hierarchical mass scales and asymmetry, similar systems have achieved nonreciprocity ~\cite{Nassar2020,Bunyan2018,Fronk2019}. 
These effects have been extended for systems with \textit{local} nonlinear gates that enable global non-reciprocity and effective diode-type features in both continuous waveguides~\cite{Grinberg2018} and discrete oscillator chains~\cite{Fu2018,Darabi2019}.
In addition to reciprocity, the concept of local gates in waveguides has recently been extended to produce effective mechanical filters for layered metamaterials with interfaces~\cite{Devaux2019} and for discrete periodic chains~\cite{Mork2022}. 

Recently, new ideas have emerged in the area of TET which explore \textit{non-resonant} energy exchanges in a directly forced primary linear structure using VI nonlinearity to redistribute modal energy within its modal space, termed \textit{inter modal targeted energy transfer}~\cite{Gzal2020}.  
This methodology was studied computationally in~\cite{Gzal2021} for a discrete mulit-DoF structure, and was later  experimentally verified in~\cite{Tempelman2022} for the case of a cantilever beam undergoing VIs.
Unlike resonant mechanisms, non-resonant mechanisms aim to \textit{scatter} energy across the underlying linear modal basis in a low-to-high frequency fashion. In a similar fashion, Theurich \textit{et al}.~studied the directed scattering of energy to higher modes in a harmonically excited beam, and found that the effectiveness of the energy scatter is dependent on the dynamic regimes of the VI system considered~\cite{Lee2009}. 

To date, non-resonant energy scattering concepts have not been extended to periodic phononic metamaterials from a wave propagation perspective. 
The most notable differences between modal and periodic acoustical systems is that the first employs a modal basis to describe stationary vibrations (and is suitable for systems of finite extent whose dynamics are governed by slow time scales), while the latter a continuous band structure to describe propagating waves (and applies to unbounded / large-scale systems whose acoustics are governed by fast time scales).
Hence, several natural questions arise when considering non-resonant TET phenomena in a phononic material. 
Namely, to what extent can the linear wave propagation be \textit{scattered} in the wave number domain across a dispersion branch, and to what capacity can energy be irreversibly \textit{transferred} from one band to another by use of localized VI nonlinearities.
This paper addresses these questions with extensive computational probing, new post-processing techniques, and physics-based reasoning of the resulting nonlinear acoustic phenomena. 

We begin by studying the effects of VI nonlinearity in a 2-band phononic lattice of diatomic resonators by extensive simulation and numerical post-processing of the acoustics.
For this, we focus on the energy scattered of energy across the frequency/wavenumber domain of the single optical band of this lattice as a function of the number of local VI unit cells and as a function of the incident wave energy grows. Next, we consider a 4-band phononic lattice, which has one acoustic and three optical bands over a relatively broad frequency/wavenumber range. 
This band structure, coupled with the strong VI nonlinearities, allows for low-to-high frequency energy generation of the impacts, as well as targeted energy transfers across bands. This brings about the new nonlinear acoustic phenomenon of  \textit{inter-band targeted energy transfer} (IBTET). 

Accordingly, the organization of this paper is as follows.
Section~\ref{SEC:WavenumberScatter} provides a system description of the unit cell of the 2-band phononic lattice, a computational framework for studying wavenumber scattering within the single optical band induced by the VIs, and quantification of the spectral disorder generated by the VIs with respect to energy.
 Section~\ref{SEC:IBTET} extends the study to a 4-band phononic lattice and presents a method for transferring energy from lower-to-higher optical bands via VIs, together with relationships between these transfers and the total system energy. 
 Section~\ref{SEC:ROM} presents a 2 DoF reduced order model (ROM) which is studied through the from the perspective of NNM analysis in order to provide a physics-based understanding of the results of Sections~\ref{SEC:WavenumberScatter} and~\ref{SEC:IBTET}, and relate the nonlinear dynamics of the ROM to the IBTET occurring in the lattice.  Lastly, Section~\ref{SEC:Conclusions} offers concluding remarks and some suggestions for further extension of this work. 

\section{Wavenumber Energy Scattering}
\label{SEC:WavenumberScatter}
We begin by studying a 1D phononic lattice in the form of a diatomic resonator chain and embed VI contact laws in select (local) resonators while preserving the global linear structure of the lattice. The system is computationally explored by performing numerical simulations with wave packet excitations over an array of excitation amplitudes and wave numbers. The resulting data sets were next post-processed with a suite of discrete signal processing methods in the spatial-temporal domain to uncover the underlying trends of energy scattering in the wavenumber domain as the excitation level (input energy) changes. 
 
\subsection{System Description and Simulations}
\label{Subsec:Sys1}

\begin{figure}[t!]
	\includegraphics[width=\linewidth]{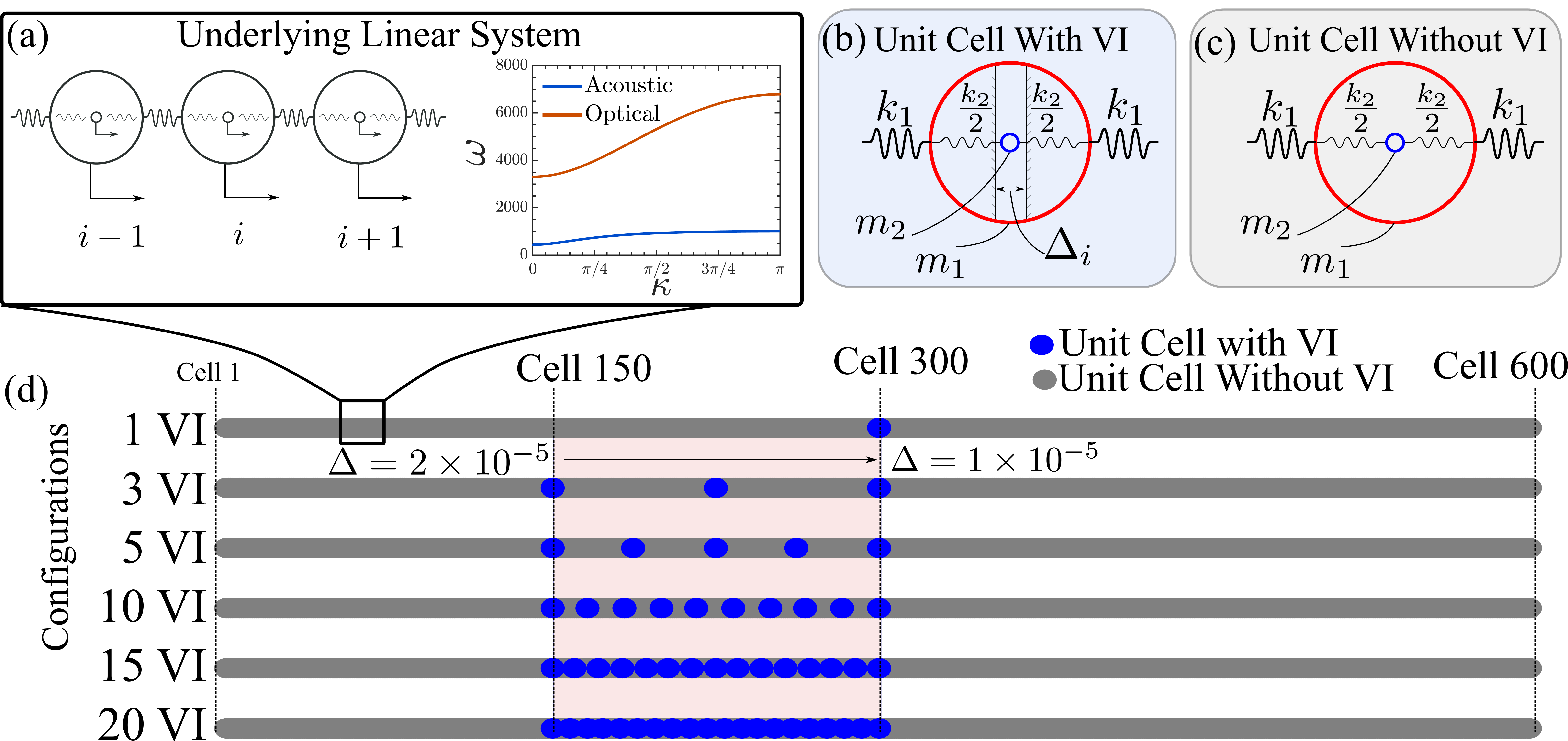}
	\caption{The linear phononic lattice composed of coupled (host) masses with embedded internal resonators which may or may not undergo vibro-impacts: (a) The primary linear periodic system with the underlying linear dispersion relation. The nominal unit cell (b) without a VI nonlinearity and (c) with the VI nonlinearity; (d) schematics of finite lattice configurations which are comprised of the linear phononic lattice with various number of embedded VI unit cells.} 
	\label{Fig:Diatomic_Cell}
\end{figure}

We consider a linear diatomic lattice constructed by the periodic tessellation of 1-D unit cells in the $x$-direction (Fig.~\ref{Fig:Diatomic_Cell}(a)). Each unit cell is composed of a host mass and within it a resonator, which depending on the existence (absence) of rigid barriers it may (may not) experience vibro-impacts (see Fig. 1).
The equations of motion for the $k$-th cell in the \textit{infinite} phononic lattice are written as:
\begin{equation}
	\begin{aligned}
	&m_1\ddot{u}_1^{k}  = k_1(u_1^{k-1}+u_1^{k+1}-2x_1^{k}) +k_2(u_2^{k}-u_1^{k}), \\
	&m_2\ddot{u}_2^{k}  = k_2(u_{1}^{k}-u_2^{k}).
	\end{aligned}
\end{equation}
Imposing the Bloch ansatz, ${\textbf{u}}(x) = \tilde{\textbf{u}}\exp(i\kappa x-i\omega t)$, recovers the linear dispersion derived from the underlying Bloch eigenvalue problem, $\tilde{\textbf{u}}(\tilde{\textbf{M}}\omega^2-\tilde{\textbf{K}}(\kappa))=\textbf{0}$, where $\tilde{\textbf{M}}$ and $\tilde{\textbf{K}}$ are the Bloch-periodic mass and stiffness matrices of a unit cell. This yields two pass bands for this lattice, namely a lower-frequency acoustic band and a higher-frequency optical band.
To computationally probe the effects of impact dynamics on the linear wave propagation, we consider six different lattice configurations, each corresponding to a unique arrangement of VI unit cells embedded in the linear lattice with the number of VIs ranging between 1 and 20.
To study the scattering of the input wave energy in the wavenumber domain accurately, a large finite system should be used for sufficient wavenumber resolution. To this end, we consider a finite configuration of 600 unit cells (1200 DoF) governed by
\begin{equation}
	\textbf{M}\ddot{\bm{u}} + \textbf{K}\bm{u} + \textbf{F}_{\rm NL}(\bm{u},\dot{\bm{u}}) = \textbf{F}_{\rm ext}(t)
	\label{EQ:finite}
\end{equation}
where $\textbf{M}$ and $\textbf{K}$ are the finite mass and stiffness matrices, $\textbf{F}_{\rm NL}(\bm{u},\dot{\bm{u}})$ the vector of nonlinear stiffness and viscous damping terms, and $\textbf{F}_{\rm ext}(t)$ the vector of excitations. 
Excitation is provided in the form of a windowed harmonic function,
\begin{equation}
	{F}_k(t) = \begin{cases} W(t)
	\sin(\Omega t) \ \ \ &k=1\\
	0,\ \ \ &\text{otherwise}
	\end{cases}
	\label{EQ:forcing}
\end{equation}
where $W(t)=A\left[H(t)-H\left(t-\frac{2\pi N_{cyc}}{\Omega}\right)\right]\left[1-\cos\left(\frac{\Omega t}{N_{cyc}}\right)\right]$ is a windowing function, $H(t)$ the Heaviside function, $A$ the forcing amplitude, $N_{cyc}$ the number of cycles in the window, and $\Omega$ the center frequency of excitation.
The nonlinear VI cells that are locally distributed through the lattice provide the following VI forces,
\begin{equation}
	F_{NL}(w_k) = k_c\left[(w_k-\Delta_i)^n_+-(-w_k-\Delta_k)^n_+\right]g(\dot{w}_k,\dot{w}_{k}^-)
	\label{EQ:contact}
\end{equation}
where $w_k(t) =u^k_2(t)-u^k_1(t)$ is the relative deflection between the resonator and its host mass, $n$ the nonlinearity coefficient which is set to $n=3/2$ to emulate Hertzian contact unless otherwise stated, $\Delta_k$ the clearance of the $k$-th VI in the lattice, and $k_c = \frac{2E_{\rm VI}\sqrt{R_{\rm VI}}}{3(1-\nu^2)}$ the stiffness parameter for Hertzian contacts, with $E_{\rm VI}$, $R_{\rm VI}$, and $\nu$ being the modulus, radius, and Poisson ratio of the VI, respectively.  The notation $(\ \ )_{+}$ indicates that only positive arguments are to be considered. We assume an inelastic contact law as derived by Hunt and Crossly~\cite{Hunt1975} which provides a hysteresis dissipation function derived from the work-energy principal in terms of the indentation depth, 
$g(\dot{w}_k,\dot{w}_k^{-}) =\left(1-\frac{3(1-r)}{2\dot{w}_k^-}\dot{w}_k\right)$, where ${\dot{w}_k^-}$ is the velocity $\dot{w}_k$ immediately before impact and $r$ the coefficient of restitution. 
Note that Eq~\eqref{EQ:contact} does not modify the underling linear band structure of the extended lattice. Moreover, for amplitudes such that $w_k<\Delta_k$ for each VI, the wave propagation remains completely linear as no VI experiences contact. 

Numerical simulations of equations~\eqref{EQ:finite} were carried out using the ODE78 routine in MATLAB. The center frequency of the excitation was selected based on the desired excitation wavenumbers, which were considered in the range between $2\pi/9\leq\kappa^\star\leq7\pi/9$ to ensure consistency in observations across the optical band structure; however we focus only on $\kappa^\star =5\pi/9$ and refer the reader to supplemental material for additional results.
 The excitation frequencies were chosen within the optical band to ensure out-of-phase motion between each resonator and host mass  and thus excite the VI (note in-phase motion, characteristic of the acoustic branch, will not excite the VI). 
 Clearances were nominally set to range between 0.0002 and 0.0001 m with a logarithmic dependence on position from the leading VI unit cell to account for the momentum loss of the wave as it passes successively through  VI cells in the lattice. 
 The mass and stiffness of the linear resonator (i.e., in the absence of rigid barriers and VIs - cf. Fig. 1)  were selected to emulate \textit{realistic} resonator systems considered in the literature~\cite{Arretche2018}. Table~\ref{TAB:params1} lists nominal parameters for stiffness, mass, and VI stiffness parameters. Within this framework, an ensemble of simulation data was constructed for 25 logarithmically increasing forcing amplitudes for each configuration and excitation wavenumber considered. 

\begin{table}[h!]
	\caption{Parameters used for the di-atomic resonator chain}
	\label{TAB:params1}
\begin{tabular}{|p{.3in}|p{.3in}|p{.4in}|p{.4in}|p{.3in}|p{.3in}|p{.3in}|p{.35in}|}
	\hline
 $m_1$ [kg]&$m_2$	[kg]&$k_1$ [kN/m]&$k_2$ [kN/m]&	$\nu$&	$r$& 	$R_{\rm VI}$ [m]&	$E_{\rm VI}$	[MPa]	\\ \hline
 0.01 & 0.08	&$90$		&$90$&			0.3&	0.7& 	0.005		&	$200$\\
	\hline	
\end{tabular}
\end{table}

\subsection{Influence of VIs on Wave Propagation}
\label{Subsec:Kscatter}
\begin{figure}[t!]
	\includegraphics[width=\linewidth]{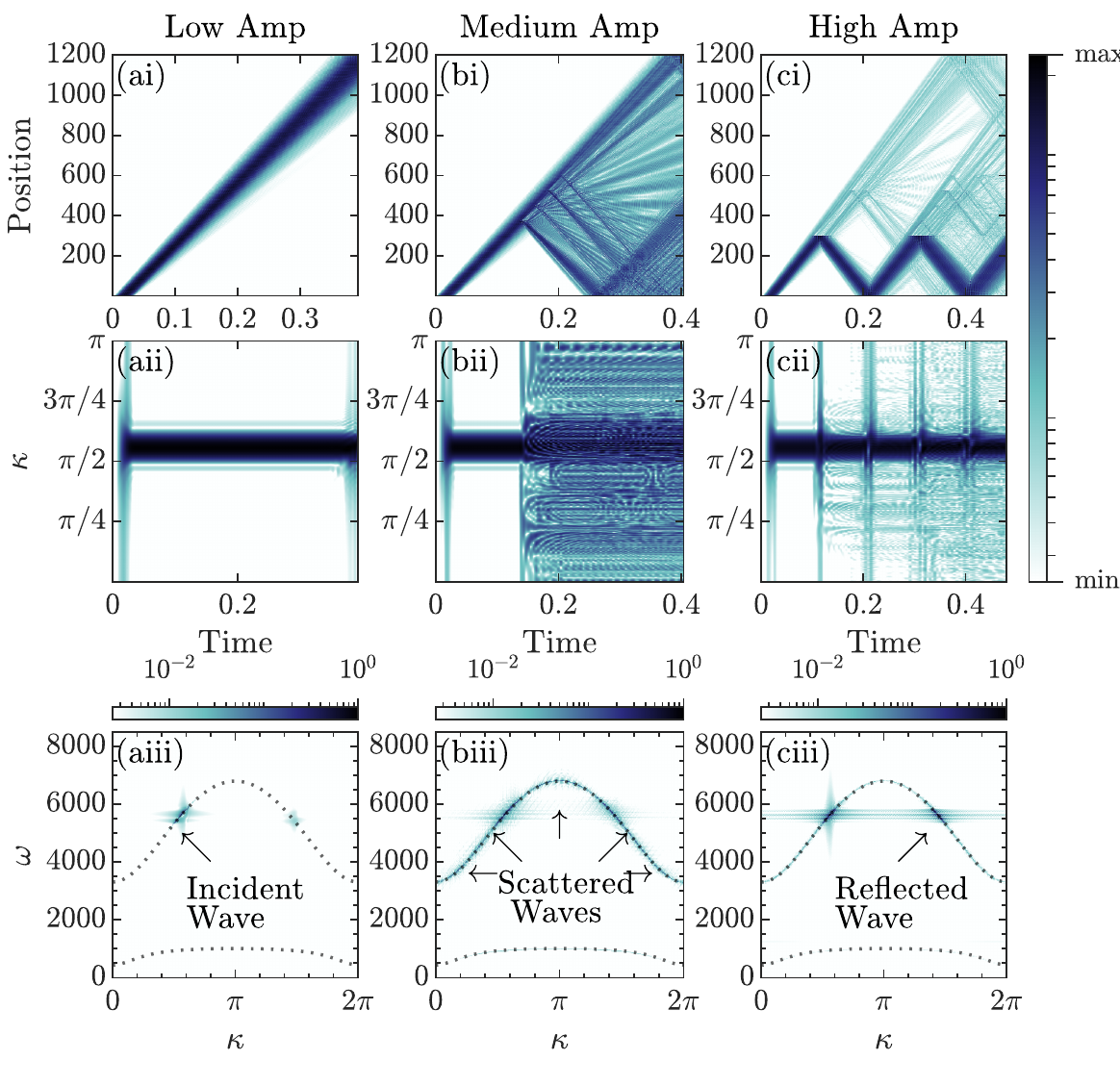}%
	\caption{Simulation results for a 5-VI configuration at excitation wavenumber $k^{\star}=5\pi/9$ (in the optical band of the linear lattice) with columns corresponding to (a) low, (b) medium, and (c) high amplitude excitations. For each amplitude, the rows depict (i)the spatio-temporal evolution of the kinetic energy of the propagating wave, (ii) the temporal variation of the wavenumber distribution in the lattice, and (iii) the numerically computed dispersion computed using the entirety of the simulation with a gray dashed line superimposed to depict the analytical dispersion of the infinite liner lattice.}
	\label{FIG:scattersims}
\end{figure}

\begin{figure}[t!]
	\includegraphics[width=\linewidth]{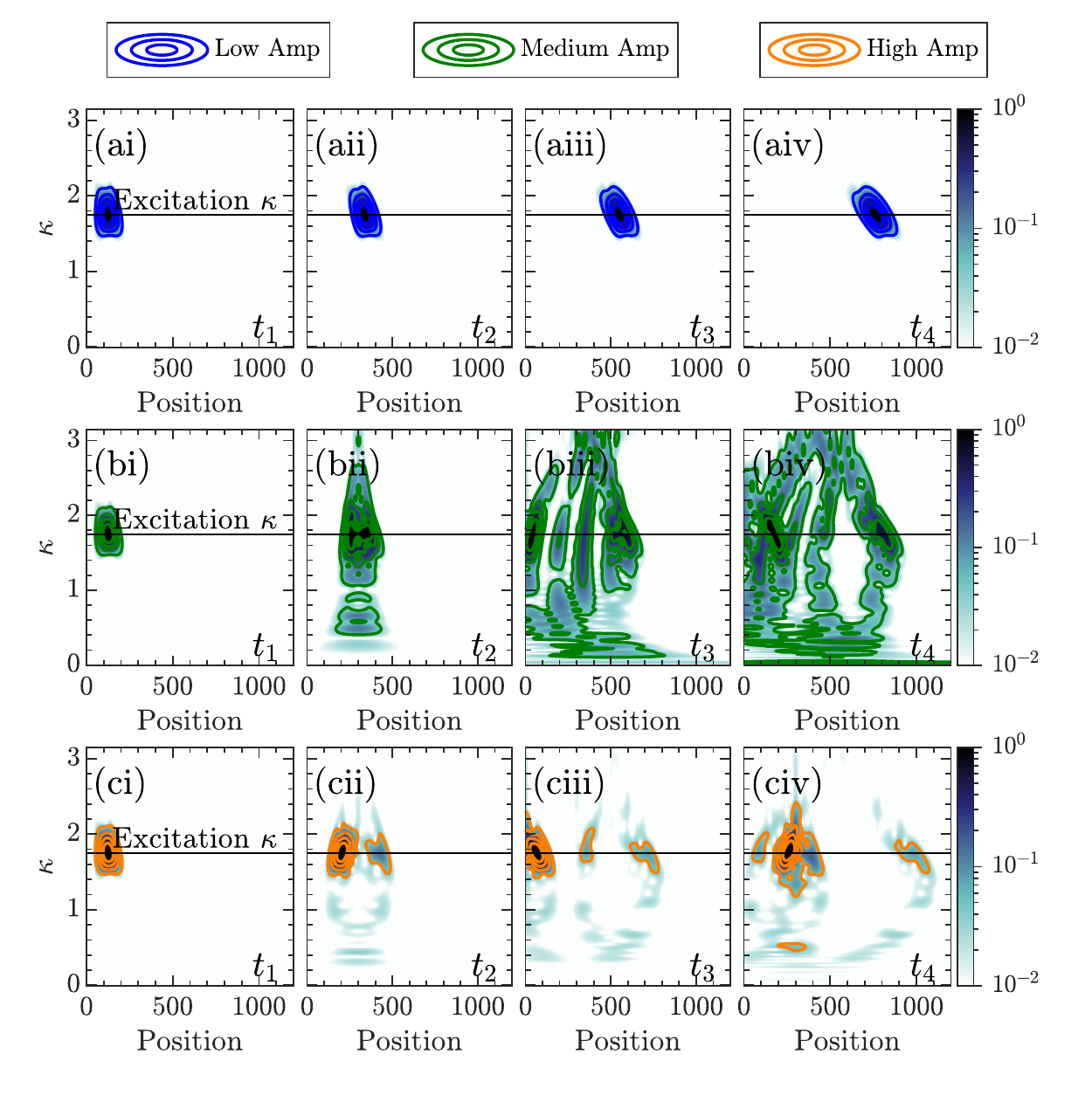}
	\caption{The spatial wavelet transformations of the propagating waves considered in Fig.~\ref{FIG:scattersims} for (a) low, (b) medium, and (c) high excitation amplitude; four time snap-shots are depicted as (i)-(iv), and the center black line depicts the wavenumber corresponding to the excitation frequency as given by the linear dispersion relation.} 
	\label{Fig:WaveletWnum}
\end{figure}

A suite of numerical post-processing tools were developed to study the influence of the VIs on wave propagation in the lattice.
The focus of the post processing was to uncover spectral content in the spatial and spatial-temporal domains with an emphasis on frequency/wavenumber scattering of the energy. 
This was achieved using Fourier and Wavelet transformations to study the energy content across the band structure in various domains including time, space, frequency, and wavenumber. 
In this section, we focus on a  narrow subset of three simulations conducted at low, medium and high forcing amplitudes in order to build intuition on the post-processing analysis procedures and a qualitative dependence on system energy. Quantitative results across all simulations will be given subsequently. 

 Fig.~\ref{FIG:scattersims} depicts the results for a representative simulation with a 5-VI configuration (cf.~Fig.~\ref{Fig:Diatomic_Cell}) for low, medium, and high forcing amplitude (equivalently low, medium, and high energy simulations) corresponding  $A=0.1, 1$, and 10 N, respectively.
 The resulting energy measures are computed directly by considering only the kinetic energies of the oscillators, which is a reasonably sufficient measure of the total energy distribution as elastic systems undergo continuous transfers from kinetic to potential energy. 
 At low amplitude, the acoustics are entirely linear as the wave does not create deflections greater than the VI clearance (Fig.~\ref{FIG:scattersims}(ai)). The interactions of the VI mechanisms come about in the medium and high amplitude simulations, whereby the energy of the propagating wave wave scatters profoundly in the space/time  domain
 (Figs.~\ref{FIG:scattersims}~(bi,ci)). 
 
In the following exposition we provide the results of post processing analysis of the measured responses of the lattices, with the aim to understand of how the VIs scatter the energy of the propagating wave in the frequency/wavenumber domain. To this end, we utilize a set of signal processing procedures which are briefly detailed in Appendix~\ref{SEC:SigProc}.
Figs.~\ref{FIG:scattersims}(aii)-(cii) depict the wavenumber distributions across the lattice computed over progressions of time snap shots for each simulation. Given the total collection of simulation data over time and space to be the matrix $\textbf{u}(x,t)$, the wavenumber domain at a given time snap shot, $t_j$, is given as $K(\kappa) = \mathcal{F}^{x}\{\textbf{u}(x,t)|_{t=t_j}\}$ where $\mathcal{F}^{x}\{\ \}$ denotes the Fourier transformation with respect to the variable $x$.
 It is clear from Figs.~\ref{FIG:scattersims}(aii)-(cii) that the linear system (corresponding to low excitation amplitude) does not affect the wavenumber distribution after excitation ends, as expected for a LTI system. In contrast, new wave numbers emerge for medium and high excitation amplitudes.
  However, for the case of high energy level, the wavenumber generation is not nearly as pronounced compared to medium energy level, indicating that the wave reflections of Fig.~\ref{FIG:scattersims}(ci) do not generate substantial  wavenumber components beyond that of the incident wave. 

 Taking the Fourier transformation across both time and space  provides the numerically resolved dispersion $\mathcal{D}(\kappa,\omega) = \mathcal{F}^{x,t}\{\textbf{u}(x,t)\}$ which is given in Figs.~\ref{FIG:scattersims}(aiii)-(ciii). Note that Figs.~\ref{FIG:scattersims}(aiii)-(ciii) consider the entire time record of the simulation from start to finish. Fig.~\ref{FIG:scattersims}(aiii) may serve as a reference since no VIs engage in the low amplitude simulations, and it is seen that only a narrow subset of the dispersion branch is energetic, corresponding directly to the excitation wavenumber.
 In the nonlinear regimes, the scattering of the energy in the $\omega$-$\kappa$ domain is  much more profound for medium energy cases, corroborating the trends established by Figs.~\ref{FIG:scattersims}(i,ii). 
Note that the spectral content generated by scattering  in Fig.~\ref{FIG:scattersims}(biii) remains bound to the  underlying linear dispersion relation.
Given that the VI nonlinearity represents a nonresonant energy scattering mechanism, this indicates that the VIs "redistribute" (scatter) wave energy across the dispersion relation of the underlying linear lattice rather than modify the dispersion altogether; this acoustical nonlinear scattering effect is directly equivalent to the nonresonant scattering mechanisms studied in modal dynamics~\cite{Tempelman2022}. 
 
Information regarding the spatial evolution of the generated wavenumber components over space and time requires a space-frequency analysis routine. To this end, we employed the continuous wavelet transformation (CWT) using the Morelet wavelet in the spatial dimension to resolve at each time snap-shot, $t_j$, a 2-D map of the wavenumber spectrum with respect to space, $X(\kappa,x) = \mathcal{W}\{\textbf{u}(x,t_j)\}$.
Fig.~\ref{Fig:WaveletWnum} depicts the evolution of the spatial wavenumber distribution tracking $X(\kappa,x)$ through four time snap-shots ($t_1$-$t_4$) for low, medium, and high amplitude simulations. 
From this, it is clear that the scattering of energy is relatively uniform with respect to wavenumber, and that the spectral energy scatters to both higher and lower wave numbers (as is also confirmed in Fig.~\ref{FIG:scattersims}(ii)). 
Moreover,  the VI-generated wavenumber components arise for both the transmitting and reflecting waves at the VI interface for medium amplitude excitations, whereas high-energy waves seemingly reflect a majority of the incident energy off the VI unit cell at the incident wavenumber. Lastly, it is apparent from Fig.~\ref{Fig:WaveletWnum}(b) that certain wavenumber components propagate much faster than others and all follow behind the incident wavenumber; this is a direct consequence of the dispersion relation of the underlying linear system (cf.~Fig.~\ref{Fig:Diatomic_Cell}) which is steepest towards the center of the optical band and therefore corresponds to larger group velocity at the incident wavenumber. Note that this is of course not the case when the excitation  wavenumber is low or high on the band, as the group velocity of the incident wave would invariably be smaller for these excitations. However, the general trends of spectral generation with respect to energy are consistent nevertheless (see supplemental information). 

\begin{figure}[t!]
	\includegraphics[width=\linewidth]{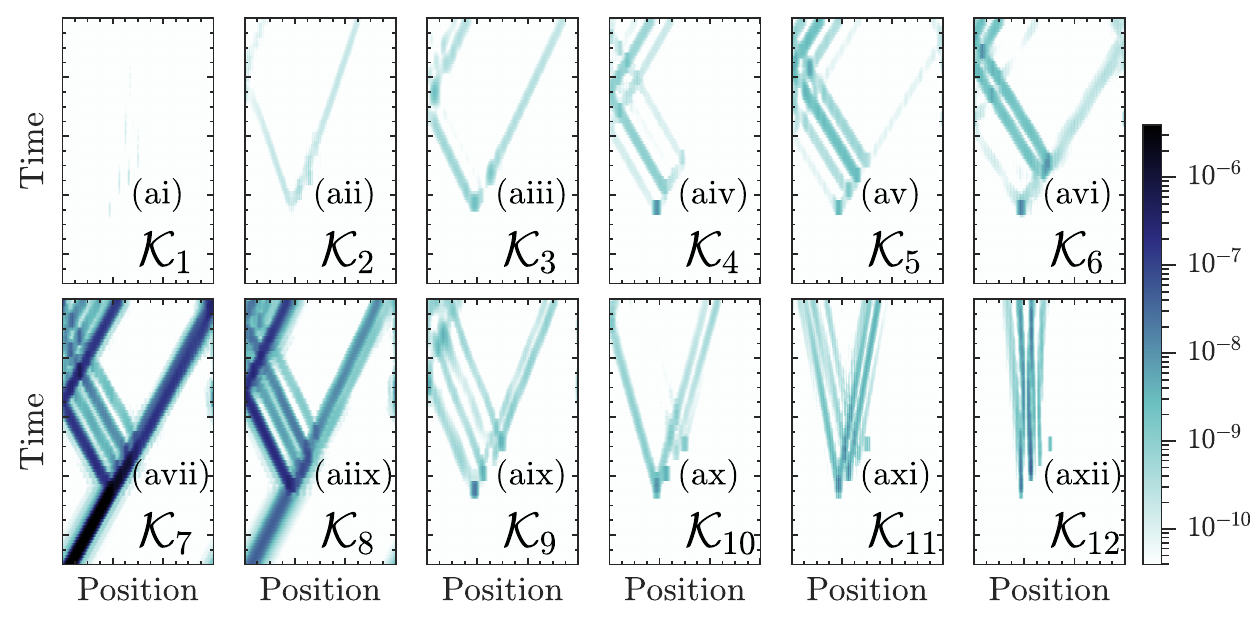}
	\includegraphics[width=\linewidth]{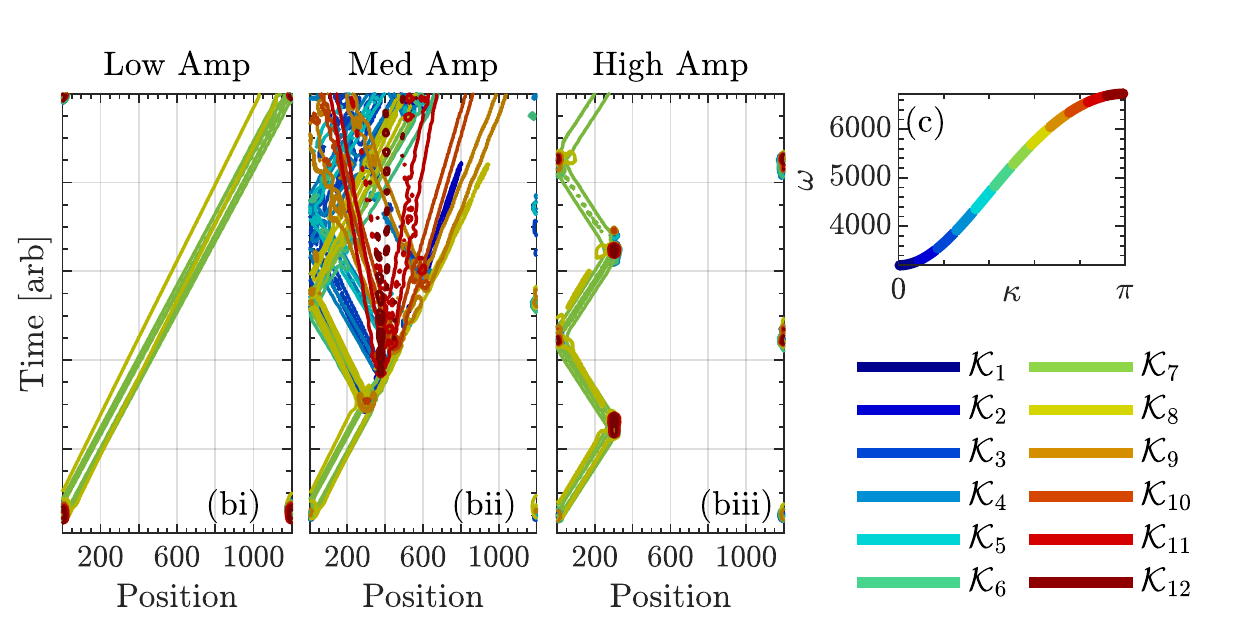}
	\caption{Propagation of wave energy at different wavenumber bands: (a) The kinetic energy versus time at each wavenumber partition for a mid-energy simulation with sub-panels (i)-(vii) plotted to the same color-scale to compare relative energies; (b) superimposition of wave propagation at each wavenumber partition depicted by contours for (i) low, (ii) medium, and (iii) high energy simulation; (c) the optical band of the linear lattice plotted with corresponding colors to the wavenumber-based energy contours of (b).}
	\label{FIG:Kpartitions} 
\end{figure}%

The spectral content of Fig.~\ref{Fig:WaveletWnum} can be mapped-back into the spatial-temporal domain by considering a spectral partitioning scheme similar to that presented in~\cite{Mojahed2021}.  The goal is to \textit{visualize} the propagation of the wave specific to different partitions of the optical band, and thus confirm that wave propagation at new wavenumbers occurs due to VI interactions.
To achieve this, the instantaneous velocities and positions over various regions of the band structure can be resolved by partitioning the wavelet space into 12 wavenumber bins and taking the inverse wavelet transform of each bin independently. 
If the 
spatial wavelet-transformed data at a time instant $t_j$ is denoted as $X(\kappa,x)\big|_{t = t_j}$, and the inverse wavelet transformation is denoted as $\mathcal{W}^{-1}$, then the dynamics of each of the optical band, $\mathcal{K}_1$-$\mathcal{K}_{12}$, are computed as the collection of binned inverse transformations of binned wavelet data over time:
\begin{equation}
	\begin{aligned}
	&\mathcal{K}_1(x,t) = \bigcup_{j}\mathcal{W}^{-1}(X(\kappa,x))\big|_{t = t_j},  \ \ 0\leq\kappa\leq\frac{\pi}{12} \\
	&\vdots \ \ \ \ \ \ \ \ \ \ \ \ \vdots\ \ \ \ \ \ \ \ \ \ \ \ \vdots\\
	&\mathcal{K}_{12}(x,t) = \bigcup_{j}\mathcal{W}^{-1}(X(\kappa,x))\big|_{t = t_j},  \ \ \frac{11\pi}{12} \leq\kappa\leq\pi.
	\end{aligned}
\end{equation}
 The kinetic energy can subsequently be computed for each spatial-spectral partition, which cannot be achieved directly in the frequency domain due to the mass dependency of the kinetic energy. Summing the energy components of each of the spectral partitions results in negligible error (1\% or less) compared to the energy computed directly from physical coordinates with no numerical integral transformations (see supplemental material), thus verifying the efficacy of the post-processing technique.
More importantly, as discussed below, the described numerical partition of the optical band enables us to study in detail the transmission of wave energy at different wavenumber bands, and, hence, can shed insight into the nonlinear physics of the scattering of the incident wave at the VI sites. 

Fig.~\ref{FIG:Kpartitions} depicts the results of the wavenumber partitioning scheme. The propagation of energy across each wavenumber partition are given by subplots~\ref{FIG:Kpartitions}(ai)-a(vii) and plotted to the same color scale in order to compare the relative energies of each wavenumber partition. The wave initiates in $\mathcal{K}_7$ and $\mathcal{K}_8$ as these posses energy from the onset of propagation while all other wavenumber partitions are dormant during the start of propagation. However, after the VIs are engaged midway through the lattice, energy begins to propagation through all partitions, and this is clear indication that the VI nonlinearity in fact generates wave propagation at wavenumbers not native to the excitation profile. To demonstrate the dependency on energy, Fig.~\ref{FIG:Kpartitions}(b) shows the wave propagation through each wavenumber band superimposed by contours for low, medium, and high profile wavenumber from which it is apparent again that wavenumber generation is far more potent at medium amplitude simulations than for high ones. Fig.~\ref{FIG:Kpartitions}(c) provides a colored depiction of the optical band to make the contours of  Fig.~\ref{FIG:Kpartitions}(b) more obvious with respect to which wavenumber components are generated; the of group velocities in Fig.~\ref{FIG:Kpartitions}(c) corresponds directly to the variable wave speeds of  Fig.~\ref{FIG:Kpartitions}(b), and this can be used to interpret the variation in spatial-spectral propagation of Fig.~\ref{Fig:WaveletWnum} as well.

\subsection{Quantifying Wavenumber Spectrum Disorder}
\label{Subsec:Kentropy}

Section~\ref{Subsec:Kscatter} established that (i) the VIs generate propagating  waves at new wavenumbers as they interact with the incident wave, and (ii) that this phenomenon is dependent on amplitude. To this point, the results have been presented in a largely qualitative manner with an emphasis on graphical interpretations (cf.~Figs.~\ref{FIG:scattersims},\ref{Fig:WaveletWnum},\ref{FIG:Kpartitions}).
We now aim to \textit{quantify} the wavenumber scattering induced by the VIs  for wave transmission over the entire domain of the lattice, based on an ensemble of simulations, in order to establish the dependence of VI induced wavenumber scattering on input amplitude. 

To this end, we make use of information theory by considering the spectral entropy of the nonlinear acoustics in the wavenumber domain.
Spectral entropy is the extension of classical Shannon entropy to the frequency domain~\cite{Boashash2013} and is a standard metric for quantifying signal complexity. We consider the wavenumber entropy generated over space at a given time snap shot as
\begin{equation}
	H(x) = - \sum_{\kappa}P(x,\kappa)\log_2P(x,\kappa),
	\label{EQ:Hx}
\end{equation}
where $P(x,\kappa) = {\mathcal{S}(x,\kappa)}\big/{\sum_{\xi}\mathcal{S}(x,\xi) }$ is the space-dependent probability distribution over wavenumber computed with the space-frequency power spectrogram $\mathcal{S}(x,\kappa)$. By computing $P(x,\kappa)$ over a progression of time snapshots, $t_j$, for each simulation, a matrix of entropy-versus-time, $\mathcal{H}(x,t)$, captures the time-evolution of wavenumber entropy as the wave propagates through the lattice.
\begin{figure}
	\includegraphics[width=\linewidth]{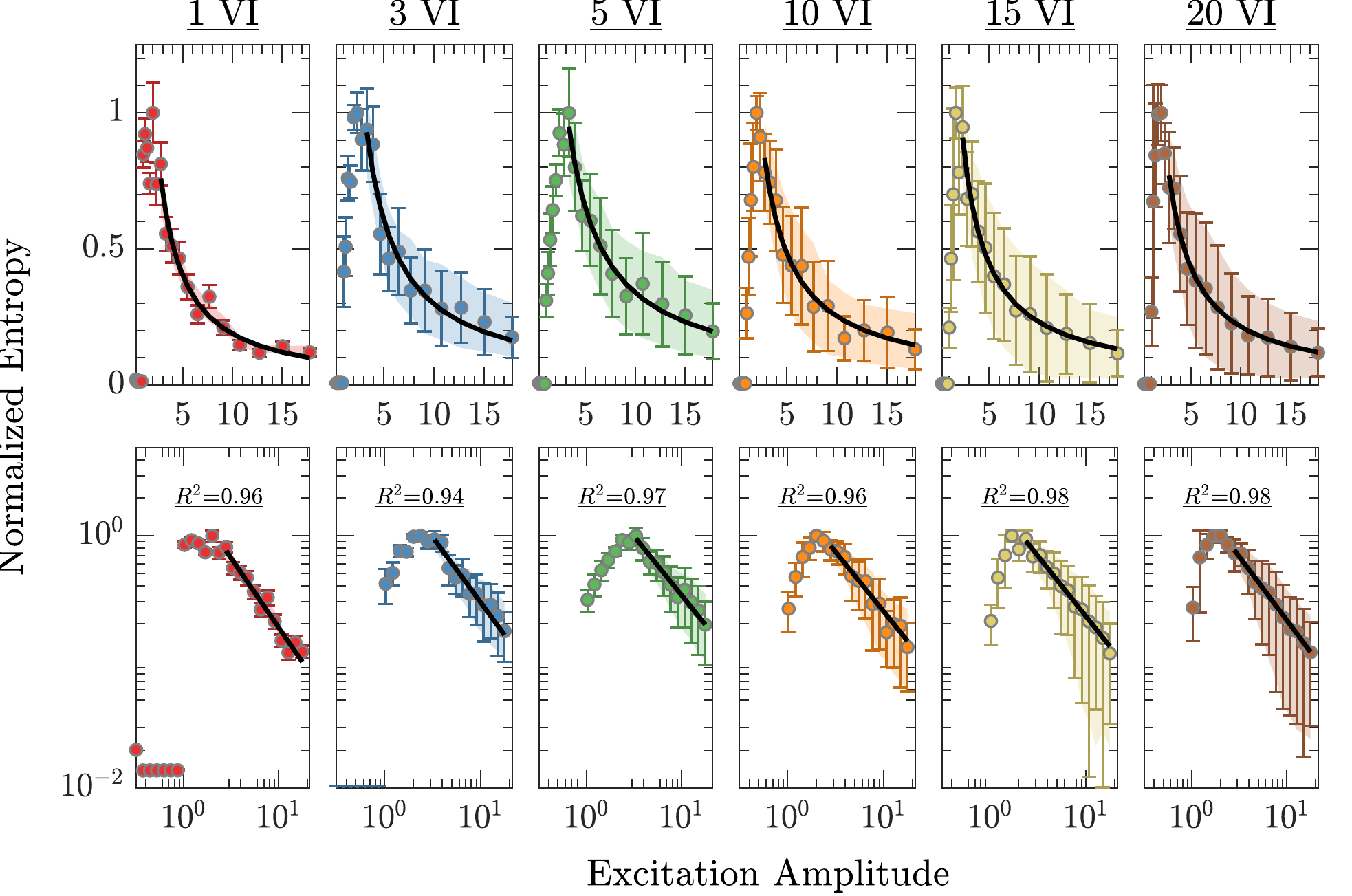}
	\caption{Mean spectral entropy in the lattice with VIs for system configurations ranging between 1 VI to 20 VI (see Fig.~\ref{Fig:Diatomic_Cell}) over an array of excitation amplitudes logarithimcally spaced from 0.1 to 20: Top and bottom plots are for the same data with the bottom plots depicting the log-log scaling; a fitted power law is denoted as a thick black line, and the adjusted $R$-squared value is listed for each configuration in the bottom plots.} 
	\label{FIG:Entropy}
\end{figure}
We compute a statistical summary of the wavenumber entropy by considering the elements of  $\mathcal{H}(x,t)$ for time intervals after the incident wave has already reached the first VI unit cell at $t = \hat{t}$. Fig.~\ref{FIG:Entropy} depicts the normalized average entropy quantity with respect to forcing amplitude for all configurations depicted in Fig.~\ref{Fig:Diatomic_Cell}. 
Normalization was performed so that the minimum and maximum entropy for each VI configuration range between 0.01 and 1. To this effect, we are capturing the \textit{relative} scattering of wavenumbers as compared to an \textit{optimal} excitation amplitude (specific to our selected configuration).
 At the lowest forcing excitation level (with no VI engagement) the wave propagation remains linear,  and so the entropy remains nearly zero as the only variation in the wavenumber comes from the intrinsic dispersive characteristics of the lattice. 
 However, once the VIs are engaged at medium and high excitation levels, the entropy rapids rises and reaches a maximum before rapidly falling  again with respect to forcing amplitude. 
 The log-log plots of Fig.~\ref{FIG:Entropy} reveal that after the maximum entropy is reached, the remainder of the data fits remarkably well with a power law with adjusted $R$-squared coefficients above 0.95 being recovered for the majority of configurations studied. Error bars in Fig.~\ref{FIG:Entropy} measure the standard deviation of entropy across the spatial extent of the lattice. This can be interpreted as a measure of how uniform the wavenumber complexity is.
 Hence, the larger error bounds at high excitation amplitudes indicate that novel wavenumber components are localized rather than distributed (or propagated) throughout the spatial extent of the lattice, and this is in direct agreement with the qualitative results of Figs.~\ref{FIG:scattersims},~\ref{Fig:WaveletWnum}, and~\ref{FIG:Kpartitions}.
 Note that the excitation wavenumber is $\kappa=5\pi/9$ for all results shown in Fig.~\ref{FIG:Entropy}; additional results given in the supplemental material confirms that the same trends hold across all incident wavenumbers. 

\section{Inter-band Targeted Energy Transfers (IBTET)}
\label{SEC:IBTET}

With section~\ref{SEC:WavenumberScatter} establishing that the VI nonlinearities can scatter energy about the optical band of a diatomic lattice, a natural next question is to what effect VI mechanisms can induce targeted energy across different bands. 
This can be considered as an acoustics-equivalent to the IMTET nonlinear mechanism established in dynamics~\cite{Gzal2020}. 
Hence, the aim of this section is to achieve \textit{inter-band targeted energy transfers} (IBTET) by irreversibly transferring energy from a lower optical band to a higher band.
Moreover, we aim to demonstrate that this phenomenon is achievable for multiple classes of VI contact laws, and introduce a bilinear version of the VI law considered previously, to be studied alongside the Hertzian model of Section~\ref{SEC:WavenumberScatter}.
This is considered in order to demonstrate that the subsequent IBTET results are reproducible for different classes of contact nonlinearity and are not particular to the Hertzian contact law utilized in section~\ref{SEC:WavenumberScatter}, hence opening a broader design space to realize the phenomenon in practice.
 To achieve IBTET requires a system with more than 2 DoF per unit cell, since the number of optical bands amenable to out-of-phase motion, and thus with the ability to interact with the VI, is dictated by $N_{\rm optical} = N_{\text{DoF}} - D$ where $ N_{\text{DoF}}$ is the degrees of freedom in the unit cell and $D$ is the unit cell dimension. Hence, to maintain the simplicity of 1D, we proceed with a 4-DoF model of the unit cell, offering two additional bands to transfer energy towards. 

\subsection{The 4-band Lattice}
\begin{figure}[t!]
	\includegraphics[width=\linewidth]{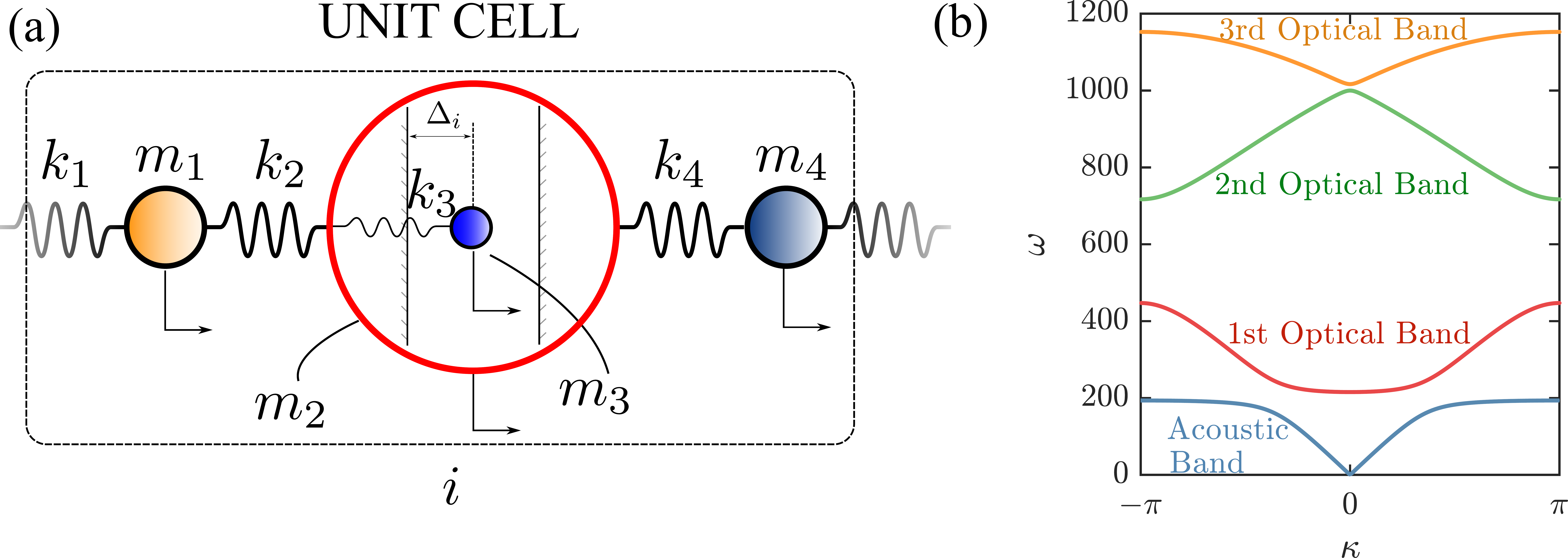}
	\caption{Increasing the bands of the lattice: (a) Schematic of the unit cell,  and (b) the corresponding dispersion diagram for parameters $\lambda = 0.1$ and $\eta = 0.5$.}
	\label{FIG:fourband}
\end{figure}
\begin{figure*}[t!]
	\includegraphics[width=\linewidth]{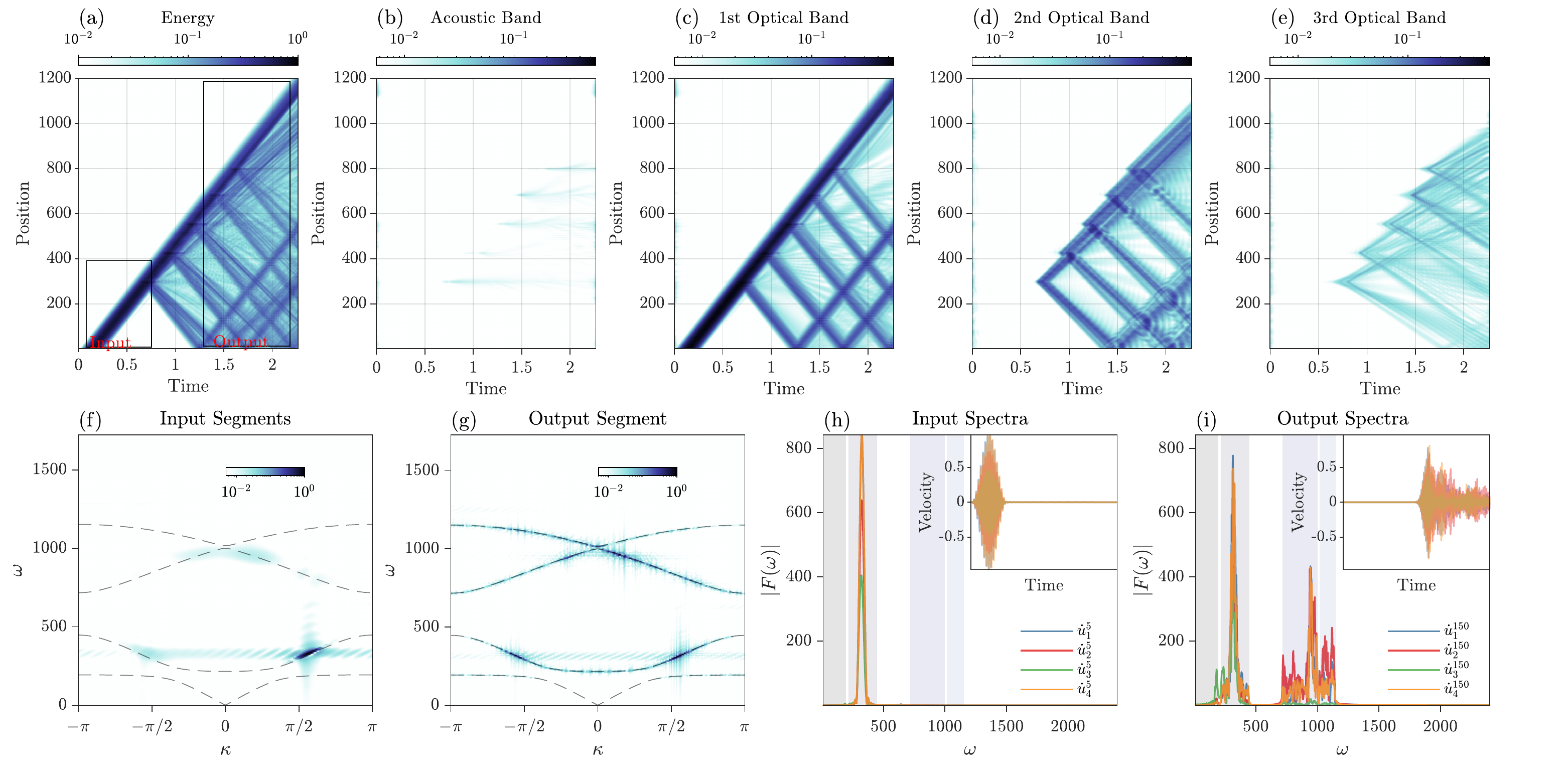}
	\caption{IBTET in the 4-band lattice with 5 VI sites: (a) shows the evolution of the propagating wave energy; (b-e) propagation of the wave energy corresponding to each band of the lattice based on the numerically recovered dispersion of the full simulation; (f,g) dispersion of the input and output segments (labeled in (a)) demonstrating the targeted energy transfer to the higher bands; (h,i) Fourier spectra corresponding to the velocity of the four unit cell DoFs selected before (5-th unit cell) and after (150-th unit cell) VI engagement, with the four band-pass regions depicted with shading and insets depicting the corresponding velocity time histories. }
	\label{FIG:Hertz_Jump}
\end{figure*}

The 4-band model emulates closely the resonator model of Fig.~\ref{Fig:Diatomic_Cell}. The main difference is that masses have been added in-series in between resonators as shown in Fig.~\ref{FIG:fourband}(a). The equations of motion for a unit cell of the \textit{infinite} 4-band phononic lattice read,
\begin{equation}
	\begin{aligned}
		&m_1\ddot{u}_1^k + k_4\left(u_1^k-u_4^{i-1}\right)+k_1\left(u_1^k-u_2^k\right) 				= 0\\
		&m_2\ddot{u}_2^k +k_2\left(u_2^k-u_1^k\right) +k_3\left(u_2^k-u_3^k\right) \\ 	&+k_4\left(u_2^k-u_4^k\right) 	 +f_{\rm NL}(w^k)	= 0\\
		&m_3\ddot{u}_3^k + k_3(u_3^k-u_2^k) -f_{\rm NL}(w^k)	= 0\\
		&m_4\ddot{u}_4^k +k_1\left(u_4^k-u_1^{i+1}\right)+k_4\left(u_4^k-u_2^k\right)			= 0
	\end{aligned}
\label{EQ:4band}
\end{equation}
which produces a 4-band dispersion relation upon application of the Bloch theorem. To maximize the potential for IBTET, the parameters of system~\eqref{EQ:4band} should be selected to satisfy the following criteria:
\begin{itemize}
	\item The displacements of the host-mass and resonator of the VI oscillator ($u_2$ and $u_3$) should be out-of-phase on the second band so that strong engagement of the VI nonlinearity can occur beneath the 2nd and 3rd optical bands (since VIs transfer energy from low-to-high frequencies~\cite{Tempelman2022}).
	\item The quantity $|\hat{w}|=|\hat{u}_3(\kappa)-\hat{u}_2(\kappa)|$ describing the resonator deflection across the second Bloch-eigenmode should be maximized over $\kappa$ on the second band.
	\item The group velocity corresponding to the second band should be as high as possible in order to minimize the dispersive effects originating from the linear band structure.
	\item The group velocities of the third and fourth bands should be maximized so as to maximize the corresponding band slopes and equivalently broaden the bandwidth that is amenable for TET from the second band.
\end{itemize}

System~\eqref{EQ:4band} is parameterized by $\eta$ and $\lambda$ which relate the mass and stiffness of the resonator cell to the nominal parameters of $m_1=m_4= m = 0.005$ kg and $k_1=k_4=k = 2\times10^4$ N/m by $m_2 = m(1-\eta)$, $m_3 = m\eta$, and $k_3 = k\lambda$ while we fix $k_2 = 10^4$ N/m. 
With these variables, the desired dispersion characteristics can be readily achieved by considering a cost-function of the form
\begin{equation}
	\begin{aligned}
	&\max_{\eta,\lambda} \left[\sum_{k=2}^4|v_g^k|\right]w, \\
	&\text{s.t.}\ \ \tilde{u}_2(\kappa)\tilde{u}_3(\kappa) <0 \ \ \forall \kappa.
	\end{aligned}
\end{equation}
We confine this search for $0.1<\lambda<1$ and $0.1<\eta<1$. With this constraint, minimizing the cost function over $(\lambda,\eta)$ is trivial and returns $\lambda= 0.1$ and $\eta = 0.5$. The resulting band structure is shown in Fig.~\ref{FIG:fourband}(b).

To simulate the system, a  finite lattice of 300 unit cells (1200 DoF) was constructed, which is one half of the total DoFs  of the resonator chain studied in section~\ref{SEC:WavenumberScatter}. 
Accordingly, we consider only a 5-VI lattice configuration (as depicted in Fig.~\ref{Fig:Diatomic_Cell}(d)) herein and refer the reader to supplemental material for the results of a 1-VI lattice configuration.
Simulations were performed similarly to section~\ref{SEC:WavenumberScatter} with
excitation provided by a windowed tone burst (Eq~\eqref{EQ:forcing}).
An input signal of 30 periods was considered, and the excitation frequency is selected based on the maximum group velocity of the optical band. Simulations were performed for 50 selections of the excitation amplitude between 1 and $10^4$ N.  

We employ the same Hertzian contact law described by Eq~\eqref{EQ:contact} for $n=3/2$, and also a bilinear contact law which takes the same form as Eq~\eqref{EQ:contact} but for $n=1$. This is performed to ensure that the subsequent results are not particular to nonlinear Hertzian contact laws but are rather a product of the contact nonlinearity.
 For the 4-band system considered, the contact stiffness parameters ($k_c$) were computed based on $E = 100$ MPa, $\nu= 0.3$, and $R_{\rm VI} = 0.005$ m, and the clearances are now varied between $10^{-2.65}$ and $10^{-2.75}$ m.

\subsection{Low-to-high band targeted energy transfer}

Fig.~\ref{FIG:Hertz_Jump} depicts an example of a wave propagating through the 4-band system with five Herzian VIs engaged. Energy clearly cascades from the main wave packet as it propagates through the lattice (Fig.~\ref{FIG:Hertz_Jump}(a)), similar to the diatomic chain (Fig.~\ref{FIG:scattersims}).
Computing the numerical dispersion at the beginning and end of the simulation clearly shows that energy in fact transfers from the lowest optical band to the higher two optical bands (Figs.~\ref{FIG:Hertz_Jump}(f,g)). This is further confirmed by Figs.~\ref{FIG:Hertz_Jump}(h,i) which shows the difference in the temporal frequency of the wave at the start versus end of the lattice and hence the low-to-high frequency targeted transfer of energy from the second band to the higher bands. 

Energy transfer between bands can be quantified by first converting the numerically measured data into the $\omega$-$\kappa$ domain with the 2-D Fourier transformation. Thereafter, the 2-D spectrum is partitioned band-by-band and also into band-gap regions. For each partition, the remainder of the spectrum is zero-padded before the inverse Fourier Transformation returns the spectral content into the spatio-temporal domain for that specific partition. This results in the propagation depicted in Figs.~\ref{FIG:Hertz_Jump}(b-e) where it can be seen that the content of the upper bands indeed corresponds to propagating waves generated by the VIs, and thereafter kinetic energy calculations over each band can be conveniently performed. 

\begin{figure}[t!]
	\includegraphics[width=\linewidth]{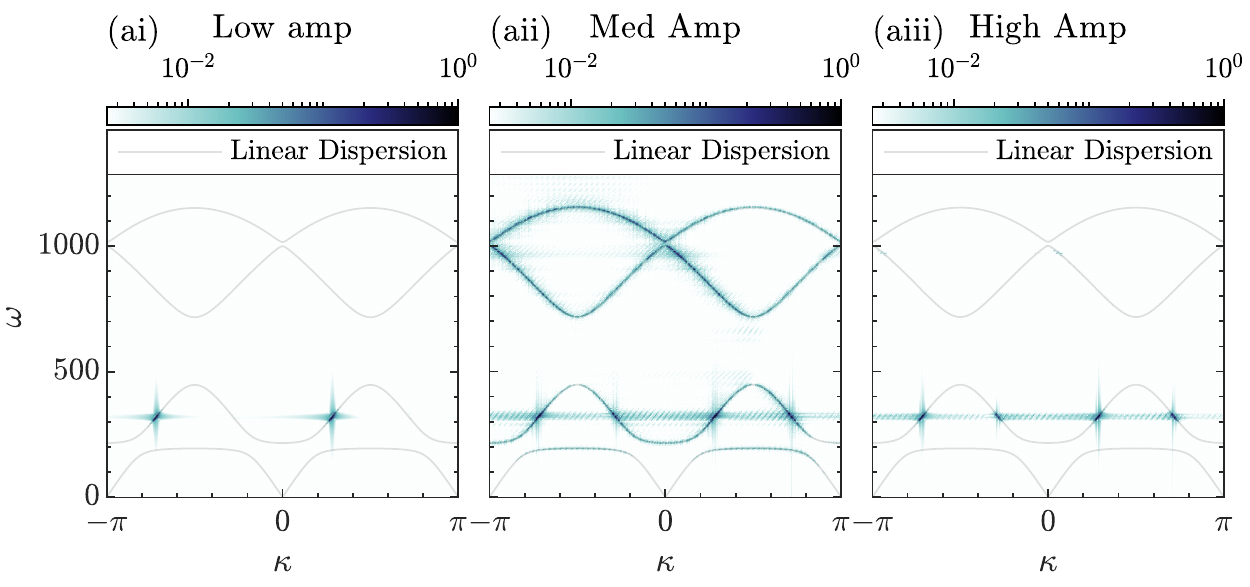}
	\includegraphics[width=\linewidth]{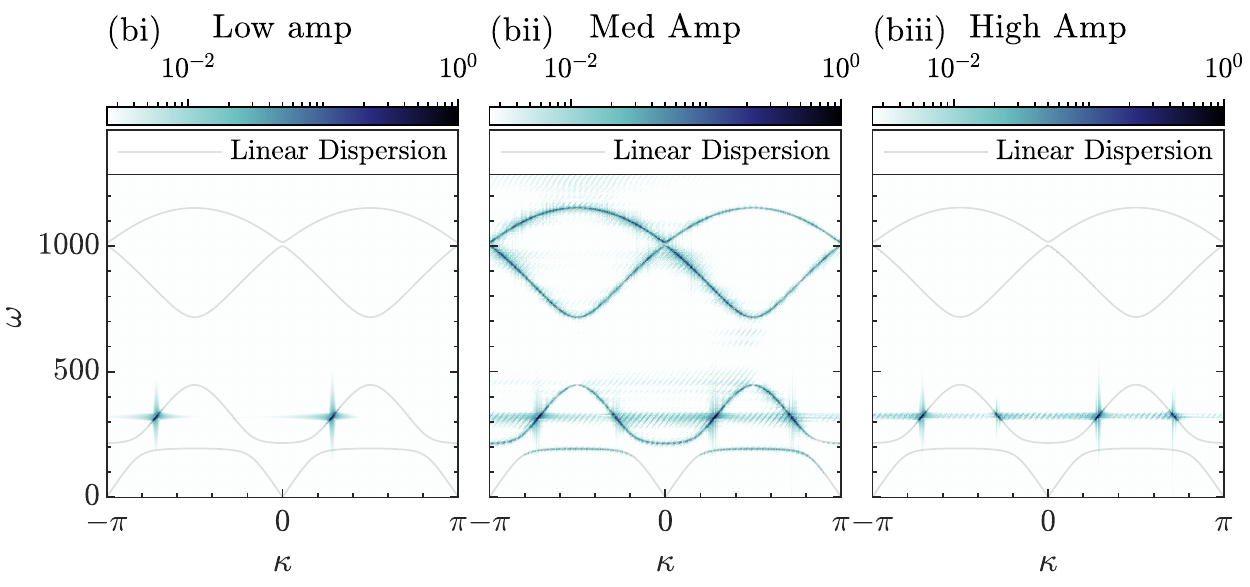}
	\caption{IBTET depicted in terms of the dispersion of the wave in the frequency/wavenumber domain for the 4-band lattice with 5 VI sites over the entire duration of the simulation for (a) Hertzian and (b) bilinear VI laws, and for (i) low, (ii) medium, and (iii) high excitation amplitudes.}
	\label{FIG:FourBandDisp} 
\end{figure}

Fig.~\ref{FIG:FourBandDisp} depicts the numerical dispersion of both the Hertzian and bilinear systems for low, medium, and high excitation amplitudes, which shows that the most profound energy transfer occurs in the medium amplitude range, much like what was seen in section~\ref{SEC:WavenumberScatter}. Note that these low, medium, and high excitation amplitudes now refer to order 1, order 10, and order 100 N. To verify and quantify the efficacy of the VIs to induce TET from low-to-high bands (i.e.,~to induce IBTET) with respect to excitation amplitude, the energy stored within the upper two optical bands is recovered and normalized per the total system energy. This normalized energy is time-averaged taking into account only the time window  after the propagating wavefront encounters  the first VI site in the lattice. 

\begin{figure}[t!]
	\includegraphics[width=\linewidth]{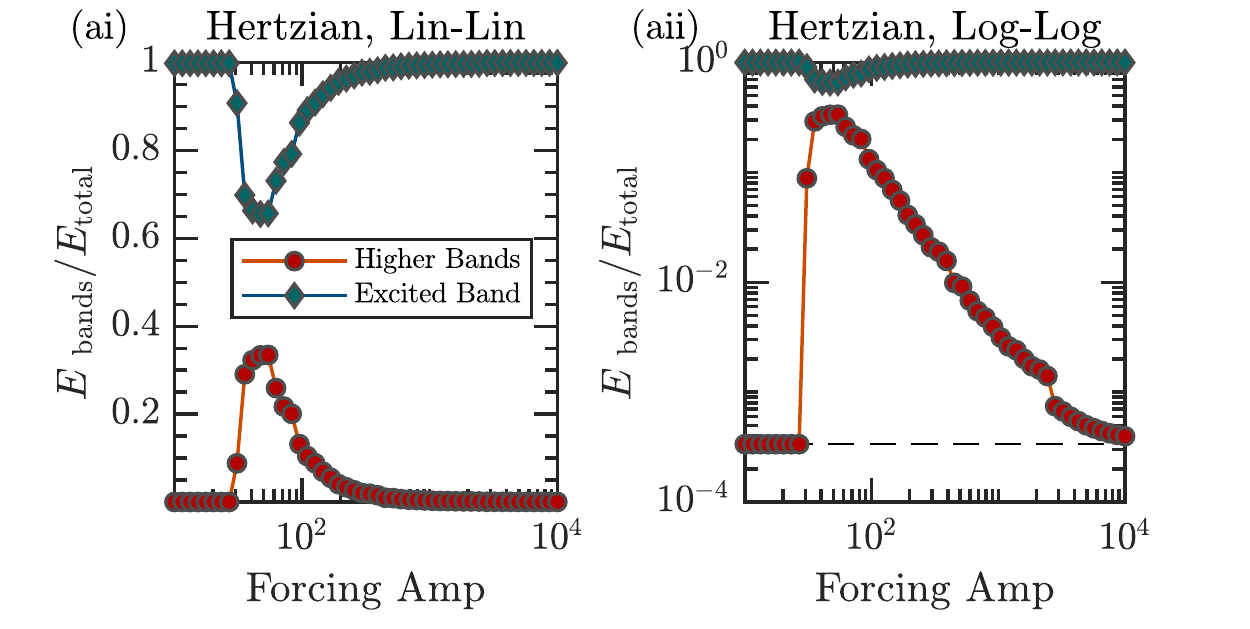}
	\includegraphics[width=\linewidth]{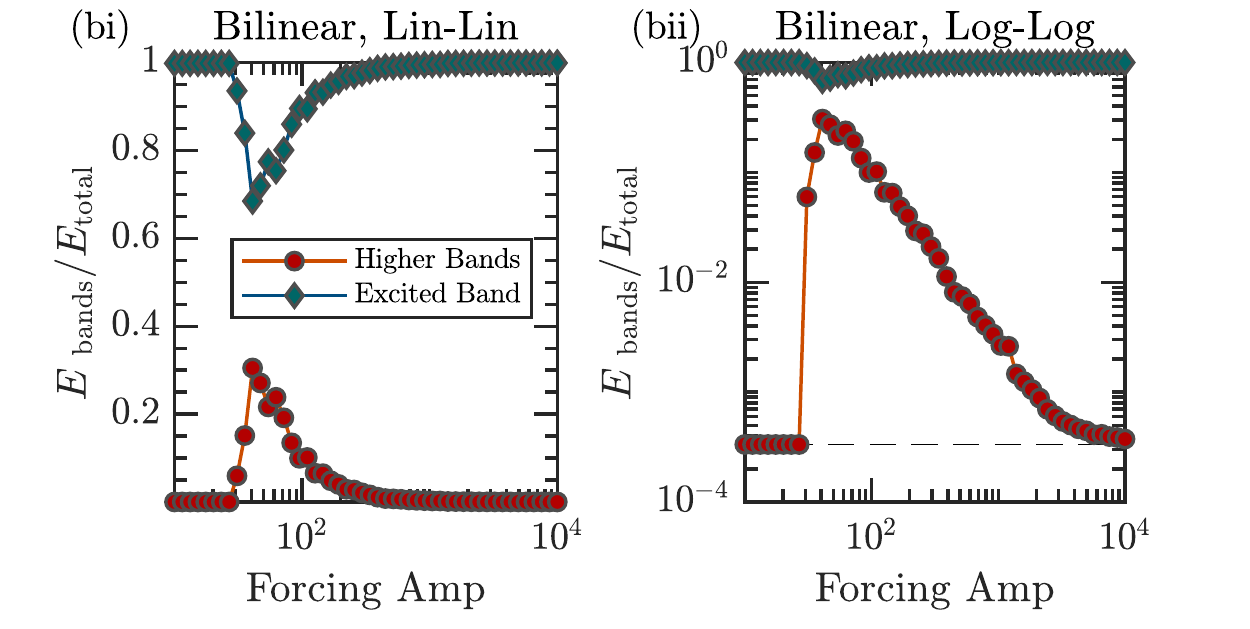}
	\caption{The portion of input energy transferred to the upper two optical bands versus forcing amplitude of the incident wave for (a) Hertzian VIs and (b) bilinear VIs in (i) depicting linear-linear and (ii) log-log scales.}
	\label{FIG:Ejump}
\end{figure}

Fig.~\ref{FIG:Ejump} depicts the results of the IBTET analysis over the ranges of forcing amplitudes considered for both Hertzian and bilinear VI laws. The log-log plots depict a very similar trend to what was observed in section~\ref{SEC:WavenumberScatter}: a sudden spike in energy transfer once the amplitude is sufficient enough to engage the VI, and a sudden decline in energy transfer as the excitation amplitudes rise thereafter. 
The portion of the energy transferred to the higher bands continues to fall until it reaches a minimum defined by the relative energy obtained by the higher bands for a completely linear system. This is on the order of 0.01 \% of the total system energy, and is of course explainable by the fact that the windowed tone burst used to excite the system assumes a Gaussian distribution in the frequency domain which invariably provides trace amounts of energy across the entirety of the spectrum due to the Fourier uncertainty principle. 

Interestingly, the same trends in IBTET are observed for both Hertzian and bilinear contacts, indicating that the nature of the contact law does not play a critical role in the energy transfer, but rather the discontinuous potential is the driving mechanism for the energy exchanges. 
This is further verified in the linearly-scaled plots of Figs.~\ref{FIG:Ejump}(aii,bii) which show that the maximum energy transferred to the higher optical bands is roughly 0.3-0.35 (30-35\%) for both the Hertzian and bilinear VIs.
Not only does this demonstrate that a substantial portion of energy may be irreversibly transferred to higher bands, but that this is achievable for a variety of VI designs, opening broader designs avenues for practical acoustic metamaterials that could exhibit IBTET. 

\section{Physical Interpretation of IBTET Mechanism}
\label{SEC:ROM}
We now seek to connect the trends established in Sections~\ref{SEC:WavenumberScatter} and~\ref{SEC:IBTET} to physics-informed arguments in order to shed physical insight into IBTET in a consistent and comprehensive way.
We do so by considering a reduced order model (ROM) of a VI-oscillator to emulate the VI unit cells embedded in the finite lattices, and then interpret IBTET by studying the nonlinear normal modes (NNMs) of the ROM.
NNMs have proven a useful tool for interpreting the responses of nonlinear dynamical systems and their passive tunability with respect to energy through either analytical or computational tools~\cite{Vakakis2008a,Kerschen2009,Avramov2013,Peeters2009}. The uses and interpretations of NNMs are quite extensive, however a direct and intelligible way of interpreting the evolution of the system's dynamics with respect to energy is with the frequency energy plot (FEP) of a given dynamical system and its bifurcating branches~\cite{Vakakis2008a}. Such methodology has been employed already for understanding the dynamical evolution of VI systems of various forms~\cite{Lee2009,Tao2019,Moussi2015}.
%

\subsection{Reduced Order Model (ROM)}

\begin{figure}[t!]
	\includegraphics[width=.5\linewidth]{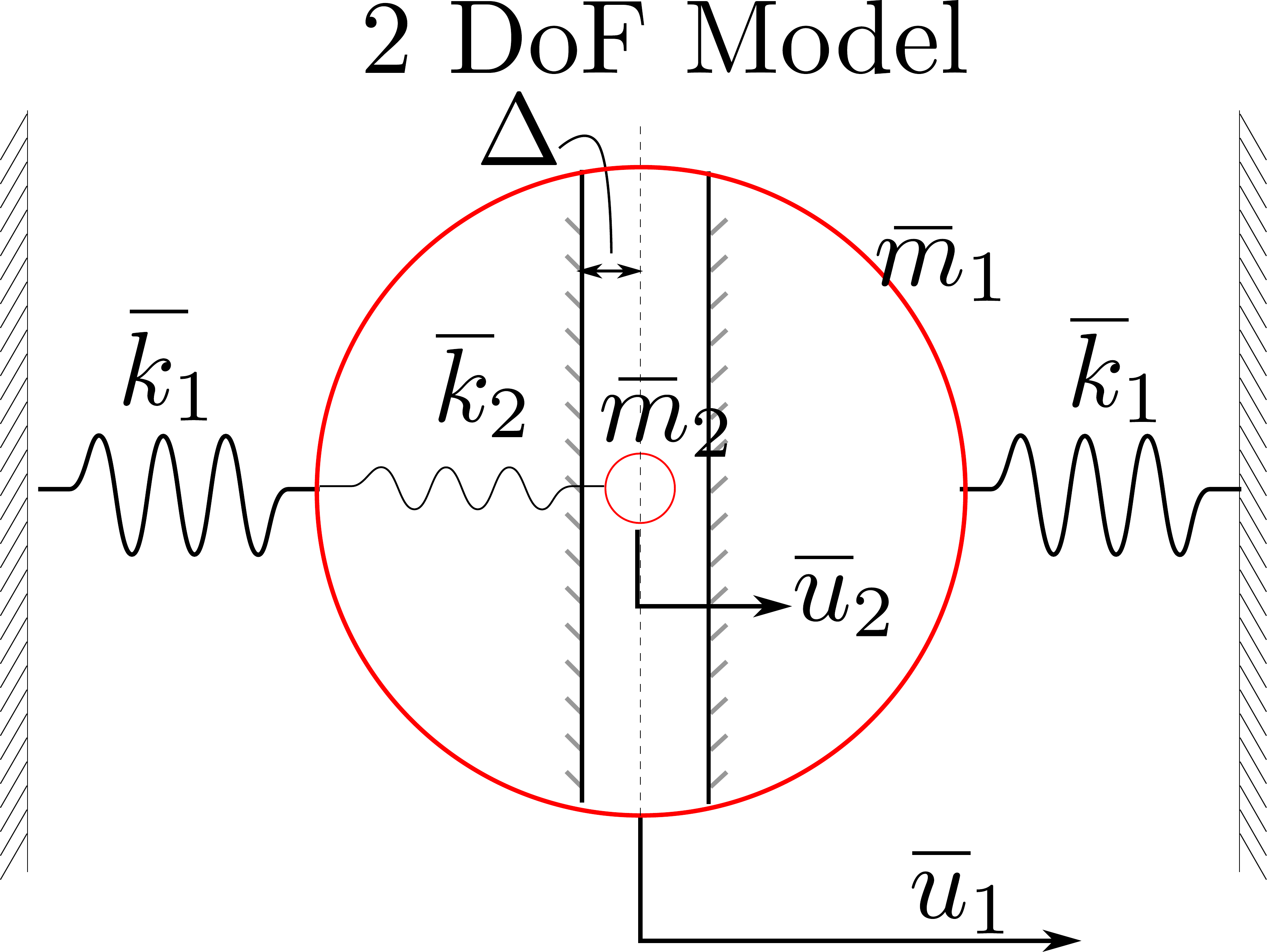}
	\caption{A 2-DoF model emulating a VI resonator cell.}
	\label{FIG:2dof}
\end{figure}
We consider a 2-DoF ROM that is designed to emulate the individual VI-resonators embedded within the 4-band lattice of section~\ref{SEC:IBTET}. Fig.~\ref{FIG:2dof} provides a schematic of the ROM whereby the parameters $\bar{k}_1=k = 2\times 10^4$ N/m, $\bar{k}_2 = 2\times 10^3$ N/m, and $\bar{m}_2=\bar{m}_2 =0.0025$ kg, which parameterize the set of equations
\begin{equation}
	\begin{aligned}
	&\bar{m}_1\ddot{\bar{u}}_1 + \bar{k}_1\bar{u}_1+k_2(\bar{u}_1-\bar{u}_2) +f_{\rm NL}(\bar{w})	= 0,\\
	&\bar{m}_2\ddot{\bar{u}}_2+ \bar{k}_2(\bar{u}_2-\bar{u}_1) -f_{\rm NL}(\bar{w})			= 0.
	\end{aligned}
	\label{EQ:ROM}
\end{equation}
where an overbar denotes that the variable is associated with the ROM and not the full phononic lattice. The nonlinear force $f_{\rm NL}(\bar{w})$ in Eq~\eqref{EQ:ROM} is taken with respect to $\bar{w} = \bar{u}_1-\bar{u}_2$, where VI nonlinearity is considered as both Hertzian and bilinear form with a contact stiffness and clearance of $10^{-2.75}$ m. 

A key difference to note is that the ROM has fixed boundaries, whereas the resonator embedded within VI unit cells of the full phononic lattice does not. 
However, we assume that the stiffness between masses in the lattice is distributed between the two mass elements, and thus the total stiffness of the ROM host mass with respect to its equilibrium position can be approximated by considering that fixed boundaries with one-half the total stiffness of the flexible boundaries of the full phononic lattice.
Moreover, the most critical component of the ROM is the internal stiffness and nonlinear VI component, which matches identically to the VI cells considered in Section~\ref{SEC:IBTET}. Hence, the ROM provides reasonable resemblance to the VI cells in the full lattice system allowing it to capture the trends of the full system with surprisingly good accuracy, as we will show. 

\subsection{Nonlinear Normal Modes as a Measure of Nonlinearity}


The energy dependencies of Figs.~\ref{FIG:Entropy} and~\ref{FIG:Ejump} make an NNM approach  a natural avenue since continuation returns an overview of the dynamics across energy scales. To this end, we compute the NNMs of the ROM by employing a continuation scheme described in~\cite{Peeters2009} with minor modifications listed (see Appendix~\ref{SEC:apx_NNM}). We provide a grossly condensed description herein and refer the reader to~\cite{Peeters2009} for full algorithmic details.
The state form of system~\eqref{EQ:ROM} is $\dot{\textbf{z}} = \textbf{g}(\textbf{z})$ where $\textbf{g}(\textbf{z})$ is a nonlinear function of the state variables. A periodic orbit (or NNM) will satisfy the two-point boundary value problem defined by the \textit{shooting function}, $\textbf{H}(\textbf{z}_{\textbf{p}_0},T) =\textbf{z}(\textbf{z}_{\textbf{p}_0},T)-\textbf{z}_{\textbf{p}_0} = \textbf{0}$.
Newton's method can be used to recover periodic solutions at low energy in the \textit{shooting stage}. We define the phase condition such that the two DoFs of the ROM have zero initial velocities.
After shooting is completed, a pseudo-arclength method is used to trace out the NNM branch in the $2n+1$ dimensional parameter space. 
In brief, this works by computing \textit{predictor} steps using the tangent vector at the most recently converged solution, and then making \textit{corrector} steps in an orthogonal direction to the tangent until convergence is achieved. This is a critical step for resolving the NNMs of the VI system since the NNM branches may have turning points that the standard Newton-Raphson algorithm cannot solve. 

\begin{figure}[t!]
	\includegraphics[width=\linewidth]{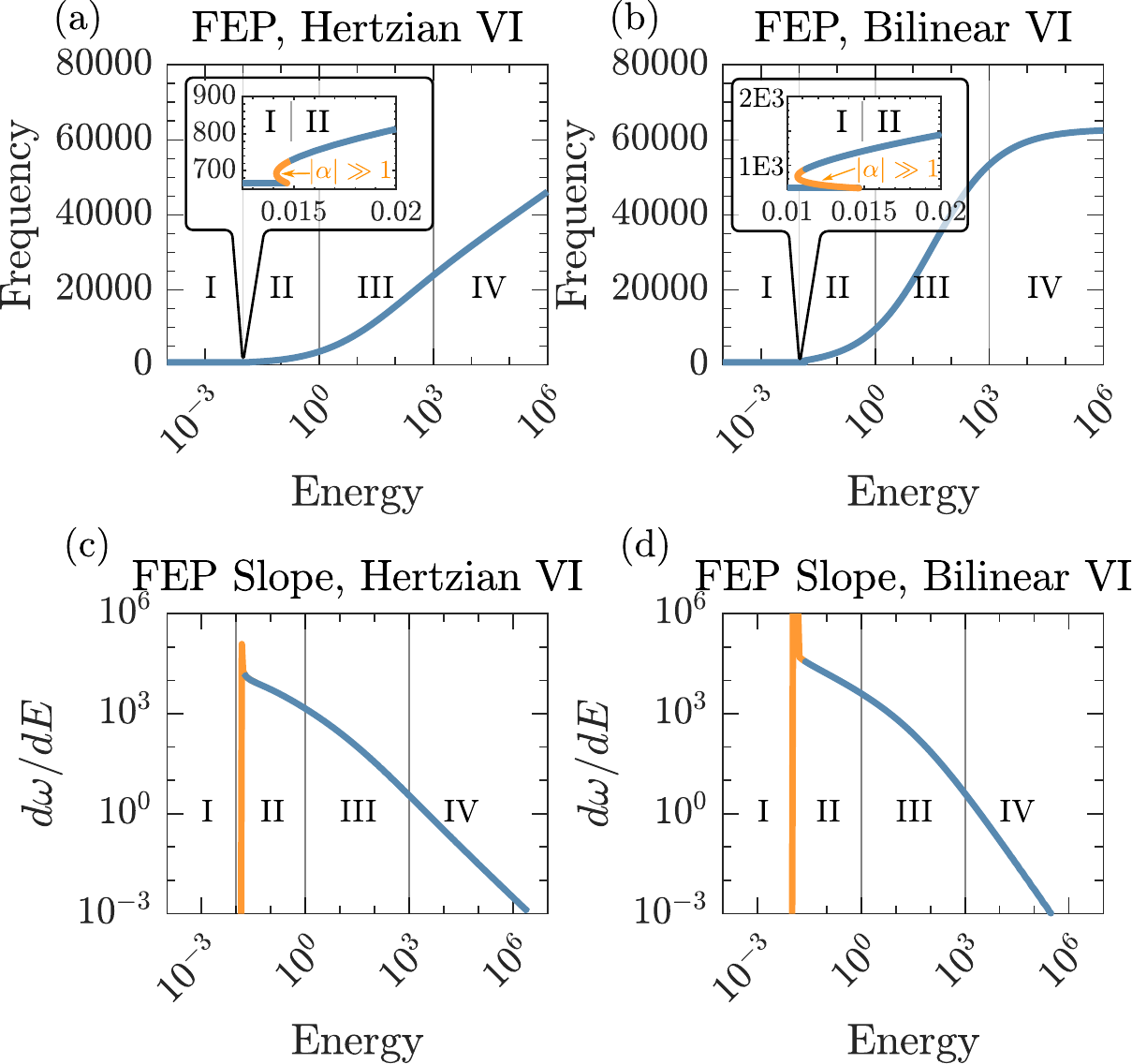}
	\includegraphics[width=\linewidth]{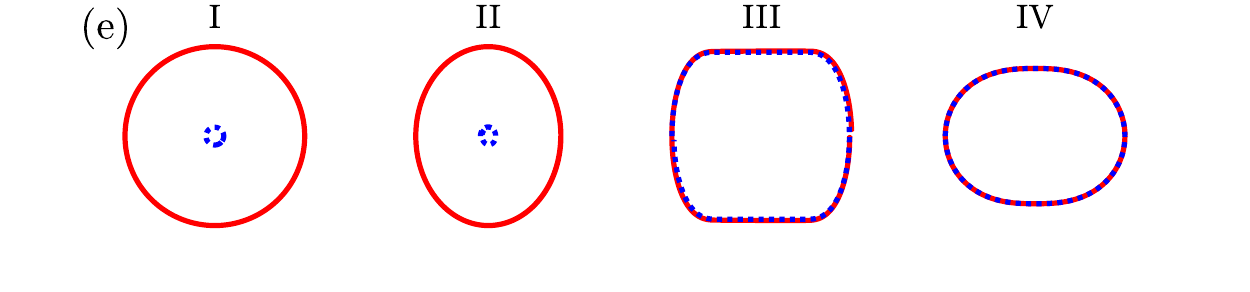}
	\includegraphics[width=\linewidth]{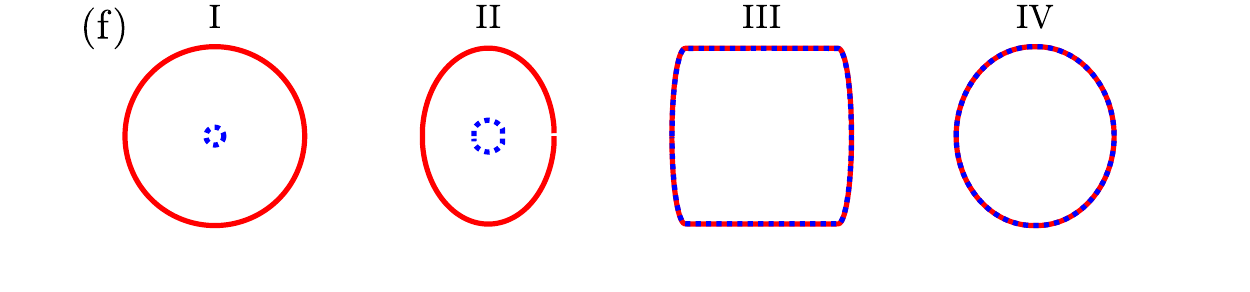}
	\caption{The FEPs of the ROMs with (a) Hertzian and (b) bilinear nonlinearity with insets zooming in on the transition from region I to II with instability denoted by orange for regions with Floquet multipliers $|\alpha|\gg1$; (c,d) slopes of the FEPs of of (a,b) with respect to energy; (e) and (f) corresponding phase trajectories of the NNMs for (a) and (b), respectively, for regions I, II, III, and IV of the FEPs.}
	\label{FIG:FEP}
\end{figure}

The result of numerical continuation is a \textit{frequency energy plot} (FEP) which describes the evolution of the NNM branch for 1:1 resonance (the so called ``backbone" branches) in the frequency-energy space. Fig.~\ref{FIG:FEP} depicts the FEPs computed for system described by equation~\eqref{EQ:ROM} for both Hertzian and bilinear contact laws. 
It is interesting to emphasize that the degree (strength) of nonlinearity of the ROM can be qualitatively interpreted by the slope of a given NNM branch~\cite{Lee2009}. The steeper the slope is of the branch is, the more sensitive the frequency-amplitude dependency of the NNM becomes, and the more intense the nonlinearity in the ROM when it responds on that NNM is.

The FEP results reveal similar trends for both Hertzian and bilinear VI ROMS, possessing four dynamical region labeled (I)-(IV) in Fig.~\ref{FIG:FEP}. The corresponding phase trajectories of the periodic orbits in each region are given in Fig.~\ref{FIG:FEP}(e) and~\ref{FIG:FEP}(f) for Hertzian and Bilinaer models, respectively.
In the low energy region (I), the VIs do not engage, and the dynamics are completely linear; this is confirmed by zero slope of the FEP.
In region (II), there is a grazing of the VI contacts, causing a sudden change in the dynamics and a rapid increase of FEP slope. 
In fact, the corresponding NNM branch folds back on itself and goes backwards in energy before re-directing again towards higher energies, with this effect being more prevalent in the bilinear model (the Hertzian nonlinearity being less prominent in the small deflection amplitude limit).
This in turn yields a small neighborhood of the NNM branch where the FEP slope is theoretically infinite, and the subplots of Figs.~\ref{FIG:FEP}(c,d) confirm that this is where to maximum is reached. 
The phase trajectories indicate that region II represents a transition where the dynamics are most sensitive to nonlinear effects. Despite the apparent smoothness of Figs.~\ref{FIG:FEP}(eII,fII) the volatile VI-grazing dynamics in region II are unstable, and hence, not physically realizable.
Computation of NNMs in this regions requires Newton predictions on a similar order of machine tolerance and results in strongly unstable NNMs as depicted in Fig.~\ref{FIG:FEP} for portions of the NNM branch with Floquet multiplier, $\alpha$, far exceeding $1$. 

After the grazing VI region in region II is surpassed with increasing energy,  the FEP gradually increases in frequency towards region III.
Region III is characterized by strong VI oscillations which is apparent by the box-like phase trajectories indicating non-smooth temporal dynamics. In this region, the linear dynamics of $\hat{k}_1$ are negligible and the VI dynamics dominant. Note that it is in region III that the slopes of the FEPs decrease in a power-law like fashion as the ROM asymptotically reaches the limiting region IV.
Region IV manifests smooth dynamics characterized by in-phase dynamics predominantly dictated the contact stiffness. In this region, the clearance is negligible and the VI contacts behave as an extremely stiff elastic spring. Hence, the dynamics of the ROM with Hertzian contacts approaches a smoothly nonlinear system with a 3/2 nonlinear coupling, whereas the dynamics of the bilinear ROM approaches a linear system at high energy, as is confirmed by the phase portraits of Figs.~\ref{FIG:FEP}(eIV,fIV). Moreover, for the bilinear system, the FEP clearly levels off as the high-energy (almost) linear limiting behavior is reached. 


\subsection{Relating the Dynamics of the ROM to the Acoustics of the Lattice}

The evolution of the FEP slope with respect to energy of the ROM (Figs.~\ref{FIG:FEP}(b,c)) posses a remarkable similarity to the observed trends of nonlinear IBTET in the full phononic lattice (Fig.~\ref{FIG:Ejump}). The two measures can be related to one another by replotting the energy transfers of Fig.~\ref{FIG:Ejump} with respect to system energy (to match the energy-dependent nature of the FEP) and superimposing the FEP slopes to compare similarities in their evolution with energy. 
To do this requires a normalization, as the maximum and minimum values of the FEP slope can be arbitrarily large or small, whereas the relative energy of the upper optical bands is lower-bounded by the amount provided by the excitation source (from the Fourier uncertainty principal), and upper-bounded by unity (since the energy in the upper bands cannot exceed the total energy of the system). 
Moreover,  the wave propagation in the 1200 DoF phononic lattice carries the energy of 30 cycles of the windowed excitation, whereas the FEP energy is parameterized by the periodic orbits of the 2 DoF ROM. 
Thus, the energy of the finite lattice must be normalized in order to be commensurate with the energy of the ROM used to generated the FEP.
These normalizations are performed as follows.
The FEP slope is divided by a scalar as to quantitatively align with the relative energy transfer in quantity so that a direct comparison can be made with respect to decay rate versus energy. 
A scalar quantity defined by the low-bound of IBTET (dashed lines of Fig.~\ref{FIG:Ejump}) is then added to the FEP slope account for the lower threshold of the energy transfer in the VI lattice.
The energy of the finite lattice is normalized so that the \textit{initiation energy}, that is, the energy required to engage the first VI site encountered by the propagating wavefront, aligns with the transition between regions I and II of the FEP. 
These normalizations preserve the slopes of both quantities since scalar multiplication results only in translations in log scaling. Hence, the previous measures can be directly compared with respect to their decrease in value with respect to increasing normalized energy. 

\begin{figure}[t!]
\includegraphics[width=\linewidth]{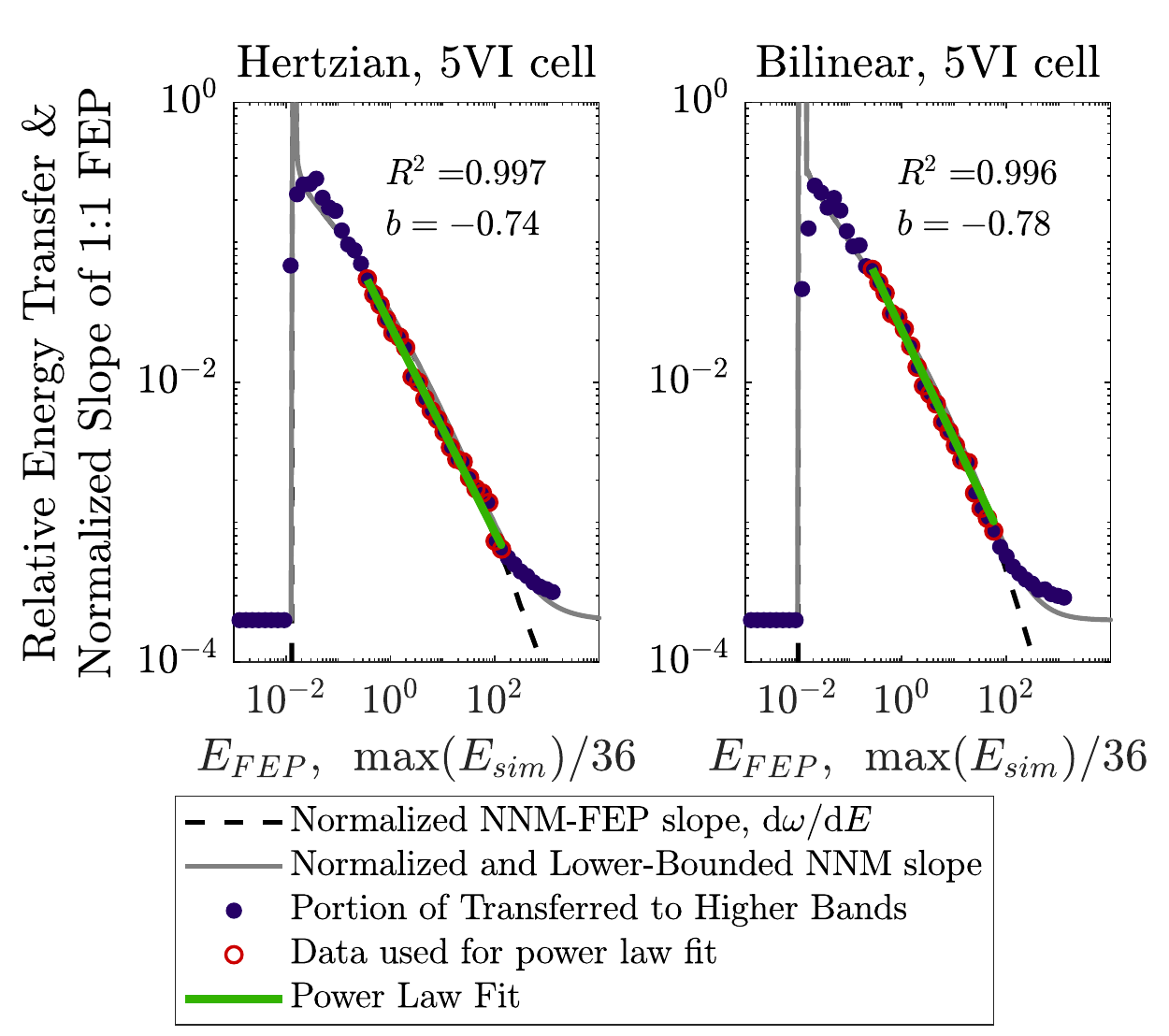}
	\caption{The relative inter-band energy transfer, with the normalized slope from the ROM-FEP superimposed for (a) Hertzian and (b) bilinear contact models; the dashed lines depict the normalized FEP slopes,  the gray lines depict the normalized FEP slopes lower-bounded by the initial (linear) energy of the higher bands, and green lines depict a power law fit to red dots, with the adjusted R-squared value shown with the inset.}
	\label{FIG:Ejump_FEP}
\end{figure}

Fig.~\ref{FIG:Ejump_FEP} displays the described superposition where a remarkable agreement is found between the trends in the slope of the FEP of the ROM and the energy transfer between bands in the lattice.
Hence, the underlying FEP of the ROM, along with the evolution of the dynamical regimes of Fig.~\ref{FIG:FEP}, clearly have a direct implication of the IBTET in the lattice. Moreover, by fitting a slope to the measured  energy transfer versus normalized system energy for data points falling in region III, a near-perfect power law is recovered as indicated by the adjusted R-squared values close to 1 (see Fig.~\ref{FIG:Ejump_FEP}). Finally, these results are in agreement with the trends observed for wavenumber spreading within the optical band of the 2-band system considered in section~\ref{SEC:WavenumberScatter}. 
Hence, the numerical results presented for the finite lattices can be understood based in terms of the underlying nonlinear dynamics of the ROM based on the single VI unit cell as it transitions between various dynamical regimes with respect to energy.
With this, a predictive tool is presented to assess the capacity for IBTET in full phononic systems based on the simplified VI ROMs which, being of low-dimensionality, are much more amenable to analysis compared to the extended nonlinear lattices considered herein. 

\section{Conclusions}
\label{SEC:Conclusions}

In this work, we have investigated the effect of local VI nonlinearities on the propagation of traveling waves in 1-D phononic lattices. 
Specifically, first a di-atomic 2-band lattice was numerically studied over a wide range of forcing amplitudes and embedded VI configurations (section~\ref{SEC:WavenumberScatter}).
It was demonstrated that wavenumber scattering in the optical band of this lattice is most profound for moderate excitation amplitudes, and decreases in effectiveness as the energy rises (Fig.~\ref{FIG:scattersims}). This was quantified by considering the spatial-spectral entropy (or wavenumber entropy), for various systems which all followed very closely to power-law decays with respect to excitation amplitude after the peak value was reached (Fig.~\ref{FIG:Entropy}).
Attention then turned to inter-band targeted energy transfer (IBTET) in a 4-band system which was parameterized in order to provide dispersion curves receptive to such energy transfers (Section~\ref{SEC:IBTET}).
Simulations were carried out over a range of excitation amplitudes with both Hertzian and bilinear contact laws. 
Numerical post-processing reconstructed the energy of each band, and it was shown that IBTET is indeed possible. 
Moreover, this phenomenon was proven effective for both Hertzian and bilinear VIs, and the trends in IBTET with respect to excitation amplitude followed closely to those observed for wavenumber scattering in the 2-band lattice (Fig.~\ref{FIG:Ejump}). 

In an attempt to shed some physical insight into the effect of the VIs on the acoustics of the lattice, a low-dimensional ROM was constructed based on the unit VI cell.  
The underlying FEP of the 2 DoF ROM was computed for the NNM family of 1:1 resonance branches which revealed four dynamic regimes that the ROM assumes with respect to energy. 
Namely, a linear low energy region, a grazing region initiated when the VI nonlinearity first enters the dynamics, a full VI-oscillator with nonsmooth temporal dynamics, and an \textit{effectively} linear or smoothly nonlinear high-energy regime, depending on the contact law (Hertzian or bilinear).
This, in turn, produced a frequency-energy slope  that directly scales to the trends of IBTET in the lattice with respect to system energy, providing the physical interpretation of the spectral scattering of sections~\ref{SEC:WavenumberScatter} and~\ref{SEC:IBTET}. Moreover, the FEP presents a means for accurately predicting energy transfer capacity of the full phononic lattice based on the low-dimensional ROM. 

Although this work focused primarily on fundamental understanding of the physics at play, the implications and potential for future developments are rather extensive. 
The low-to-high energy transfers directly correspond to a reduction in magnitude, since the energy must be preserved in the frequency transfer. Moreover, the evolution of the VI dynamics with respect to energy corresponds to an effective filter that can greatly alter transmissibility of incident waves (cf.~Fig.~\ref{Fig:WaveletWnum}). These attributes alone make VI-based methods attractive for wave transmission  tuning (or tailoring) with respect to amplitude. Moreover, while we have targeted low-to-high energy transfers between bands, future works could explore the potential for targeting specific bands and specific sub-regions of bands of phononic lattices by optimizing the distribution and parameters of local VIs in lattices through methods such as genetic programming or machine learning. 

\appendix
 \begin{acknowledgments}
This work was supported in part by the National Science Foundation Graduate Research Fellowship Program under Grant No. DGE – 1746047. Any opinions, findings, and conclusions or recommendations expressed in this material are those of the authors and do not necessarily reflect the views of the National Science Foundation.
 \end{acknowledgments}
 
\section{Details on Signal Processing Procedures}
\label{SEC:SigProc}
\subsubsection{Continuous Wavelet Transformation (CWT)}
In this section, we provide a brief discussion of the wavelet transformation algorithm employed in this work in order to clarify the mathematical details pertinent for performing the wavelet-based wavenumber partition analysis of section~\ref{SEC:WavenumberScatter} (cf. Fig.~\ref{FIG:Kpartitions}). A similar discourse may be found in~\cite{Mojahed2021}.
The CWT is traditionally used as a time-frequency analysis tool by transforming the signal from the time domain to the time-frequency domain. To the same effect, one can consider the space-wavenumber domain. 
For 1D systems the standard definition of the CWT with respect to the spatial variable $x$ is,
\begin{equation}
	X(x,\kappa) = \sqrt{\frac{\kappa}{\kappa_c}} \int_{-\infty}^{\infty}u(\xi)\psi^*\left(\frac{\xi-x}{\kappa_c}\right){\rm d}\xi
\end{equation}
where $\psi^*(\xi)$ is the complex conjugate of the mother wavelet function and $\kappa_c$ the center frequency,
\begin{equation}
	 \kappa_c = \left[ \frac{\int_0^\infty \kappa^2|\Psi(\kappa)|^2{\rm d}\kappa}{\int_0^\infty \Psi(\kappa)|^2{\rm d}\kappa} \right]^{1/2}.
\end{equation}
 We consider the Morelet wavelet for all transformations in this work:
\begin{equation}
	\psi(x)= \frac{1}{\pi^{1/4}}\left(e^{i\kappa_c x}-e^{-\kappa_c^2/2}\right)e^{-x^2/2}.
\end{equation}
For the scale and quantities of datasets considered in this work, computational efficiency is a requirement. To this end, the Fast Fourier Transform is employed to speed up wavelet computations. Taking $\Psi(\kappa)$ as the analytical Fourier Transform of the mother wavelet, 
\begin{equation}
	\Psi(\kappa) = e^{-(\kappa-\kappa_c)^2/2},
\end{equation}
and $\tilde{x}(\kappa)$ the FFT of the signal,
the wavelet transformation  can be written equivalently as:
\begin{equation}
	X(\kappa,x) = \sqrt{\frac{\kappa_c}{\kappa}}\int_{-\infty}^{\infty}\tilde{x}(\eta)\Psi^*(\eta\kappa/\kappa_c)e^{i\eta x}{\rm d}\eta.
\end{equation}
\begin{figure}[t!]
	\includegraphics[width=\linewidth]{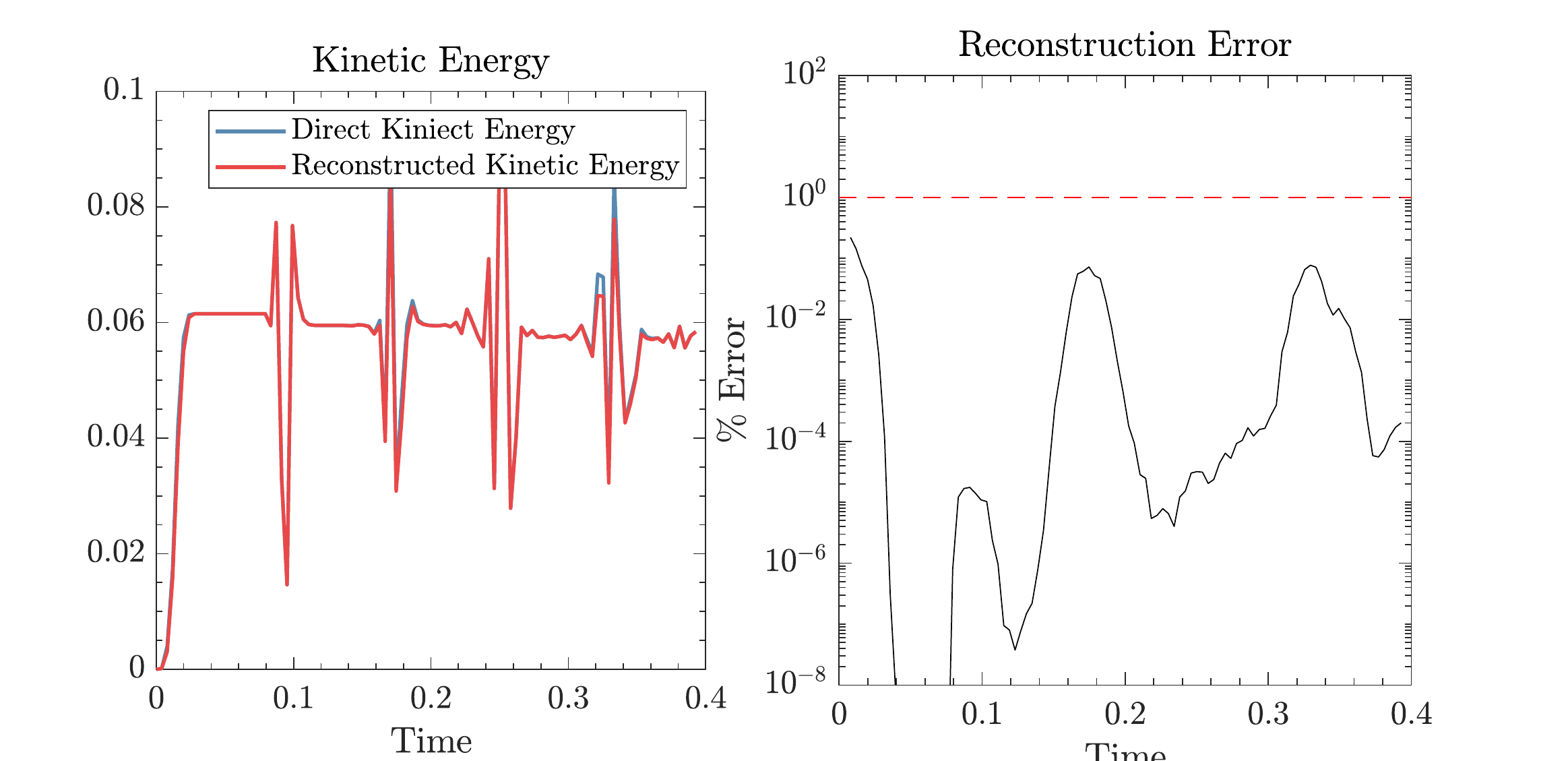}
	\caption{The reconstructed kinetic energy and corresponding reconstruction error for the described wavelet partition scheme; red dashed line indicates 1 percent error.}
	\label{FIG:recerror}
\end{figure}
Each wavelet transformation can be partitioned over space and wavenumber.
The spectral partitions are defined over 12 regions spanning between $\kappa= 0$ and $\kappa =\pi$ to account for 12 different wavelet-domain representations of the spatial signal at each time instant. The $k$-th wavenumber partition is defined as:
\begin{equation}
	\begin{aligned}
&	X_k(\kappa,x) = X(\kappa,x) h_k(\kappa), \\
&	h_k(\kappa)  = H\left(\kappa -\frac{(k-1)\pi}{12} \right)- H\left(\kappa -\frac{k\pi}{12} \right).
	\end{aligned}
\end{equation}

The inverse wavelet transformation can be applied at each time snap shot to each wavenumber partition, $u_k(x) = \mathcal{W}^{-1}\left\{X_k(\kappa,x)\right\}$, which is computed as:
\begin{equation}
	u_k(x) = \frac{\sqrt{\kappa}}{\kappa^{3/2}_c C}\int_{0}^{\infty}\int_{-\infty}^{\infty} \hat{X}_k(\kappa,\xi) \Psi\left(\frac{\xi\kappa}{\kappa_c }\right) {\rm d}\xi{\rm d}\kappa.
\end{equation}
where $\hat{X}_k(\kappa,\xi)$ is the Fourier transformation of $X_k(\kappa,x)$ with respect to $x$.  Fig.~\ref{FIG:recerror} depicts the reconstructed kinetic energy of the lattice, $KE_{rec}$, as well as the directly computed (exact) kinetic energy from the numerical simulations  $KE_{phys}$, with the error between the two quantities computed by: 
\begin{equation}
	e(t) = \frac{ ||KE_{rec}(t)-KE_{phys}(T)||}{||KE_{phys}(t)||}.
\end{equation}

\subsubsection{Spectral Entropy}
\begin{figure}[t!]
	\includegraphics[width=\linewidth]{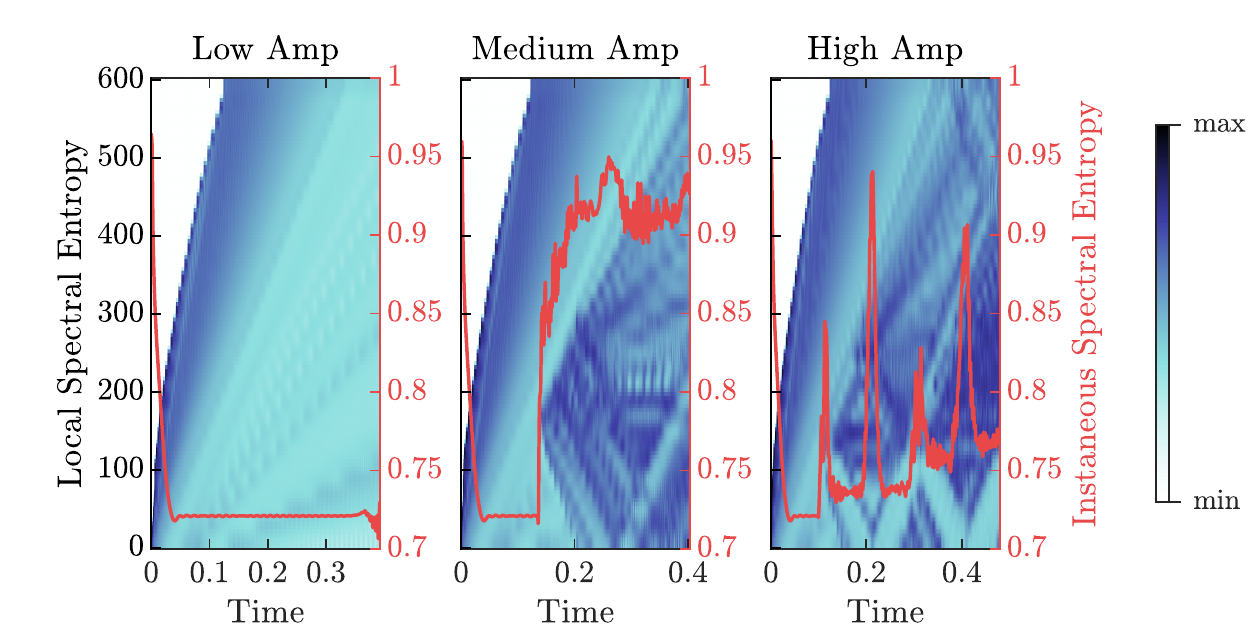}
	\includegraphics[width=\linewidth]{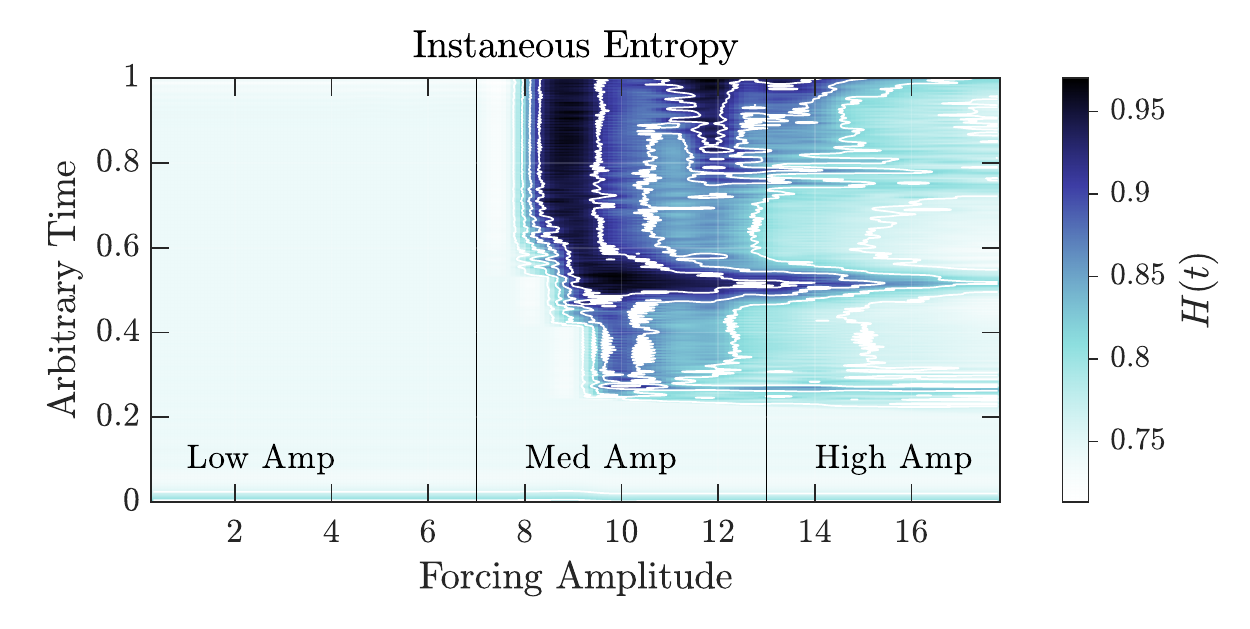}
	\caption{Contours of the instantaneous wavenumber entropy across the time-entropy domain for low, medium, and high amplitude simulations(top), and the summary contours of the instantaneous entropy H(t) (bottom).}
	\label{FIG:ent_detail}
\end{figure}

Here, we provide more details pertaining to the spectral entropy plots displayed in Fig.~\ref{FIG:Entropy}. Fig.~\ref{FIG:ent_detail} depicts the distribution of entropy using Eq~\eqref{EQ:Hx} to recover $H(x)$ for each $t$. The resulting matrix $\mathcal{H}(x,t)$ is plotted as an image for low, medium, and high excitation amplitudes. The distribution of high-entropy regions is clearly seen in the medium and high excitation amplitude simulations as the VIs engage the incoming wave. Superimposed on each image is the \textit{instantaneous} spectral entropy, which summarizes $\mathcal{H}(x,t)$ over space to render time-dependent measures $H(t)$. 

A data set storing $H(t)$ for each excitation amplitude in the simulation ensemble can then be generated and plotted in the form of an image to study how the wavenumber entropy varies in time with respect to the forcing amplitude for a given lattice configuration. This is depicted in the bottom plot of Fig.~\ref{FIG:ent_detail}. In the low-amplitude region with no VI engagement, no entropy is generated after excitation (as expected). For medium amplitudes, regions of \textit{sustained} high wavenumber entropy are realized after the VIs engage the incident wave. In contrast, only \textit{localized} patches of high entropy are seen for high-amplitude simulations, indicating that the VIs do not affect the global wavenumber of the lattice after the incident wave passes through (or reflects off of) the unit cells with embedded VIs. 

\subsubsection{Computing energy on each band}
\begin{figure}[t!]
	\includegraphics[width=\linewidth]{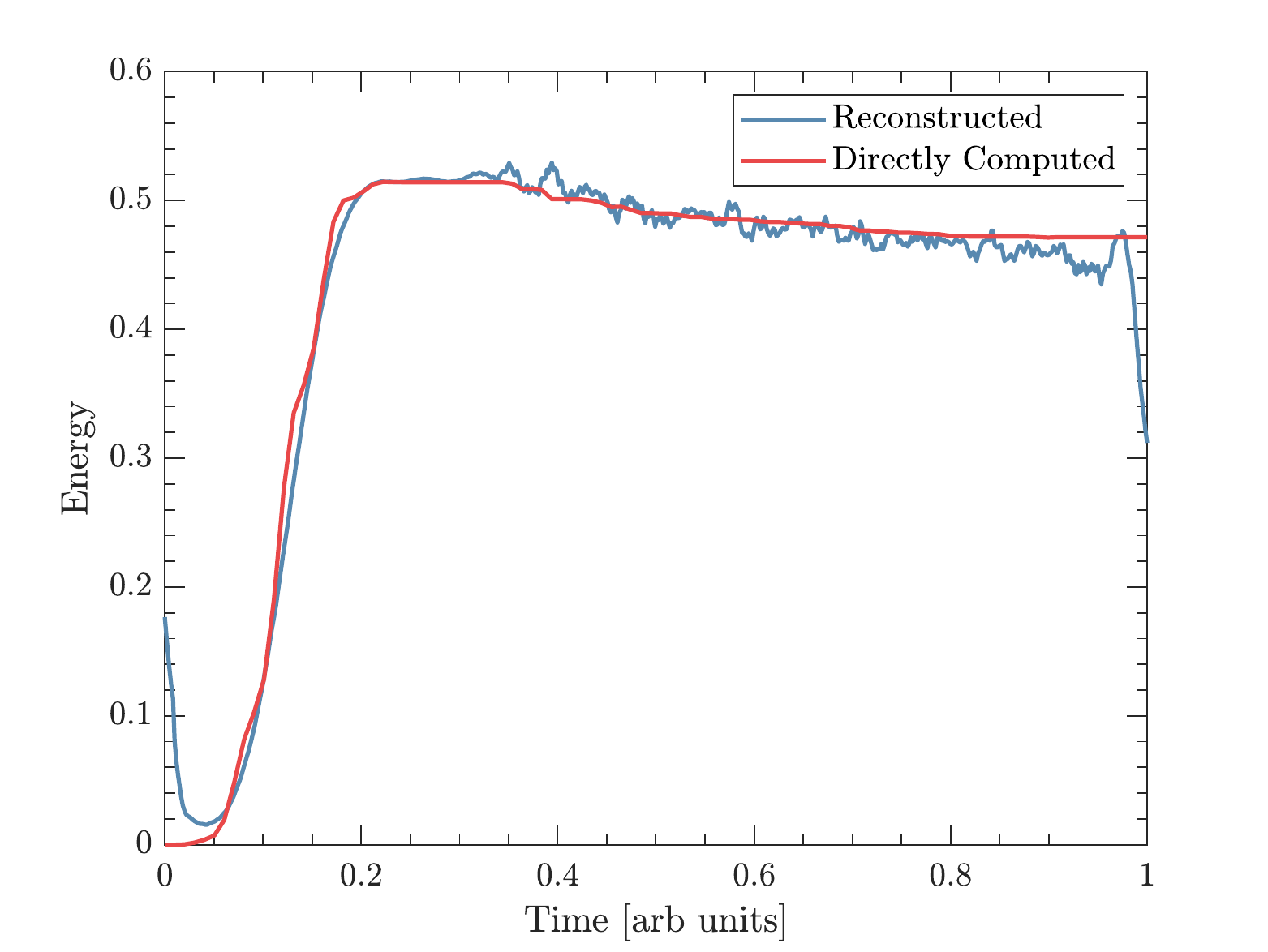}%
	\caption{Energy Reconstruction of band-partitioning decomposition.}
\end{figure}

The computation of wave energy over each band in section~\ref{SEC:IBTET} is performed as follows. The data matrix for a given simulation is mapped to the Fourier domain using the 2D FFT algorithm $\mathcal{D}(\kappa,\omega)=\mathcal{F}^{x,t}\{\textbf{u}(x,t)\}$. Next, frequency filters are constructed as follows,
\begin{equation}
G_k(\kappa,\omega) = \begin{cases}
				1 \ \ \ &\omega\in \, \mathcal{B}_k, \ -\pi\leq \kappa \leq\pi \\
				0 \ \   &\text{otherwise}
				\end{cases}
\end{equation}
were the first four ranges of frequencies $\mathcal{B}_k$ are defined over the temporal frequency limits of the four pass-bands (${\rm PB}$),
\begin{equation}
	\begin{aligned}
		&\mathcal{B}_1 = \min({{\rm PB}_1})\leq\omega\leq\max({{\rm PB}_1}) \\
		&\mathcal{B}_2 = \min({{\rm PB}_2})\leq\omega\leq\max({{\rm PB}_2}) \\
		&\mathcal{B}_3 = \min({{\rm PB}_3})\leq\omega\leq\max({{\rm PB}_3}) \\
		&\mathcal{B}_4 = \min({{\rm PB}_4})\leq\omega\leq\max({{\rm PB}_4}) \\
	\end{aligned}
\end{equation}
A remaining two filter banks are constructed for the band gap between the acoustic band and first optical band (${\rm BG}_1$), and of for the band gap between the upper two optical bands (${\rm BG}_2$),
\begin{equation}
	\begin{aligned}
		&\mathcal{B}_5 = \min({{\rm BG}_1})\leq\omega\leq\max({{\rm BG}_1}) \\
		&\mathcal{B}_6= \min({{\rm BG}_2})\leq\omega\leq\max({{\rm BG}_2}). \\
	\end{aligned}
\end{equation}
The spatial-temporal dynamics corresponding to each pass band and band gap regions are then given as,
\[
\textbf{u}_k(x,t) = \mathcal{F}^{-x,-t}\{ G_k(\kappa,\omega) \cdot \mathcal{D}(\kappa,\omega)\}
\]
where $\mathcal{F}^{-x,-t}\{\ \}$ indicates the 2D inverse FFT with respect to $x$ and $t$.
The rigid boundaries of the filters in Fourier space inevitably results in minute numerical artifacts in the inverse transformation for each partition taking the form of ripples along the space-time boundaries. However, the reconstruction of energies computed by summing the energy over each band matched nearly identically to the energies computed for the direct numerical simulations, and hence these numerical artifacts are negligible.

\section{Nonlinear Normal Mode Computations}
\label{SEC:apx_NNM}
The recipe for NNM calculations follows very closely to the procedure outlined in \cite{Peeters2009}. For all FEP calculations, the shooting method used a prescribed initial step size of $1\time10^{-5}$ and a tolerance of $\varepsilon = 1\times10^{-6}$. For low energy orbits, Newmark integration was employed with 2000 steps per period, and Jacobian calculations of predictor-corrector steps were computed using the  sensitivity analysis in~\cite{Peeters2009}. In region II, the unstable dynamics proved to be challenging for the computation of the corresponding NNM branch. Hence, sufficiently small predictor steps were required for convergence, with the residual reduction being varied from $10^{-12}$ to $10^{-10}$. Sensitivity analysis was employed again to compute Jacobian terms in region II. 
	
Once the dynamics of the NNMs stabilized to that of a definitive VI oscillator in region III, and moreover to smoothly stable NNMs in region IV, the finite difference method sufficiently approximated Jacobian terms allowing for the implementation of fast and accurate Runge-Kutta based methods such as ODE78. The nonsmooth nature of dynamics in region III would require still a great number of Newmark iterations to achieve the same accuracy as the ODE78 routine, and therefore the transition was made to a finite-difference Jacobian calculation scheme based on ODE78 for energies beyond region II to increase computational speed and reduce the number of steps required to resolve the high-energy regions of the FEP branch.
%
%
%

\bibliography{VI_Lattice}

\begin{thebibliography}{81}%
\makeatletter
\providecommand \@ifxundefined [1]{%
 \@ifx{#1\undefined}
}%
\providecommand \@ifnum [1]{%
 \ifnum #1\expandafter \@firstoftwo
 \else \expandafter \@secondoftwo
 \fi
}%
\providecommand \@ifx [1]{%
 \ifx #1\expandafter \@firstoftwo
 \else \expandafter \@secondoftwo
 \fi
}%
\providecommand \natexlab [1]{#1}%
\providecommand \enquote  [1]{``#1''}%
\providecommand \bibnamefont  [1]{#1}%
\providecommand \bibfnamefont [1]{#1}%
\providecommand \citenamefont [1]{#1}%
\providecommand \href@noop [0]{\@secondoftwo}%
\providecommand \href [0]{\begingroup \@sanitize@url \@href}%
\providecommand \@href[1]{\@@startlink{#1}\@@href}%
\providecommand \@@href[1]{\endgroup#1\@@endlink}%
\providecommand \@sanitize@url [0]{\catcode `\\12\catcode `\$12\catcode
  `\&12\catcode `\#12\catcode `\^12\catcode `\_12\catcode `\%12\relax}%
\providecommand \@@startlink[1]{}%
\providecommand \@@endlink[0]{}%
\providecommand \url  [0]{\begingroup\@sanitize@url \@url }%
\providecommand \@url [1]{\endgroup\@href {#1}{\urlprefix }}%
\providecommand \urlprefix  [0]{URL }%
\providecommand \Eprint [0]{\href }%
\providecommand \doibase [0]{https://doi.org/}%
\providecommand \selectlanguage [0]{\@gobble}%
\providecommand \bibinfo  [0]{\@secondoftwo}%
\providecommand \bibfield  [0]{\@secondoftwo}%
\providecommand \translation [1]{[#1]}%
\providecommand \BibitemOpen [0]{}%
\providecommand \bibitemStop [0]{}%
\providecommand \bibitemNoStop [0]{.\EOS\space}%
\providecommand \EOS [0]{\spacefactor3000\relax}%
\providecommand \BibitemShut  [1]{\csname bibitem#1\endcsname}%
\let\auto@bib@innerbib\@empty
\bibitem [{\citenamefont {Cummer}\ \emph {et~al.}(2016)\citenamefont {Cummer},
  \citenamefont {Christensen},\ and\ \citenamefont {Al{\`{u}}}}]{Cummer2016}%
  \BibitemOpen
  \bibfield  {author} {\bibinfo {author} {\bibfnamefont {S.~A.}\ \bibnamefont
  {Cummer}}, \bibinfo {author} {\bibfnamefont {J.}~\bibnamefont
  {Christensen}},\ and\ \bibinfo {author} {\bibfnamefont {A.}~\bibnamefont
  {Al{\`{u}}}},\ }\bibfield  {title} {\bibinfo {title} {Controlling sound with
  acoustic metamaterials},\ }\bibfield  {journal} {\bibinfo  {journal} {Nature
  Reviews Materials}\ }\textbf {\bibinfo {volume} {1}},\ \href
  {https://doi.org/10.1038/natrevmats.2016.1} {10.1038/natrevmats.2016.1}
  (\bibinfo {year} {2016})\BibitemShut {NoStop}%
\bibitem [{\citenamefont {Surjadi}\ \emph {et~al.}(2019)\citenamefont
  {Surjadi}, \citenamefont {Gao}, \citenamefont {Du}, \citenamefont {Li},
  \citenamefont {Xiong}, \citenamefont {Fang},\ and\ \citenamefont
  {Lu}}]{Surjadi2019}%
  \BibitemOpen
  \bibfield  {author} {\bibinfo {author} {\bibfnamefont {J.~U.}\ \bibnamefont
  {Surjadi}}, \bibinfo {author} {\bibfnamefont {L.}~\bibnamefont {Gao}},
  \bibinfo {author} {\bibfnamefont {H.}~\bibnamefont {Du}}, \bibinfo {author}
  {\bibfnamefont {X.}~\bibnamefont {Li}}, \bibinfo {author} {\bibfnamefont
  {X.}~\bibnamefont {Xiong}}, \bibinfo {author} {\bibfnamefont {N.~X.}\
  \bibnamefont {Fang}},\ and\ \bibinfo {author} {\bibfnamefont
  {Y.}~\bibnamefont {Lu}},\ }\bibfield  {title} {\bibinfo {title} {Mechanical
  metamaterials and their engineering applications},\ }\href
  {https://doi.org/10.1002/adem.201800864} {\bibfield  {journal} {\bibinfo
  {journal} {Advanced Engineering Materials}\ }\textbf {\bibinfo {volume}
  {21}},\ \bibinfo {pages} {1800864} (\bibinfo {year} {2019})}\BibitemShut
  {NoStop}%
\bibitem [{\citenamefont {Hussein}\ \emph {et~al.}(2014)\citenamefont
  {Hussein}, \citenamefont {Leamy},\ and\ \citenamefont
  {Ruzzene}}]{Hussein2014}%
  \BibitemOpen
  \bibfield  {author} {\bibinfo {author} {\bibfnamefont {M.~I.}\ \bibnamefont
  {Hussein}}, \bibinfo {author} {\bibfnamefont {M.~J.}\ \bibnamefont {Leamy}},\
  and\ \bibinfo {author} {\bibfnamefont {M.}~\bibnamefont {Ruzzene}},\
  }\bibfield  {title} {\bibinfo {title} {Dynamics of phononic materials and
  structures: Historical origins, recent progress, and future outlook},\
  }\bibfield  {journal} {\bibinfo  {journal} {Applied Mechanics Reviews}\
  }\textbf {\bibinfo {volume} {66}},\ \href {https://doi.org/10.1115/1.4026911}
  {10.1115/1.4026911} (\bibinfo {year} {2014})\BibitemShut {NoStop}%
\bibitem [{\citenamefont {Tol}\ \emph {et~al.}(2019)\citenamefont {Tol},
  \citenamefont {Degertekin},\ and\ \citenamefont {Erturk}}]{Tol2019}%
  \BibitemOpen
  \bibfield  {author} {\bibinfo {author} {\bibfnamefont {S.}~\bibnamefont
  {Tol}}, \bibinfo {author} {\bibfnamefont {F.}~\bibnamefont {Degertekin}},\
  and\ \bibinfo {author} {\bibfnamefont {A.}~\bibnamefont {Erturk}},\
  }\bibfield  {title} {\bibinfo {title} {3d-printed phononic crystal lens for
  elastic wave focusing and energy harvesting},\ }\href
  {https://doi.org/10.1016/j.addma.2019.100780} {\bibfield  {journal} {\bibinfo
   {journal} {Additive Manufacturing}\ }\textbf {\bibinfo {volume} {29}},\
  \bibinfo {pages} {100780} (\bibinfo {year} {2019})}\BibitemShut {NoStop}%
\bibitem [{\citenamefont {Chen}\ \emph {et~al.}(2014)\citenamefont {Chen},
  \citenamefont {Guo}, \citenamefont {Yang},\ and\ \citenamefont
  {Cheng}}]{Chen2014}%
  \BibitemOpen
  \bibfield  {author} {\bibinfo {author} {\bibfnamefont {Z.}~\bibnamefont
  {Chen}}, \bibinfo {author} {\bibfnamefont {B.}~\bibnamefont {Guo}}, \bibinfo
  {author} {\bibfnamefont {Y.}~\bibnamefont {Yang}},\ and\ \bibinfo {author}
  {\bibfnamefont {C.}~\bibnamefont {Cheng}},\ }\bibfield  {title} {\bibinfo
  {title} {Metamaterials-based enhanced energy harvesting: A review},\ }\href
  {https://doi.org/10.1016/j.physb.2013.12.040} {\bibfield  {journal} {\bibinfo
   {journal} {Physica B: Condensed Matter}\ }\textbf {\bibinfo {volume}
  {438}},\ \bibinfo {pages} {1} (\bibinfo {year} {2014})}\BibitemShut {NoStop}%
\bibitem [{\citenamefont {Qi}\ \emph {et~al.}(2016)\citenamefont {Qi},
  \citenamefont {Oudich}, \citenamefont {Li},\ and\ \citenamefont
  {Assouar}}]{Qi2016}%
  \BibitemOpen
  \bibfield  {author} {\bibinfo {author} {\bibfnamefont {S.}~\bibnamefont
  {Qi}}, \bibinfo {author} {\bibfnamefont {M.}~\bibnamefont {Oudich}}, \bibinfo
  {author} {\bibfnamefont {Y.}~\bibnamefont {Li}},\ and\ \bibinfo {author}
  {\bibfnamefont {B.}~\bibnamefont {Assouar}},\ }\bibfield  {title} {\bibinfo
  {title} {Acoustic energy harvesting based on a planar acoustic
  metamaterial},\ }\href {https://doi.org/10.1063/1.4954987} {\bibfield
  {journal} {\bibinfo  {journal} {Applied Physics Letters}\ }\textbf {\bibinfo
  {volume} {108}},\ \bibinfo {pages} {263501} (\bibinfo {year}
  {2016})}\BibitemShut {NoStop}%
\bibitem [{\citenamefont {Carrara}\ \emph {et~al.}(2013)\citenamefont
  {Carrara}, \citenamefont {Cacan}, \citenamefont {Toussaint}, \citenamefont
  {Leamy}, \citenamefont {Ruzzene},\ and\ \citenamefont
  {Erturk}}]{Carrara2013}%
  \BibitemOpen
  \bibfield  {author} {\bibinfo {author} {\bibfnamefont {M.}~\bibnamefont
  {Carrara}}, \bibinfo {author} {\bibfnamefont {M.~R.}\ \bibnamefont {Cacan}},
  \bibinfo {author} {\bibfnamefont {J.}~\bibnamefont {Toussaint}}, \bibinfo
  {author} {\bibfnamefont {M.~J.}\ \bibnamefont {Leamy}}, \bibinfo {author}
  {\bibfnamefont {M.}~\bibnamefont {Ruzzene}},\ and\ \bibinfo {author}
  {\bibfnamefont {A.}~\bibnamefont {Erturk}},\ }\bibfield  {title} {\bibinfo
  {title} {Metamaterial-inspired structures and concepts for elastoacoustic
  wave energy harvesting},\ }\href
  {https://doi.org/10.1088/0964-1726/22/6/065004} {\bibfield  {journal}
  {\bibinfo  {journal} {Smart Materials and Structures}\ }\textbf {\bibinfo
  {volume} {22}},\ \bibinfo {pages} {065004} (\bibinfo {year}
  {2013})}\BibitemShut {NoStop}%
\bibitem [{\citenamefont {Reichl}\ and\ \citenamefont
  {Inman}(2017)}]{Reichl2017}%
  \BibitemOpen
  \bibfield  {author} {\bibinfo {author} {\bibfnamefont {K.~K.}\ \bibnamefont
  {Reichl}}\ and\ \bibinfo {author} {\bibfnamefont {D.~J.}\ \bibnamefont
  {Inman}},\ }\bibfield  {title} {\bibinfo {title} {Lumped mass model of a 1d
  metastructure for vibration suppression with no additional mass},\ }\href
  {https://doi.org/10.1016/j.jsv.2017.05.026} {\bibfield  {journal} {\bibinfo
  {journal} {Journal of Sound and Vibration}\ }\textbf {\bibinfo {volume}
  {403}},\ \bibinfo {pages} {75} (\bibinfo {year} {2017})}\BibitemShut
  {NoStop}%
\bibitem [{\citenamefont {Matlack}\ \emph {et~al.}(2016)\citenamefont
  {Matlack}, \citenamefont {Bauhofer}, \citenamefont {Krödel}, \citenamefont
  {Palermo},\ and\ \citenamefont {Daraio}}]{Matlack2016}%
  \BibitemOpen
  \bibfield  {author} {\bibinfo {author} {\bibfnamefont {K.~H.}\ \bibnamefont
  {Matlack}}, \bibinfo {author} {\bibfnamefont {A.}~\bibnamefont {Bauhofer}},
  \bibinfo {author} {\bibfnamefont {S.}~\bibnamefont {Krödel}}, \bibinfo
  {author} {\bibfnamefont {A.}~\bibnamefont {Palermo}},\ and\ \bibinfo {author}
  {\bibfnamefont {C.}~\bibnamefont {Daraio}},\ }\bibfield  {title} {\bibinfo
  {title} {Composite 3d-printed metastructures for low-frequency and broadband
  vibration absorption},\ }\href {https://doi.org/10.1073/pnas.1600171113}
  {\bibfield  {journal} {\bibinfo  {journal} {Proceedings of the National
  Academy of Sciences}\ }\textbf {\bibinfo {volume} {113}},\ \bibinfo {pages}
  {8386} (\bibinfo {year} {2016})}\BibitemShut {NoStop}%
\bibitem [{\citenamefont {Mueller}\ \emph {et~al.}(2019)\citenamefont
  {Mueller}, \citenamefont {Matlack}, \citenamefont {Shea},\ and\ \citenamefont
  {Daraio}}]{Mueller2019}%
  \BibitemOpen
  \bibfield  {author} {\bibinfo {author} {\bibfnamefont {J.}~\bibnamefont
  {Mueller}}, \bibinfo {author} {\bibfnamefont {K.~H.}\ \bibnamefont
  {Matlack}}, \bibinfo {author} {\bibfnamefont {K.}~\bibnamefont {Shea}},\ and\
  \bibinfo {author} {\bibfnamefont {C.}~\bibnamefont {Daraio}},\ }\bibfield
  {title} {\bibinfo {title} {Energy absorption properties of periodic and
  stochastic 3d lattice materials},\ }\href
  {https://doi.org/10.1002/adts.201900081} {\bibfield  {journal} {\bibinfo
  {journal} {Advanced Theory and Simulations}\ }\textbf {\bibinfo {volume}
  {2}},\ \bibinfo {pages} {1900081} (\bibinfo {year} {2019})}\BibitemShut
  {NoStop}%
\bibitem [{\citenamefont {Sirota}\ \emph {et~al.}(2021)\citenamefont {Sirota},
  \citenamefont {Sabsovich}, \citenamefont {Lahini}, \citenamefont {Ilan},\
  and\ \citenamefont {Shokef}}]{Sirota2021}%
  \BibitemOpen
  \bibfield  {author} {\bibinfo {author} {\bibfnamefont {L.}~\bibnamefont
  {Sirota}}, \bibinfo {author} {\bibfnamefont {D.}~\bibnamefont {Sabsovich}},
  \bibinfo {author} {\bibfnamefont {Y.}~\bibnamefont {Lahini}}, \bibinfo
  {author} {\bibfnamefont {R.}~\bibnamefont {Ilan}},\ and\ \bibinfo {author}
  {\bibfnamefont {Y.}~\bibnamefont {Shokef}},\ }\bibfield  {title} {\bibinfo
  {title} {Real-time steering of curved sound beams in a feedback-based
  topological acoustic metamaterial},\ }\href
  {https://doi.org/10.1016/j.ymssp.2020.107479} {\bibfield  {journal} {\bibinfo
   {journal} {Mechanical Systems and Signal Processing}\ }\textbf {\bibinfo
  {volume} {153}},\ \bibinfo {pages} {107479} (\bibinfo {year}
  {2021})}\BibitemShut {NoStop}%
\bibitem [{\citenamefont {Xia}\ \emph {et~al.}(2018)\citenamefont {Xia},
  \citenamefont {Jia}, \citenamefont {Sun}, \citenamefont {Yuan}, \citenamefont
  {Ge}, \citenamefont {Si},\ and\ \citenamefont {Liu}}]{Xia2018}%
  \BibitemOpen
  \bibfield  {author} {\bibinfo {author} {\bibfnamefont {J.-P.}\ \bibnamefont
  {Xia}}, \bibinfo {author} {\bibfnamefont {D.}~\bibnamefont {Jia}}, \bibinfo
  {author} {\bibfnamefont {H.-X.}\ \bibnamefont {Sun}}, \bibinfo {author}
  {\bibfnamefont {S.-Q.}\ \bibnamefont {Yuan}}, \bibinfo {author}
  {\bibfnamefont {Y.}~\bibnamefont {Ge}}, \bibinfo {author} {\bibfnamefont
  {Q.-R.}\ \bibnamefont {Si}},\ and\ \bibinfo {author} {\bibfnamefont {X.-J.}\
  \bibnamefont {Liu}},\ }\bibfield  {title} {\bibinfo {title} {Programmable
  coding acoustic topological insulator},\ }\href
  {https://doi.org/10.1002/adma.201805002} {\bibfield  {journal} {\bibinfo
  {journal} {Advanced Materials}\ }\textbf {\bibinfo {volume} {30}},\ \bibinfo
  {pages} {1805002} (\bibinfo {year} {2018})}\BibitemShut {NoStop}%
\bibitem [{\citenamefont {Zangeneh-Nejad}\ and\ \citenamefont
  {Fleury}(2019)}]{ZangenehNejad2019}%
  \BibitemOpen
  \bibfield  {author} {\bibinfo {author} {\bibfnamefont {F.}~\bibnamefont
  {Zangeneh-Nejad}}\ and\ \bibinfo {author} {\bibfnamefont {R.}~\bibnamefont
  {Fleury}},\ }\bibfield  {title} {\bibinfo {title} {Topological analog signal
  processing},\ }\bibfield  {journal} {\bibinfo  {journal} {Nature
  Communications}\ }\textbf {\bibinfo {volume} {10}},\ \href
  {https://doi.org/10.1038/s41467-019-10086-3} {10.1038/s41467-019-10086-3}
  (\bibinfo {year} {2019})\BibitemShut {NoStop}%
\bibitem [{\citenamefont {Pal}\ and\ \citenamefont {Ruzzene}(2017)}]{Pal2017}%
  \BibitemOpen
  \bibfield  {author} {\bibinfo {author} {\bibfnamefont {R.~K.}\ \bibnamefont
  {Pal}}\ and\ \bibinfo {author} {\bibfnamefont {M.}~\bibnamefont {Ruzzene}},\
  }\bibfield  {title} {\bibinfo {title} {Edge waves in plates with resonators:
  an elastic analogue of the quantum valley hall effect},\ }\href
  {https://doi.org/10.1088/1367-2630/aa56a2} {\bibfield  {journal} {\bibinfo
  {journal} {New Journal of Physics}\ }\textbf {\bibinfo {volume} {19}},\
  \bibinfo {pages} {025001} (\bibinfo {year} {2017})}\BibitemShut {NoStop}%
\bibitem [{\citenamefont {Chen}\ \emph {et~al.}(2018)\citenamefont {Chen},
  \citenamefont {Nassar},\ and\ \citenamefont {Huang}}]{Chen2018}%
  \BibitemOpen
  \bibfield  {author} {\bibinfo {author} {\bibfnamefont {H.}~\bibnamefont
  {Chen}}, \bibinfo {author} {\bibfnamefont {H.}~\bibnamefont {Nassar}},\ and\
  \bibinfo {author} {\bibfnamefont {G.}~\bibnamefont {Huang}},\ }\bibfield
  {title} {\bibinfo {title} {A study of topological effects in 1d and 2d
  mechanical lattices},\ }\href {https://doi.org/10.1016/j.jmps.2018.04.013}
  {\bibfield  {journal} {\bibinfo  {journal} {Journal of the Mechanics and
  Physics of Solids}\ }\textbf {\bibinfo {volume} {117}},\ \bibinfo {pages}
  {22} (\bibinfo {year} {2018})}\BibitemShut {NoStop}%
\bibitem [{\citenamefont {Tempelman}\ \emph
  {et~al.}(2022{\natexlab{a}})\citenamefont {Tempelman}, \citenamefont
  {Vakakis},\ and\ \citenamefont {Matlack}}]{Tempelman2022a}%
  \BibitemOpen
  \bibfield  {author} {\bibinfo {author} {\bibfnamefont {J.~R.}\ \bibnamefont
  {Tempelman}}, \bibinfo {author} {\bibfnamefont {A.~F.}\ \bibnamefont
  {Vakakis}},\ and\ \bibinfo {author} {\bibfnamefont {K.~H.}\ \bibnamefont
  {Matlack}},\ }\bibfield  {title} {\bibinfo {title} {A modal decomposition
  approach to topological wave propagation},\ }\href@noop {} {\  (\bibinfo
  {year} {2022}{\natexlab{a}})},\ \Eprint {https://arxiv.org/abs/2210.11665}
  {arXiv:2210.11665 [physics.app-ph]} \BibitemShut {NoStop}%
\bibitem [{\citenamefont {Patil}\ and\ \citenamefont
  {Matlack}(2021)}]{Patil2021}%
  \BibitemOpen
  \bibfield  {author} {\bibinfo {author} {\bibfnamefont {G.~U.}\ \bibnamefont
  {Patil}}\ and\ \bibinfo {author} {\bibfnamefont {K.~H.}\ \bibnamefont
  {Matlack}},\ }\bibfield  {title} {\bibinfo {title} {Review of exploiting
  nonlinearity in phononic materials to enable nonlinear wave responses},\
  }\href {https://doi.org/10.1007/s00707-021-03089-z} {\bibfield  {journal}
  {\bibinfo  {journal} {Acta Mechanica}\ }\textbf {\bibinfo {volume} {233}},\
  \bibinfo {pages} {1} (\bibinfo {year} {2021})}\BibitemShut {NoStop}%
\bibitem [{\citenamefont {Vasseur}\ \emph {et~al.}(2011)\citenamefont
  {Vasseur}, \citenamefont {Matar}, \citenamefont {Robillard}, \citenamefont
  {Hladky-Hennion},\ and\ \citenamefont {Deymier}}]{Vasseur2011}%
  \BibitemOpen
  \bibfield  {author} {\bibinfo {author} {\bibfnamefont {J.~O.}\ \bibnamefont
  {Vasseur}}, \bibinfo {author} {\bibfnamefont {O.~B.}\ \bibnamefont {Matar}},
  \bibinfo {author} {\bibfnamefont {J.~F.}\ \bibnamefont {Robillard}}, \bibinfo
  {author} {\bibfnamefont {A.-C.}\ \bibnamefont {Hladky-Hennion}},\ and\
  \bibinfo {author} {\bibfnamefont {P.~A.}\ \bibnamefont {Deymier}},\
  }\bibfield  {title} {\bibinfo {title} {Band structures tunability of bulk 2d
  phononic crystals made of magneto-elastic materials},\ }\href
  {https://doi.org/10.1063/1.3676172} {\bibfield  {journal} {\bibinfo
  {journal} {{AIP} Advances}\ }\textbf {\bibinfo {volume} {1}},\ \bibinfo
  {pages} {041904} (\bibinfo {year} {2011})}\BibitemShut {NoStop}%
\bibitem [{\citenamefont {Allein}\ \emph {et~al.}(2016)\citenamefont {Allein},
  \citenamefont {Tournat}, \citenamefont {Gusev},\ and\ \citenamefont
  {Theocharis}}]{Allein2016}%
  \BibitemOpen
  \bibfield  {author} {\bibinfo {author} {\bibfnamefont {F.}~\bibnamefont
  {Allein}}, \bibinfo {author} {\bibfnamefont {V.}~\bibnamefont {Tournat}},
  \bibinfo {author} {\bibfnamefont {V.~E.}\ \bibnamefont {Gusev}},\ and\
  \bibinfo {author} {\bibfnamefont {G.}~\bibnamefont {Theocharis}},\ }\bibfield
   {title} {\bibinfo {title} {Tunable magneto-granular phononic crystals},\
  }\href {https://doi.org/10.1063/1.4947192} {\bibfield  {journal} {\bibinfo
  {journal} {Applied Physics Letters}\ }\textbf {\bibinfo {volume} {108}},\
  \bibinfo {pages} {161903} (\bibinfo {year} {2016})}\BibitemShut {NoStop}%
\bibitem [{\citenamefont {Pierce}\ \emph {et~al.}(2020)\citenamefont {Pierce},
  \citenamefont {Willey}, \citenamefont {Chen}, \citenamefont {Hardin},
  \citenamefont {Berrigan}, \citenamefont {Juhl},\ and\ \citenamefont
  {Matlack}}]{Pierce2020}%
  \BibitemOpen
  \bibfield  {author} {\bibinfo {author} {\bibfnamefont {C.~D.}\ \bibnamefont
  {Pierce}}, \bibinfo {author} {\bibfnamefont {C.~L.}\ \bibnamefont {Willey}},
  \bibinfo {author} {\bibfnamefont {V.~W.}\ \bibnamefont {Chen}}, \bibinfo
  {author} {\bibfnamefont {J.~O.}\ \bibnamefont {Hardin}}, \bibinfo {author}
  {\bibfnamefont {J.~D.}\ \bibnamefont {Berrigan}}, \bibinfo {author}
  {\bibfnamefont {A.~T.}\ \bibnamefont {Juhl}},\ and\ \bibinfo {author}
  {\bibfnamefont {K.~H.}\ \bibnamefont {Matlack}},\ }\bibfield  {title}
  {\bibinfo {title} {Adaptive elastic metastructures from magneto-active
  elastomers},\ }\href {https://doi.org/10.1088/1361-665x/ab80e4} {\bibfield
  {journal} {\bibinfo  {journal} {Smart Materials and Structures}\ }\textbf
  {\bibinfo {volume} {29}},\ \bibinfo {pages} {065004} (\bibinfo {year}
  {2020})}\BibitemShut {NoStop}%
\bibitem [{\citenamefont {Cheng}\ \emph {et~al.}(2011)\citenamefont {Cheng},
  \citenamefont {Liu},\ and\ \citenamefont {Wu}}]{Cheng2011}%
  \BibitemOpen
  \bibfield  {author} {\bibinfo {author} {\bibfnamefont {Y.}~\bibnamefont
  {Cheng}}, \bibinfo {author} {\bibfnamefont {X.~J.}\ \bibnamefont {Liu}},\
  and\ \bibinfo {author} {\bibfnamefont {D.~J.}\ \bibnamefont {Wu}},\
  }\bibfield  {title} {\bibinfo {title} {Temperature effects on the band gaps
  of lamb waves in a one-dimensional phononic-crystal plate (l)},\ }\href
  {https://doi.org/10.1121/1.3543970} {\bibfield  {journal} {\bibinfo
  {journal} {The Journal of the Acoustical Society of America}\ }\textbf
  {\bibinfo {volume} {129}},\ \bibinfo {pages} {1157} (\bibinfo {year}
  {2011})}\BibitemShut {NoStop}%
\bibitem [{\citenamefont {Yao}\ \emph {et~al.}(2011)\citenamefont {Yao},
  \citenamefont {Wu}, \citenamefont {Zhang},\ and\ \citenamefont
  {Hou}}]{Yao2011}%
  \BibitemOpen
  \bibfield  {author} {\bibinfo {author} {\bibfnamefont {Y.}~\bibnamefont
  {Yao}}, \bibinfo {author} {\bibfnamefont {F.}~\bibnamefont {Wu}}, \bibinfo
  {author} {\bibfnamefont {X.}~\bibnamefont {Zhang}},\ and\ \bibinfo {author}
  {\bibfnamefont {Z.}~\bibnamefont {Hou}},\ }\bibfield  {title} {\bibinfo
  {title} {Thermal tuning of lamb wave band structure in a two-dimensional
  phononic crystal plate},\ }\href {https://doi.org/10.1063/1.3669391}
  {\bibfield  {journal} {\bibinfo  {journal} {Journal of Applied Physics}\
  }\textbf {\bibinfo {volume} {110}},\ \bibinfo {pages} {123503} (\bibinfo
  {year} {2011})}\BibitemShut {NoStop}%
\bibitem [{\citenamefont {Zhao}\ \emph {et~al.}(2022)\citenamefont {Zhao},
  \citenamefont {Cui}, \citenamefont {Yin}, \citenamefont {Li},\ and\
  \citenamefont {Li}}]{Zhao2022}%
  \BibitemOpen
  \bibfield  {author} {\bibinfo {author} {\bibfnamefont {Z.}~\bibnamefont
  {Zhao}}, \bibinfo {author} {\bibfnamefont {X.}~\bibnamefont {Cui}}, \bibinfo
  {author} {\bibfnamefont {Y.}~\bibnamefont {Yin}}, \bibinfo {author}
  {\bibfnamefont {Y.}~\bibnamefont {Li}},\ and\ \bibinfo {author}
  {\bibfnamefont {M.}~\bibnamefont {Li}},\ }\bibfield  {title} {\bibinfo
  {title} {Thermal tuning of vibration band gaps in homogenous metamaterial
  plate},\ }\href {https://doi.org/10.1016/j.ijmecsci.2022.107374} {\bibfield
  {journal} {\bibinfo  {journal} {International Journal of Mechanical
  Sciences}\ }\textbf {\bibinfo {volume} {225}},\ \bibinfo {pages} {107374}
  (\bibinfo {year} {2022})}\BibitemShut {NoStop}%
\bibitem [{\citenamefont {Denz}\ \emph {et~al.}(2010)\citenamefont {Denz},
  \citenamefont {Flach},\ and\ \citenamefont {Kivshar}}]{Denz2010}%
  \BibitemOpen
  \bibinfo {editor} {\bibfnamefont {C.}~\bibnamefont {Denz}}, \bibinfo {editor}
  {\bibfnamefont {S.}~\bibnamefont {Flach}},\ and\ \bibinfo {editor}
  {\bibfnamefont {Y.~S.}\ \bibnamefont {Kivshar}},\ eds.,\ \href
  {https://doi.org/10.1007/978-3-642-02066-7} {\emph {\bibinfo {title}
  {Nonlinearities in Periodic Structures and Metamaterials}}}\ (\bibinfo
  {publisher} {Springer Berlin Heidelberg},\ \bibinfo {year}
  {2010})\BibitemShut {NoStop}%
\bibitem [{\citenamefont {Li}\ \emph {et~al.}(2012)\citenamefont {Li},
  \citenamefont {Ngo}, \citenamefont {Yang},\ and\ \citenamefont
  {Daraio}}]{Li2012}%
  \BibitemOpen
  \bibfield  {author} {\bibinfo {author} {\bibfnamefont {F.}~\bibnamefont
  {Li}}, \bibinfo {author} {\bibfnamefont {D.}~\bibnamefont {Ngo}}, \bibinfo
  {author} {\bibfnamefont {J.}~\bibnamefont {Yang}},\ and\ \bibinfo {author}
  {\bibfnamefont {C.}~\bibnamefont {Daraio}},\ }\bibfield  {title} {\bibinfo
  {title} {Tunable phononic crystals based on cylindrical hertzian contact},\
  }\href {https://doi.org/10.1063/1.4762832} {\bibfield  {journal} {\bibinfo
  {journal} {Applied Physics Letters}\ }\textbf {\bibinfo {volume} {101}},\
  \bibinfo {pages} {171903} (\bibinfo {year} {2012})}\BibitemShut {NoStop}%
\bibitem [{\citenamefont {Chaunsali}\ \emph {et~al.}(2016)\citenamefont
  {Chaunsali}, \citenamefont {Li},\ and\ \citenamefont {Yang}}]{Chaunsali2016}%
  \BibitemOpen
  \bibfield  {author} {\bibinfo {author} {\bibfnamefont {R.}~\bibnamefont
  {Chaunsali}}, \bibinfo {author} {\bibfnamefont {F.}~\bibnamefont {Li}},\ and\
  \bibinfo {author} {\bibfnamefont {J.}~\bibnamefont {Yang}},\ }\bibfield
  {title} {\bibinfo {title} {Stress wave isolation by purely mechanical
  topological phononic crystals},\ }\bibfield  {journal} {\bibinfo  {journal}
  {Scientific Reports}\ }\textbf {\bibinfo {volume} {6}},\ \href
  {https://doi.org/10.1038/srep30662} {10.1038/srep30662} (\bibinfo {year}
  {2016})\BibitemShut {NoStop}%
\bibitem [{\citenamefont {Chaunsali}\ \emph {et~al.}(2017)\citenamefont
  {Chaunsali}, \citenamefont {Kim}, \citenamefont {Thakkar}, \citenamefont
  {Kevrekidis},\ and\ \citenamefont {Yang}}]{Chaunsali2017}%
  \BibitemOpen
  \bibfield  {author} {\bibinfo {author} {\bibfnamefont {R.}~\bibnamefont
  {Chaunsali}}, \bibinfo {author} {\bibfnamefont {E.}~\bibnamefont {Kim}},
  \bibinfo {author} {\bibfnamefont {A.}~\bibnamefont {Thakkar}}, \bibinfo
  {author} {\bibfnamefont {P.}~\bibnamefont {Kevrekidis}},\ and\ \bibinfo
  {author} {\bibfnamefont {J.}~\bibnamefont {Yang}},\ }\bibfield  {title}
  {\bibinfo {title} {Demonstrating an lessigreaterin situless/igreater
  topological band transition in cylindrical granular chains},\ }\href
  {https://doi.org/10.1103/physrevlett.119.024301} {\bibfield  {journal}
  {\bibinfo  {journal} {Physical Review Letters}\ }\textbf {\bibinfo {volume}
  {119}},\ \bibinfo {pages} {024301} (\bibinfo {year} {2017})}\BibitemShut
  {NoStop}%
\bibitem [{\citenamefont {Narisetti}\ \emph {et~al.}(2010)\citenamefont
  {Narisetti}, \citenamefont {Leamy},\ and\ \citenamefont
  {Ruzzene}}]{Narisetti2010}%
  \BibitemOpen
  \bibfield  {author} {\bibinfo {author} {\bibfnamefont {R.~K.}\ \bibnamefont
  {Narisetti}}, \bibinfo {author} {\bibfnamefont {M.~J.}\ \bibnamefont
  {Leamy}},\ and\ \bibinfo {author} {\bibfnamefont {M.}~\bibnamefont
  {Ruzzene}},\ }\bibfield  {title} {\bibinfo {title} {A perturbation approach
  for predicting wave propagation in one-dimensional nonlinear periodic
  structures},\ }\bibfield  {journal} {\bibinfo  {journal} {Journal of
  Vibration and Acoustics}\ }\textbf {\bibinfo {volume} {132}},\ \href
  {https://doi.org/10.1115/1.4000775} {10.1115/1.4000775} (\bibinfo {year}
  {2010})\BibitemShut {NoStop}%
\bibitem [{\citenamefont {Tempelman}\ \emph {et~al.}(2021)\citenamefont
  {Tempelman}, \citenamefont {Matlack},\ and\ \citenamefont
  {Vakakis}}]{Tempelman2021}%
  \BibitemOpen
  \bibfield  {author} {\bibinfo {author} {\bibfnamefont {J.~R.}\ \bibnamefont
  {Tempelman}}, \bibinfo {author} {\bibfnamefont {K.~H.}\ \bibnamefont
  {Matlack}},\ and\ \bibinfo {author} {\bibfnamefont {A.~F.}\ \bibnamefont
  {Vakakis}},\ }\bibfield  {title} {\bibinfo {title} {Topological protection in
  a strongly nonlinear interface lattice},\ }\href
  {https://doi.org/10.1103/physrevb.104.174306} {\bibfield  {journal} {\bibinfo
   {journal} {Physical Review B}\ }\textbf {\bibinfo {volume} {104}},\ \bibinfo
  {pages} {174306} (\bibinfo {year} {2021})}\BibitemShut {NoStop}%
\bibitem [{\citenamefont {Jayaprakash}\ \emph {et~al.}(2010)\citenamefont
  {Jayaprakash}, \citenamefont {Starosvetsky}, \citenamefont {Vakakis},
  \citenamefont {Peeters},\ and\ \citenamefont {Kerschen}}]{Jayaprakash2010}%
  \BibitemOpen
  \bibfield  {author} {\bibinfo {author} {\bibfnamefont {K.~R.}\ \bibnamefont
  {Jayaprakash}}, \bibinfo {author} {\bibfnamefont {Y.}~\bibnamefont
  {Starosvetsky}}, \bibinfo {author} {\bibfnamefont {A.~F.}\ \bibnamefont
  {Vakakis}}, \bibinfo {author} {\bibfnamefont {M.}~\bibnamefont {Peeters}},\
  and\ \bibinfo {author} {\bibfnamefont {G.}~\bibnamefont {Kerschen}},\
  }\bibfield  {title} {\bibinfo {title} {Nonlinear normal modes and band zones
  in granular chains with no pre-compression},\ }\href
  {https://doi.org/10.1007/s11071-010-9809-0} {\bibfield  {journal} {\bibinfo
  {journal} {Nonlinear Dynamics}\ }\textbf {\bibinfo {volume} {63}},\ \bibinfo
  {pages} {359} (\bibinfo {year} {2010})}\BibitemShut {NoStop}%
\bibitem [{\citenamefont {Mojahed}\ and\ \citenamefont
  {Vakakis}(2019)}]{Mojahed2019}%
  \BibitemOpen
  \bibfield  {author} {\bibinfo {author} {\bibfnamefont {A.}~\bibnamefont
  {Mojahed}}\ and\ \bibinfo {author} {\bibfnamefont {A.~F.}\ \bibnamefont
  {Vakakis}},\ }\bibfield  {title} {\bibinfo {title} {Certain aspects of the
  acoustics of a strongly nonlinear discrete lattice},\ }\href
  {https://doi.org/10.1007/s11071-019-05080-9} {\bibfield  {journal} {\bibinfo
  {journal} {Nonlinear Dynamics}\ }\textbf {\bibinfo {volume} {99}},\ \bibinfo
  {pages} {643} (\bibinfo {year} {2019})}\BibitemShut {NoStop}%
\bibitem [{\citenamefont {Mojahed}\ \emph {et~al.}(2019)\citenamefont
  {Mojahed}, \citenamefont {Gendelman},\ and\ \citenamefont
  {Vakakis}}]{Mojahed2019a}%
  \BibitemOpen
  \bibfield  {author} {\bibinfo {author} {\bibfnamefont {A.}~\bibnamefont
  {Mojahed}}, \bibinfo {author} {\bibfnamefont {O.~V.}\ \bibnamefont
  {Gendelman}},\ and\ \bibinfo {author} {\bibfnamefont {A.~F.}\ \bibnamefont
  {Vakakis}},\ }\bibfield  {title} {\bibinfo {title} {Breather arrest,
  localization, and acoustic non-reciprocity in dissipative nonlinear
  lattices},\ }\href {https://doi.org/10.1121/1.5114915} {\bibfield  {journal}
  {\bibinfo  {journal} {The Journal of the Acoustical Society of America}\
  }\textbf {\bibinfo {volume} {146}},\ \bibinfo {pages} {826} (\bibinfo {year}
  {2019})}\BibitemShut {NoStop}%
\bibitem [{\citenamefont {Patil}\ and\ \citenamefont
  {Matlack}(2022)}]{Patil2022}%
  \BibitemOpen
  \bibfield  {author} {\bibinfo {author} {\bibfnamefont {G.~U.}\ \bibnamefont
  {Patil}}\ and\ \bibinfo {author} {\bibfnamefont {K.~H.}\ \bibnamefont
  {Matlack}},\ }\bibfield  {title} {\bibinfo {title} {Strongly nonlinear wave
  dynamics of continuum phononic materials with periodic rough contacts},\
  }\href {https://doi.org/10.1103/physreve.105.024201} {\bibfield  {journal}
  {\bibinfo  {journal} {Physical Review E}\ }\textbf {\bibinfo {volume}
  {105}},\ \bibinfo {pages} {024201} (\bibinfo {year} {2022})}\BibitemShut
  {NoStop}%
\bibitem [{\citenamefont {Zaera}\ \emph {et~al.}(2018)\citenamefont {Zaera},
  \citenamefont {Vila}, \citenamefont {Fernandez-Saez},\ and\ \citenamefont
  {Ruzzene}}]{Zaera2018}%
  \BibitemOpen
  \bibfield  {author} {\bibinfo {author} {\bibfnamefont {R.}~\bibnamefont
  {Zaera}}, \bibinfo {author} {\bibfnamefont {J.}~\bibnamefont {Vila}},
  \bibinfo {author} {\bibfnamefont {J.}~\bibnamefont {Fernandez-Saez}},\ and\
  \bibinfo {author} {\bibfnamefont {M.}~\bibnamefont {Ruzzene}},\ }\bibfield
  {title} {\bibinfo {title} {Propagation of solitons in a two-dimensional
  nonlinear square lattice},\ }\href
  {https://doi.org/10.1016/j.ijnonlinmec.2018.08.002} {\bibfield  {journal}
  {\bibinfo  {journal} {International Journal of Non-Linear Mechanics}\
  }\textbf {\bibinfo {volume} {106}},\ \bibinfo {pages} {188} (\bibinfo {year}
  {2018})}\BibitemShut {NoStop}%
\bibitem [{\citenamefont {Gendelman}\ and\ \citenamefont
  {Manevitch}(2008)}]{Gendelman2008}%
  \BibitemOpen
  \bibfield  {author} {\bibinfo {author} {\bibfnamefont {O.~V.}\ \bibnamefont
  {Gendelman}}\ and\ \bibinfo {author} {\bibfnamefont {L.~I.}\ \bibnamefont
  {Manevitch}},\ }\bibfield  {title} {\bibinfo {title} {Discrete breathers in
  vibroimpact chains: Analytic solutions},\ }\href
  {https://doi.org/10.1103/physreve.78.026609} {\bibfield  {journal} {\bibinfo
  {journal} {Physical Review E}\ }\textbf {\bibinfo {volume} {78}},\ \bibinfo
  {pages} {026609} (\bibinfo {year} {2008})}\BibitemShut {NoStop}%
\bibitem [{\citenamefont {Gendelman}(2013)}]{Gendelman2013}%
  \BibitemOpen
  \bibfield  {author} {\bibinfo {author} {\bibfnamefont {O.~V.}\ \bibnamefont
  {Gendelman}},\ }\bibfield  {title} {\bibinfo {title} {Exact solutions for
  discrete breathers in a forced-damped chain},\ }\href
  {https://doi.org/10.1103/physreve.87.062911} {\bibfield  {journal} {\bibinfo
  {journal} {Physical Review E}\ }\textbf {\bibinfo {volume} {87}},\ \bibinfo
  {pages} {062911} (\bibinfo {year} {2013})}\BibitemShut {NoStop}%
\bibitem [{\citenamefont {Vakakis}\ \emph {et~al.}(2022)\citenamefont
  {Vakakis}, \citenamefont {Gendelman}, \citenamefont {Bergman}, \citenamefont
  {Mojahed},\ and\ \citenamefont {Gzal}}]{Vakakis2022}%
  \BibitemOpen
  \bibfield  {author} {\bibinfo {author} {\bibfnamefont {A.~F.}\ \bibnamefont
  {Vakakis}}, \bibinfo {author} {\bibfnamefont {O.~V.}\ \bibnamefont
  {Gendelman}}, \bibinfo {author} {\bibfnamefont {L.~A.}\ \bibnamefont
  {Bergman}}, \bibinfo {author} {\bibfnamefont {A.}~\bibnamefont {Mojahed}},\
  and\ \bibinfo {author} {\bibfnamefont {M.}~\bibnamefont {Gzal}},\ }\bibfield
  {title} {\bibinfo {title} {Nonlinear targeted energy transfer: state of the
  art and new perspectives},\ }\href
  {https://doi.org/10.1007/s11071-022-07216-w} {\bibfield  {journal} {\bibinfo
  {journal} {Nonlinear Dynamics}\ }\textbf {\bibinfo {volume} {108}},\ \bibinfo
  {pages} {711} (\bibinfo {year} {2022})}\BibitemShut {NoStop}%
\bibitem [{\citenamefont {Wang}\ \emph {et~al.}(2015)\citenamefont {Wang},
  \citenamefont {Wierschem}, \citenamefont {Spencer},\ and\ \citenamefont
  {Lu}}]{Wang2015}%
  \BibitemOpen
  \bibfield  {author} {\bibinfo {author} {\bibfnamefont {J.}~\bibnamefont
  {Wang}}, \bibinfo {author} {\bibfnamefont {N.}~\bibnamefont {Wierschem}},
  \bibinfo {author} {\bibfnamefont {B.~F.}\ \bibnamefont {Spencer}},\ and\
  \bibinfo {author} {\bibfnamefont {X.}~\bibnamefont {Lu}},\ }\bibfield
  {title} {\bibinfo {title} {Numerical and experimental study of the
  performance of a single-sided vibro-impact track nonlinear energy sink},\
  }\href {https://doi.org/10.1002/eqe.2677} {\bibfield  {journal} {\bibinfo
  {journal} {Earthquake Engineering \& Structural Dynamics}\ }\textbf {\bibinfo
  {volume} {45}},\ \bibinfo {pages} {635} (\bibinfo {year} {2015})}\BibitemShut
  {NoStop}%
\bibitem [{\citenamefont {Vakakis}\ and\ \citenamefont
  {Gendelman}(2000)}]{Vakakis2000}%
  \BibitemOpen
  \bibfield  {author} {\bibinfo {author} {\bibfnamefont {A.~F.}\ \bibnamefont
  {Vakakis}}\ and\ \bibinfo {author} {\bibfnamefont {O.}~\bibnamefont
  {Gendelman}},\ }\bibfield  {title} {\bibinfo {title} {Energy pumping in
  nonlinear mechanical oscillators: Part {II}{\textemdash}resonance capture},\
  }\href {https://doi.org/10.1115/1.1345525} {\bibfield  {journal} {\bibinfo
  {journal} {Journal of Applied Mechanics}\ }\textbf {\bibinfo {volume} {68}},\
  \bibinfo {pages} {42} (\bibinfo {year} {2000})}\BibitemShut {NoStop}%
\bibitem [{\citenamefont {Vakakis}(2001)}]{Vakakis2001}%
  \BibitemOpen
  \bibfield  {author} {\bibinfo {author} {\bibfnamefont {A.~F.}\ \bibnamefont
  {Vakakis}},\ }\bibfield  {title} {\bibinfo {title} {Inducing passive
  nonlinear energy sinks in vibrating systems},\ }\href
  {https://doi.org/10.1115/1.1368883} {\bibfield  {journal} {\bibinfo
  {journal} {Journal of Vibration and Acoustics}\ }\textbf {\bibinfo {volume}
  {123}},\ \bibinfo {pages} {324} (\bibinfo {year} {2001})}\BibitemShut
  {NoStop}%
\bibitem [{\citenamefont {Taleshi}\ \emph {et~al.}(2016)\citenamefont
  {Taleshi}, \citenamefont {Dardel},\ and\ \citenamefont
  {Pashaie}}]{Taleshi2016}%
  \BibitemOpen
  \bibfield  {author} {\bibinfo {author} {\bibfnamefont {M.}~\bibnamefont
  {Taleshi}}, \bibinfo {author} {\bibfnamefont {M.}~\bibnamefont {Dardel}},\
  and\ \bibinfo {author} {\bibfnamefont {M.~H.}\ \bibnamefont {Pashaie}},\
  }\bibfield  {title} {\bibinfo {title} {Passive targeted energy transfer in
  the steady state dynamics of a nonlinear plate with nonlinear absorber},\
  }\href {https://doi.org/10.1016/j.chaos.2016.09.017} {\bibfield  {journal}
  {\bibinfo  {journal} {Chaos, Solitons \& Fractals}\ }\textbf {\bibinfo
  {volume} {92}},\ \bibinfo {pages} {56} (\bibinfo {year} {2016})}\BibitemShut
  {NoStop}%
\bibitem [{\citenamefont {Starosvetsky}\ and\ \citenamefont
  {Gendelman}(2007)}]{Starosvetsky2007}%
  \BibitemOpen
  \bibfield  {author} {\bibinfo {author} {\bibfnamefont {Y.}~\bibnamefont
  {Starosvetsky}}\ and\ \bibinfo {author} {\bibfnamefont {O.~V.}\ \bibnamefont
  {Gendelman}},\ }\bibfield  {title} {\bibinfo {title} {Attractors of
  harmonically forced linear oscillator with attached nonlinear energy sink.
  {II}: Optimization of a nonlinear vibration absorber},\ }\href
  {https://doi.org/10.1007/s11071-006-9168-z} {\bibfield  {journal} {\bibinfo
  {journal} {Nonlinear Dynamics}\ }\textbf {\bibinfo {volume} {51}},\ \bibinfo
  {pages} {47} (\bibinfo {year} {2007})}\BibitemShut {NoStop}%
\bibitem [{\citenamefont {Starosvetsky}\ and\ \citenamefont
  {Gendelman}(2010)}]{Starosvetsky2010}%
  \BibitemOpen
  \bibfield  {author} {\bibinfo {author} {\bibfnamefont {Y.}~\bibnamefont
  {Starosvetsky}}\ and\ \bibinfo {author} {\bibfnamefont {O.}~\bibnamefont
  {Gendelman}},\ }\bibfield  {title} {\bibinfo {title} {Interaction of
  nonlinear energy sink with a two degrees of freedom linear system: Internal
  resonance},\ }\href {https://doi.org/10.1016/j.jsv.2009.11.025} {\bibfield
  {journal} {\bibinfo  {journal} {Journal of Sound and Vibration}\ }\textbf
  {\bibinfo {volume} {329}},\ \bibinfo {pages} {1836} (\bibinfo {year}
  {2010})}\BibitemShut {NoStop}%
\bibitem [{\citenamefont {Qiu}\ \emph {et~al.}(2017)\citenamefont {Qiu},
  \citenamefont {Seguy},\ and\ \citenamefont {Paredes}}]{Qiu2017}%
  \BibitemOpen
  \bibfield  {author} {\bibinfo {author} {\bibfnamefont {D.}~\bibnamefont
  {Qiu}}, \bibinfo {author} {\bibfnamefont {S.}~\bibnamefont {Seguy}},\ and\
  \bibinfo {author} {\bibfnamefont {M.}~\bibnamefont {Paredes}},\ }\bibfield
  {title} {\bibinfo {title} {Tuned nonlinear energy sink with conical spring:
  Design theory and sensitivity analysis},\ }\bibfield  {journal} {\bibinfo
  {journal} {Journal of Mechanical Design}\ }\textbf {\bibinfo {volume}
  {140}},\ \href {https://doi.org/10.1115/1.4038304} {10.1115/1.4038304}
  (\bibinfo {year} {2017})\BibitemShut {NoStop}%
\bibitem [{\citenamefont {Huang}\ \emph {et~al.}(2019)\citenamefont {Huang},
  \citenamefont {Li},\ and\ \citenamefont {Yang}}]{Huang2019}%
  \BibitemOpen
  \bibfield  {author} {\bibinfo {author} {\bibfnamefont {D.}~\bibnamefont
  {Huang}}, \bibinfo {author} {\bibfnamefont {R.}~\bibnamefont {Li}},\ and\
  \bibinfo {author} {\bibfnamefont {G.}~\bibnamefont {Yang}},\ }\bibfield
  {title} {\bibinfo {title} {On the dynamic response regimes of a viscoelastic
  isolation system integrated with a nonlinear energy sink},\ }\href
  {https://doi.org/10.1016/j.cnsns.2019.104916} {\bibfield  {journal} {\bibinfo
   {journal} {Communications in Nonlinear Science and Numerical Simulation}\
  }\textbf {\bibinfo {volume} {79}},\ \bibinfo {pages} {104916} (\bibinfo
  {year} {2019})}\BibitemShut {NoStop}%
\bibitem [{\citenamefont {Georgiades}\ and\ \citenamefont
  {Vakakis}(2007)}]{Georgiades2007}%
  \BibitemOpen
  \bibfield  {author} {\bibinfo {author} {\bibfnamefont {F.}~\bibnamefont
  {Georgiades}}\ and\ \bibinfo {author} {\bibfnamefont {A.}~\bibnamefont
  {Vakakis}},\ }\bibfield  {title} {\bibinfo {title} {Dynamics of a linear beam
  with an attached local nonlinear energy sink},\ }\href
  {https://doi.org/10.1016/j.cnsns.2005.07.003} {\bibfield  {journal} {\bibinfo
   {journal} {Communications in Nonlinear Science and Numerical Simulation}\
  }\textbf {\bibinfo {volume} {12}},\ \bibinfo {pages} {643} (\bibinfo {year}
  {2007})}\BibitemShut {NoStop}%
\bibitem [{\citenamefont {Gendelman}\ \emph {et~al.}(2000)\citenamefont
  {Gendelman}, \citenamefont {Manevitch}, \citenamefont {Vakakis},\ and\
  \citenamefont {M'Closkey}}]{Gendelman2000}%
  \BibitemOpen
  \bibfield  {author} {\bibinfo {author} {\bibfnamefont {O.}~\bibnamefont
  {Gendelman}}, \bibinfo {author} {\bibfnamefont {L.~I.}\ \bibnamefont
  {Manevitch}}, \bibinfo {author} {\bibfnamefont {A.~F.}\ \bibnamefont
  {Vakakis}},\ and\ \bibinfo {author} {\bibfnamefont {R.}~\bibnamefont
  {M'Closkey}},\ }\bibfield  {title} {\bibinfo {title} {Energy pumping in
  nonlinear mechanical oscillators: Part {I}{\textemdash}dynamics of the
  underlying hamiltonian systems},\ }\href {https://doi.org/10.1115/1.1345524}
  {\bibfield  {journal} {\bibinfo  {journal} {Journal of Applied Mechanics}\
  }\textbf {\bibinfo {volume} {68}},\ \bibinfo {pages} {34} (\bibinfo {year}
  {2000})}\BibitemShut {NoStop}%
\bibitem [{\citenamefont {Gendelman}(2001)}]{Gendelman2001}%
  \BibitemOpen
  \bibfield  {author} {\bibinfo {author} {\bibfnamefont {O.~V.}\ \bibnamefont
  {Gendelman}},\ }\bibfield  {title} {\bibinfo {title} {Transition of energy to
  a nonlinear localized mode in a highly asymmetric system of two
  oscillators},\ }\href {https://doi.org/10.1023/a:1012967003477} {\bibfield
  {journal} {\bibinfo  {journal} {Nonlinear Dynamics}\ }\textbf {\bibinfo
  {volume} {25}},\ \bibinfo {pages} {237} (\bibinfo {year} {2001})}\BibitemShut
  {NoStop}%
\bibitem [{\citenamefont {Gendelman}\ \emph {et~al.}(2007)\citenamefont
  {Gendelman}, \citenamefont {Starosvetsky},\ and\ \citenamefont
  {Feldman}}]{Gendelman2007}%
  \BibitemOpen
  \bibfield  {author} {\bibinfo {author} {\bibfnamefont {O.~V.}\ \bibnamefont
  {Gendelman}}, \bibinfo {author} {\bibfnamefont {Y.}~\bibnamefont
  {Starosvetsky}},\ and\ \bibinfo {author} {\bibfnamefont {M.}~\bibnamefont
  {Feldman}},\ }\bibfield  {title} {\bibinfo {title} {Attractors of
  harmonically forced linear oscillator with attached nonlinear energy sink
  {I}: Description of response regimes},\ }\href
  {https://doi.org/10.1007/s11071-006-9167-0} {\bibfield  {journal} {\bibinfo
  {journal} {Nonlinear Dynamics}\ }\textbf {\bibinfo {volume} {51}},\ \bibinfo
  {pages} {31} (\bibinfo {year} {2007})}\BibitemShut {NoStop}%
\bibitem [{\citenamefont {Darabi}\ and\ \citenamefont
  {Leamy}(2016)}]{Darabi2016}%
  \BibitemOpen
  \bibfield  {author} {\bibinfo {author} {\bibfnamefont {A.}~\bibnamefont
  {Darabi}}\ and\ \bibinfo {author} {\bibfnamefont {M.~J.}\ \bibnamefont
  {Leamy}},\ }\bibfield  {title} {\bibinfo {title} {Clearance-type nonlinear
  energy sinks for enhancing performance in electroacoustic wave energy
  harvesting},\ }\href {https://doi.org/10.1007/s11071-016-3177-3} {\bibfield
  {journal} {\bibinfo  {journal} {Nonlinear Dynamics}\ }\textbf {\bibinfo
  {volume} {87}},\ \bibinfo {pages} {2127} (\bibinfo {year}
  {2016})}\BibitemShut {NoStop}%
\bibitem [{\citenamefont {Dai}\ \emph {et~al.}(2017)\citenamefont {Dai},
  \citenamefont {Abdelkefi},\ and\ \citenamefont {Wang}}]{Dai2017}%
  \BibitemOpen
  \bibfield  {author} {\bibinfo {author} {\bibfnamefont {H.}~\bibnamefont
  {Dai}}, \bibinfo {author} {\bibfnamefont {A.}~\bibnamefont {Abdelkefi}},\
  and\ \bibinfo {author} {\bibfnamefont {L.}~\bibnamefont {Wang}},\ }\bibfield
  {title} {\bibinfo {title} {Vortex-induced vibrations mitigation through a
  nonlinear energy sink},\ }\href {https://doi.org/10.1016/j.cnsns.2016.05.014}
  {\ \textbf {\bibinfo {volume} {42}},\ \bibinfo {pages} {22} (\bibinfo {year}
  {2017})}\BibitemShut {NoStop}%
\bibitem [{\citenamefont {Bergman}\ \emph {et~al.}(2008)\citenamefont
  {Bergman}, \citenamefont {Gendelman}, \citenamefont {Kerschen}, \citenamefont
  {Lee},\ and\ \citenamefont {McFarland}}]{Bergman2008}%
  \BibitemOpen
  \bibfield  {author} {\bibinfo {author} {\bibfnamefont {L.~A.}\ \bibnamefont
  {Bergman}}, \bibinfo {author} {\bibfnamefont {O.~V.}\ \bibnamefont
  {Gendelman}}, \bibinfo {author} {\bibfnamefont {G.}~\bibnamefont {Kerschen}},
  \bibinfo {author} {\bibfnamefont {Y.~S.}\ \bibnamefont {Lee}},\ and\ \bibinfo
  {author} {\bibfnamefont {D.~M.}\ \bibnamefont {McFarland}},\ }\href@noop {}
  {\emph {\bibinfo {title} {Nonlinear Targeted Energy Transfer in Mechanical
  and Structural Systems}}}\ (\bibinfo  {publisher} {Springer-Verlag GmbH},\
  \bibinfo {year} {2008})\BibitemShut {NoStop}%
\bibitem [{\citenamefont {Manevitch}(2015)}]{Manevitch2015}%
  \BibitemOpen
  \bibfield  {author} {\bibinfo {author} {\bibfnamefont {L.~I.}\ \bibnamefont
  {Manevitch}},\ }\bibfield  {title} {\bibinfo {title} {A concept of limiting
  phase trajectories and description of highly non-stationary resonance
  processes},\ }\href {https://doi.org/10.12988/ams.2015.55378} {\bibfield
  {journal} {\bibinfo  {journal} {Applied Mathematical Sciences}\ }\textbf
  {\bibinfo {volume} {9}},\ \bibinfo {pages} {4269} (\bibinfo {year}
  {2015})}\BibitemShut {NoStop}%
\bibitem [{\citenamefont {Nucera}\ \emph {et~al.}(2007)\citenamefont {Nucera},
  \citenamefont {Vakakis}, \citenamefont {McFarland}, \citenamefont {Bergman},\
  and\ \citenamefont {Kerschen}}]{Nucera2007}%
  \BibitemOpen
  \bibfield  {author} {\bibinfo {author} {\bibfnamefont {F.}~\bibnamefont
  {Nucera}}, \bibinfo {author} {\bibfnamefont {A.~F.}\ \bibnamefont {Vakakis}},
  \bibinfo {author} {\bibfnamefont {D.~M.}\ \bibnamefont {McFarland}}, \bibinfo
  {author} {\bibfnamefont {L.~A.}\ \bibnamefont {Bergman}},\ and\ \bibinfo
  {author} {\bibfnamefont {G.}~\bibnamefont {Kerschen}},\ }\bibfield  {title}
  {\bibinfo {title} {Targeted energy transfers in vibro-impact oscillators for
  seismic mitigation},\ }\href {https://doi.org/10.1007/s11071-006-9189-7}
  {\bibfield  {journal} {\bibinfo  {journal} {Nonlinear Dynamics}\ }\textbf
  {\bibinfo {volume} {50}},\ \bibinfo {pages} {651} (\bibinfo {year}
  {2007})}\BibitemShut {NoStop}%
\bibitem [{\citenamefont {Gendelman}(2012)}]{Gendelman2012}%
  \BibitemOpen
  \bibfield  {author} {\bibinfo {author} {\bibfnamefont {O.}~\bibnamefont
  {Gendelman}},\ }\bibfield  {title} {\bibinfo {title} {Analytic treatment of a
  system with a vibro-impact nonlinear energy sink},\ }\href
  {https://doi.org/10.1016/j.jsv.2012.05.021} {\bibfield  {journal} {\bibinfo
  {journal} {Journal of Sound and Vibration}\ }\textbf {\bibinfo {volume}
  {331}},\ \bibinfo {pages} {4599} (\bibinfo {year} {2012})}\BibitemShut
  {NoStop}%
\bibitem [{\citenamefont {Li}\ \emph {et~al.}(2016)\citenamefont {Li},
  \citenamefont {Seguy},\ and\ \citenamefont {Berlioz}}]{Li2016}%
  \BibitemOpen
  \bibfield  {author} {\bibinfo {author} {\bibfnamefont {T.}~\bibnamefont
  {Li}}, \bibinfo {author} {\bibfnamefont {S.}~\bibnamefont {Seguy}},\ and\
  \bibinfo {author} {\bibfnamefont {A.}~\bibnamefont {Berlioz}},\ }\bibfield
  {title} {\bibinfo {title} {Optimization mechanism of targeted energy transfer
  with vibro-impact energy sink under periodic and transient excitation},\
  }\href {https://doi.org/10.1007/s11071-016-3200-8} {\bibfield  {journal}
  {\bibinfo  {journal} {Nonlinear Dynamics}\ }\textbf {\bibinfo {volume}
  {87}},\ \bibinfo {pages} {2415} (\bibinfo {year} {2016})}\BibitemShut
  {NoStop}%
\bibitem [{\citenamefont {Gendelman}\ and\ \citenamefont
  {Vakakis}(2018)}]{Gendelman2018}%
  \BibitemOpen
  \bibfield  {author} {\bibinfo {author} {\bibfnamefont {O.~V.}\ \bibnamefont
  {Gendelman}}\ and\ \bibinfo {author} {\bibfnamefont {A.~F.}\ \bibnamefont
  {Vakakis}},\ }\bibfield  {title} {\bibinfo {title} {Introduction to a topical
  issue `nonlinear energy transfer in dynamical and acoustical
  systems{\textquotesingle}},\ }\href {https://doi.org/10.1098/rsta.2017.0129}
  {\bibfield  {journal} {\bibinfo  {journal} {Philosophical Transactions of the
  Royal Society A: Mathematical, Physical and Engineering Sciences}\ }\textbf
  {\bibinfo {volume} {376}},\ \bibinfo {pages} {20170129} (\bibinfo {year}
  {2018})}\BibitemShut {NoStop}%
\bibitem [{\citenamefont {Rothos}\ and\ \citenamefont
  {Vakakis}(2009)}]{Rothos2009}%
  \BibitemOpen
  \bibfield  {author} {\bibinfo {author} {\bibfnamefont {V.}~\bibnamefont
  {Rothos}}\ and\ \bibinfo {author} {\bibfnamefont {A.}~\bibnamefont
  {Vakakis}},\ }\bibfield  {title} {\bibinfo {title} {Dynamic interactions of
  traveling waves propagating in a linear chain with an local essentially
  nonlinear attachment},\ }\href
  {https://doi.org/10.1016/j.wavemoti.2008.10.004} {\bibfield  {journal}
  {\bibinfo  {journal} {Wave Motion}\ }\textbf {\bibinfo {volume} {46}},\
  \bibinfo {pages} {174} (\bibinfo {year} {2009})}\BibitemShut {NoStop}%
\bibitem [{\citenamefont {Vakakis}\ \emph {et~al.}(2014)\citenamefont
  {Vakakis}, \citenamefont {AL-Shudeifat},\ and\ \citenamefont
  {Hasan}}]{Vakakis2014}%
  \BibitemOpen
  \bibfield  {author} {\bibinfo {author} {\bibfnamefont {A.~F.}\ \bibnamefont
  {Vakakis}}, \bibinfo {author} {\bibfnamefont {M.~A.}\ \bibnamefont
  {AL-Shudeifat}},\ and\ \bibinfo {author} {\bibfnamefont {M.~A.}\ \bibnamefont
  {Hasan}},\ }\bibfield  {title} {\bibinfo {title} {Interactions of propagating
  waves in a one-dimensional chain of linear oscillators with a strongly
  nonlinear local attachment},\ }\href
  {https://doi.org/10.1007/s11012-014-0008-9} {\bibfield  {journal} {\bibinfo
  {journal} {Meccanica}\ }\textbf {\bibinfo {volume} {49}},\ \bibinfo {pages}
  {2375} (\bibinfo {year} {2014})}\BibitemShut {NoStop}%
\bibitem [{\citenamefont {Nassar}\ \emph {et~al.}(2020)\citenamefont {Nassar},
  \citenamefont {Yousefzadeh}, \citenamefont {Fleury}, \citenamefont {Ruzzene},
  \citenamefont {Al{\`{u}}}, \citenamefont {Daraio}, \citenamefont {Norris},
  \citenamefont {Huang},\ and\ \citenamefont {Haberman}}]{Nassar2020}%
  \BibitemOpen
  \bibfield  {author} {\bibinfo {author} {\bibfnamefont {H.}~\bibnamefont
  {Nassar}}, \bibinfo {author} {\bibfnamefont {B.}~\bibnamefont {Yousefzadeh}},
  \bibinfo {author} {\bibfnamefont {R.}~\bibnamefont {Fleury}}, \bibinfo
  {author} {\bibfnamefont {M.}~\bibnamefont {Ruzzene}}, \bibinfo {author}
  {\bibfnamefont {A.}~\bibnamefont {Al{\`{u}}}}, \bibinfo {author}
  {\bibfnamefont {C.}~\bibnamefont {Daraio}}, \bibinfo {author} {\bibfnamefont
  {A.~N.}\ \bibnamefont {Norris}}, \bibinfo {author} {\bibfnamefont
  {G.}~\bibnamefont {Huang}},\ and\ \bibinfo {author} {\bibfnamefont {M.~R.}\
  \bibnamefont {Haberman}},\ }\bibfield  {title} {\bibinfo {title}
  {Nonreciprocity in acoustic and elastic materials},\ }\href
  {https://doi.org/10.1038/s41578-020-0206-0} {\bibfield  {journal} {\bibinfo
  {journal} {Nature Reviews Materials}\ }\textbf {\bibinfo {volume} {5}},\
  \bibinfo {pages} {667} (\bibinfo {year} {2020})}\BibitemShut {NoStop}%
\bibitem [{\citenamefont {Bunyan}\ \emph {et~al.}(2018)\citenamefont {Bunyan},
  \citenamefont {Moore}, \citenamefont {Mojahed}, \citenamefont {Fronk},
  \citenamefont {Leamy}, \citenamefont {Tawfick},\ and\ \citenamefont
  {Vakakis}}]{Bunyan2018}%
  \BibitemOpen
  \bibfield  {author} {\bibinfo {author} {\bibfnamefont {J.}~\bibnamefont
  {Bunyan}}, \bibinfo {author} {\bibfnamefont {K.~J.}\ \bibnamefont {Moore}},
  \bibinfo {author} {\bibfnamefont {A.}~\bibnamefont {Mojahed}}, \bibinfo
  {author} {\bibfnamefont {M.~D.}\ \bibnamefont {Fronk}}, \bibinfo {author}
  {\bibfnamefont {M.}~\bibnamefont {Leamy}}, \bibinfo {author} {\bibfnamefont
  {S.}~\bibnamefont {Tawfick}},\ and\ \bibinfo {author} {\bibfnamefont {A.~F.}\
  \bibnamefont {Vakakis}},\ }\bibfield  {title} {\bibinfo {title} {Acoustic
  nonreciprocity in a lattice incorporating nonlinearity, asymmetry, and
  internal scale hierarchy: Experimental study},\ }\href
  {https://doi.org/10.1103/physreve.97.052211} {\bibfield  {journal} {\bibinfo
  {journal} {Physical Review E}\ }\textbf {\bibinfo {volume} {97}},\ \bibinfo
  {pages} {052211} (\bibinfo {year} {2018})}\BibitemShut {NoStop}%
\bibitem [{\citenamefont {Fronk}\ \emph {et~al.}(2019)\citenamefont {Fronk},
  \citenamefont {Tawfick}, \citenamefont {Daraio}, \citenamefont {Li},
  \citenamefont {Vakakis},\ and\ \citenamefont {Leamy}}]{Fronk2019}%
  \BibitemOpen
  \bibfield  {author} {\bibinfo {author} {\bibfnamefont {M.~D.}\ \bibnamefont
  {Fronk}}, \bibinfo {author} {\bibfnamefont {S.}~\bibnamefont {Tawfick}},
  \bibinfo {author} {\bibfnamefont {C.}~\bibnamefont {Daraio}}, \bibinfo
  {author} {\bibfnamefont {S.}~\bibnamefont {Li}}, \bibinfo {author}
  {\bibfnamefont {A.}~\bibnamefont {Vakakis}},\ and\ \bibinfo {author}
  {\bibfnamefont {M.~J.}\ \bibnamefont {Leamy}},\ }\bibfield  {title} {\bibinfo
  {title} {Acoustic non-reciprocity in lattices with nonlinearity, internal
  hierarchy, and asymmetry: Computational study},\ }\bibfield  {journal}
  {\bibinfo  {journal} {Journal of Vibration and Acoustics}\ }\textbf {\bibinfo
  {volume} {141}},\ \href {https://doi.org/10.1115/1.4043783}
  {10.1115/1.4043783} (\bibinfo {year} {2019})\BibitemShut {NoStop}%
\bibitem [{\citenamefont {Grinberg}\ \emph {et~al.}(2018)\citenamefont
  {Grinberg}, \citenamefont {Vakakis},\ and\ \citenamefont
  {Gendelman}}]{Grinberg2018}%
  \BibitemOpen
  \bibfield  {author} {\bibinfo {author} {\bibfnamefont {I.}~\bibnamefont
  {Grinberg}}, \bibinfo {author} {\bibfnamefont {A.~F.}\ \bibnamefont
  {Vakakis}},\ and\ \bibinfo {author} {\bibfnamefont {O.~V.}\ \bibnamefont
  {Gendelman}},\ }\bibfield  {title} {\bibinfo {title} {Acoustic diode: Wave
  non-reciprocity in nonlinearly coupled waveguides},\ }\href
  {https://doi.org/10.1016/j.wavemoti.2018.08.005} {\bibfield  {journal}
  {\bibinfo  {journal} {Wave Motion}\ }\textbf {\bibinfo {volume} {83}},\
  \bibinfo {pages} {49} (\bibinfo {year} {2018})}\BibitemShut {NoStop}%
\bibitem [{\citenamefont {Fu}\ \emph {et~al.}(2018)\citenamefont {Fu},
  \citenamefont {Wang}, \citenamefont {Zhao},\ and\ \citenamefont
  {Chen}}]{Fu2018}%
  \BibitemOpen
  \bibfield  {author} {\bibinfo {author} {\bibfnamefont {C.}~\bibnamefont
  {Fu}}, \bibinfo {author} {\bibfnamefont {B.}~\bibnamefont {Wang}}, \bibinfo
  {author} {\bibfnamefont {T.}~\bibnamefont {Zhao}},\ and\ \bibinfo {author}
  {\bibfnamefont {C.~Q.}\ \bibnamefont {Chen}},\ }\bibfield  {title} {\bibinfo
  {title} {High efficiency and broadband acoustic diodes},\ }\href
  {https://doi.org/10.1063/1.5020698} {\bibfield  {journal} {\bibinfo
  {journal} {Applied Physics Letters}\ }\textbf {\bibinfo {volume} {112}},\
  \bibinfo {pages} {051902} (\bibinfo {year} {2018})}\BibitemShut {NoStop}%
\bibitem [{\citenamefont {Darabi}\ \emph {et~al.}(2019)\citenamefont {Darabi},
  \citenamefont {Fang}, \citenamefont {Mojahed}, \citenamefont {Fronk},
  \citenamefont {Vakakis},\ and\ \citenamefont {Leamy}}]{Darabi2019}%
  \BibitemOpen
  \bibfield  {author} {\bibinfo {author} {\bibfnamefont {A.}~\bibnamefont
  {Darabi}}, \bibinfo {author} {\bibfnamefont {L.}~\bibnamefont {Fang}},
  \bibinfo {author} {\bibfnamefont {A.}~\bibnamefont {Mojahed}}, \bibinfo
  {author} {\bibfnamefont {M.~D.}\ \bibnamefont {Fronk}}, \bibinfo {author}
  {\bibfnamefont {A.~F.}\ \bibnamefont {Vakakis}},\ and\ \bibinfo {author}
  {\bibfnamefont {M.~J.}\ \bibnamefont {Leamy}},\ }\bibfield  {title} {\bibinfo
  {title} {Broadband passive nonlinear acoustic diode},\ }\href
  {https://doi.org/10.1103/physrevb.99.214305} {\bibfield  {journal} {\bibinfo
  {journal} {Physical Review B}\ }\textbf {\bibinfo {volume} {99}},\ \bibinfo
  {pages} {214305} (\bibinfo {year} {2019})}\BibitemShut {NoStop}%
\bibitem [{\citenamefont {Devaux}\ \emph {et~al.}(2019)\citenamefont {Devaux},
  \citenamefont {Cebrecos}, \citenamefont {Richoux}, \citenamefont {Pagneux},\
  and\ \citenamefont {Tournat}}]{Devaux2019}%
  \BibitemOpen
  \bibfield  {author} {\bibinfo {author} {\bibfnamefont {T.}~\bibnamefont
  {Devaux}}, \bibinfo {author} {\bibfnamefont {A.}~\bibnamefont {Cebrecos}},
  \bibinfo {author} {\bibfnamefont {O.}~\bibnamefont {Richoux}}, \bibinfo
  {author} {\bibfnamefont {V.}~\bibnamefont {Pagneux}},\ and\ \bibinfo {author}
  {\bibfnamefont {V.}~\bibnamefont {Tournat}},\ }\bibfield  {title} {\bibinfo
  {title} {Acoustic radiation pressure for nonreciprocal transmission and
  switch effects},\ }\bibfield  {journal} {\bibinfo  {journal} {Nature
  Communications}\ }\textbf {\bibinfo {volume} {10}},\ \href
  {https://doi.org/10.1038/s41467-019-11305-7} {10.1038/s41467-019-11305-7}
  (\bibinfo {year} {2019})\BibitemShut {NoStop}%
\bibitem [{\citenamefont {Mork}\ \emph {et~al.}(2022)\citenamefont {Mork},
  \citenamefont {Fronk}, \citenamefont {Sinclair},\ and\ \citenamefont
  {Leamy}}]{Mork2022}%
  \BibitemOpen
  \bibfield  {author} {\bibinfo {author} {\bibfnamefont {N.}~\bibnamefont
  {Mork}}, \bibinfo {author} {\bibfnamefont {M.~D.}\ \bibnamefont {Fronk}},
  \bibinfo {author} {\bibfnamefont {M.~B.}\ \bibnamefont {Sinclair}},\ and\
  \bibinfo {author} {\bibfnamefont {M.~J.}\ \bibnamefont {Leamy}},\ }\bibfield
  {title} {\bibinfo {title} {Nonlinear hierarchical unit cell for passive,
  amplitude-dependent filtering of acoustic waves},\ }\href
  {https://doi.org/10.1016/j.eml.2022.101915} {\bibfield  {journal} {\bibinfo
  {journal} {Extreme Mechanics Letters}\ ,\ \bibinfo {pages} {101915}}
  (\bibinfo {year} {2022})}\BibitemShut {NoStop}%
\bibitem [{\citenamefont {Gzal}\ \emph {et~al.}(2020)\citenamefont {Gzal},
  \citenamefont {Fang}, \citenamefont {Vakakis}, \citenamefont {Bergman},\ and\
  \citenamefont {Gendelman}}]{Gzal2020}%
  \BibitemOpen
  \bibfield  {author} {\bibinfo {author} {\bibfnamefont {M.}~\bibnamefont
  {Gzal}}, \bibinfo {author} {\bibfnamefont {B.}~\bibnamefont {Fang}}, \bibinfo
  {author} {\bibfnamefont {A.~F.}\ \bibnamefont {Vakakis}}, \bibinfo {author}
  {\bibfnamefont {L.~A.}\ \bibnamefont {Bergman}},\ and\ \bibinfo {author}
  {\bibfnamefont {O.~V.}\ \bibnamefont {Gendelman}},\ }\bibfield  {title}
  {\bibinfo {title} {Rapid non-resonant intermodal targeted energy transfer
  ({IMTET}) caused by vibro-impact nonlinearity},\ }\href
  {https://doi.org/10.1007/s11071-020-05909-8} {\bibfield  {journal} {\bibinfo
  {journal} {Nonlinear Dynamics}\ }\textbf {\bibinfo {volume} {101}},\ \bibinfo
  {pages} {2087} (\bibinfo {year} {2020})}\BibitemShut {NoStop}%
\bibitem [{\citenamefont {Gzal}\ \emph {et~al.}(2021)\citenamefont {Gzal},
  \citenamefont {Vakakis}, \citenamefont {Bergman},\ and\ \citenamefont
  {Gendelman}}]{Gzal2021}%
  \BibitemOpen
  \bibfield  {author} {\bibinfo {author} {\bibfnamefont {M.}~\bibnamefont
  {Gzal}}, \bibinfo {author} {\bibfnamefont {A.~F.}\ \bibnamefont {Vakakis}},
  \bibinfo {author} {\bibfnamefont {L.~A.}\ \bibnamefont {Bergman}},\ and\
  \bibinfo {author} {\bibfnamefont {O.~V.}\ \bibnamefont {Gendelman}},\
  }\bibfield  {title} {\bibinfo {title} {Extreme intermodal energy transfers
  through vibro-impacts for highly effective and rapid blast mitigation},\
  }\href {https://doi.org/10.1016/j.cnsns.2021.106012} {\bibfield  {journal}
  {\bibinfo  {journal} {Communications in Nonlinear Science and Numerical
  Simulation}\ }\textbf {\bibinfo {volume} {103}},\ \bibinfo {pages} {106012}
  (\bibinfo {year} {2021})}\BibitemShut {NoStop}%
\bibitem [{\citenamefont {Tempelman}\ \emph
  {et~al.}(2022{\natexlab{b}})\citenamefont {Tempelman}, \citenamefont
  {Mojahed}, \citenamefont {Gzal}, \citenamefont {Matlack}, \citenamefont
  {Gendelman}, \citenamefont {Bergman},\ and\ \citenamefont
  {Vakakis}}]{Tempelman2022}%
  \BibitemOpen
  \bibfield  {author} {\bibinfo {author} {\bibfnamefont {J.~R.}\ \bibnamefont
  {Tempelman}}, \bibinfo {author} {\bibfnamefont {A.}~\bibnamefont {Mojahed}},
  \bibinfo {author} {\bibfnamefont {M.}~\bibnamefont {Gzal}}, \bibinfo {author}
  {\bibfnamefont {K.~H.}\ \bibnamefont {Matlack}}, \bibinfo {author}
  {\bibfnamefont {O.~V.}\ \bibnamefont {Gendelman}}, \bibinfo {author}
  {\bibfnamefont {L.~A.}\ \bibnamefont {Bergman}},\ and\ \bibinfo {author}
  {\bibfnamefont {A.~F.}\ \bibnamefont {Vakakis}},\ }\bibfield  {title}
  {\bibinfo {title} {Experimental inter-modal targeted energy transfer in a
  cantilever beam undergoing vibro-impacts},\ }\href
  {https://doi.org/10.1016/j.jsv.2022.117212} {\bibfield  {journal} {\bibinfo
  {journal} {Journal of Sound and Vibration}\ }\textbf {\bibinfo {volume}
  {539}},\ \bibinfo {pages} {117212} (\bibinfo {year}
  {2022}{\natexlab{b}})}\BibitemShut {NoStop}%
\bibitem [{\citenamefont {Lee}\ \emph {et~al.}(2009)\citenamefont {Lee},
  \citenamefont {Nucera}, \citenamefont {Vakakis}, \citenamefont {McFarland},\
  and\ \citenamefont {Bergman}}]{Lee2009}%
  \BibitemOpen
  \bibfield  {author} {\bibinfo {author} {\bibfnamefont {Y.~S.}\ \bibnamefont
  {Lee}}, \bibinfo {author} {\bibfnamefont {F.}~\bibnamefont {Nucera}},
  \bibinfo {author} {\bibfnamefont {A.~F.}\ \bibnamefont {Vakakis}}, \bibinfo
  {author} {\bibfnamefont {D.~M.}\ \bibnamefont {McFarland}},\ and\ \bibinfo
  {author} {\bibfnamefont {L.~A.}\ \bibnamefont {Bergman}},\ }\bibfield
  {title} {\bibinfo {title} {Periodic orbits, damped transitions and targeted
  energy transfers in oscillators with vibro-impact attachments},\ }\href
  {https://doi.org/10.1016/j.physd.2009.06.013} {\bibfield  {journal} {\bibinfo
   {journal} {Physica D: Nonlinear Phenomena}\ }\textbf {\bibinfo {volume}
  {238}},\ \bibinfo {pages} {1868} (\bibinfo {year} {2009})}\BibitemShut
  {NoStop}%
\bibitem [{\citenamefont {Hunt}\ and\ \citenamefont
  {Crossley}(1975)}]{Hunt1975}%
  \BibitemOpen
  \bibfield  {author} {\bibinfo {author} {\bibfnamefont {K.~H.}\ \bibnamefont
  {Hunt}}\ and\ \bibinfo {author} {\bibfnamefont {F.~R.~E.}\ \bibnamefont
  {Crossley}},\ }\bibfield  {title} {\bibinfo {title} {Coefficient of
  restitution interpreted as damping in vibroimpact},\ }\href
  {https://doi.org/10.1115/1.3423596} {\bibfield  {journal} {\bibinfo
  {journal} {Journal of Applied Mechanics}\ }\textbf {\bibinfo {volume} {42}},\
  \bibinfo {pages} {440} (\bibinfo {year} {1975})}\BibitemShut {NoStop}%
\bibitem [{\citenamefont {Arretche}\ and\ \citenamefont
  {Matlack}(2018)}]{Arretche2018}%
  \BibitemOpen
  \bibfield  {author} {\bibinfo {author} {\bibfnamefont {I.}~\bibnamefont
  {Arretche}}\ and\ \bibinfo {author} {\bibfnamefont {K.~H.}\ \bibnamefont
  {Matlack}},\ }\bibfield  {title} {\bibinfo {title} {On the interrelationship
  between static and vibration mitigation properties of architected
  metastructures},\ }\bibfield  {journal} {\bibinfo  {journal} {Frontiers in
  Materials}\ }\textbf {\bibinfo {volume} {5}},\ \href
  {https://doi.org/10.3389/fmats.2018.00068} {10.3389/fmats.2018.00068}
  (\bibinfo {year} {2018})\BibitemShut {NoStop}%
\bibitem [{\citenamefont {Mojahed}\ \emph {et~al.}(2021)\citenamefont
  {Mojahed}, \citenamefont {Bergman},\ and\ \citenamefont
  {Vakakis}}]{Mojahed2021}%
  \BibitemOpen
  \bibfield  {author} {\bibinfo {author} {\bibfnamefont {A.}~\bibnamefont
  {Mojahed}}, \bibinfo {author} {\bibfnamefont {L.~A.}\ \bibnamefont
  {Bergman}},\ and\ \bibinfo {author} {\bibfnamefont {A.~F.}\ \bibnamefont
  {Vakakis}},\ }\bibfield  {title} {\bibinfo {title} {New inverse wavelet
  transform method with broad application in dynamics},\ }\href
  {https://doi.org/10.1016/j.ymssp.2021.107691} {\bibfield  {journal} {\bibinfo
   {journal} {Mechanical Systems and Signal Processing}\ }\textbf {\bibinfo
  {volume} {156}},\ \bibinfo {pages} {107691} (\bibinfo {year}
  {2021})}\BibitemShut {NoStop}%
\bibitem [{\citenamefont {Boashash}(2013)}]{Boashash2013}%
  \BibitemOpen
  \bibfield  {author} {\bibinfo {author} {\bibfnamefont {B.}~\bibnamefont
  {Boashash}},\ }\href@noop {} {\emph {\bibinfo {title} {Timefrequency Signal
  Analysis And Processing A Comprehensive Review}}}\ (\bibinfo  {publisher}
  {Academic Press},\ \bibinfo {year} {2013})\BibitemShut {NoStop}%
\bibitem [{\citenamefont {Vakakis}\ \emph {et~al.}(2008)\citenamefont
  {Vakakis}, \citenamefont {Manevitch}, \citenamefont {Mikhlin}, \citenamefont
  {Pilipchuk},\ and\ \citenamefont {Zevin}}]{Vakakis2008a}%
  \BibitemOpen
  \bibfield  {author} {\bibinfo {author} {\bibfnamefont {A.~F.}\ \bibnamefont
  {Vakakis}}, \bibinfo {author} {\bibfnamefont {L.~I.}\ \bibnamefont
  {Manevitch}}, \bibinfo {author} {\bibfnamefont {Y.~V.}\ \bibnamefont
  {Mikhlin}}, \bibinfo {author} {\bibfnamefont {V.~N.}\ \bibnamefont
  {Pilipchuk}},\ and\ \bibinfo {author} {\bibfnamefont {A.~A.}\ \bibnamefont
  {Zevin}},\ }\href@noop {} {\emph {\bibinfo {title} {Normal Modes and
  Localization in Nonlinear Systems}}}\ (\bibinfo  {publisher} {Wiley \& Sons,
  Incorporated, John},\ \bibinfo {year} {2008})\ p.\ \bibinfo {pages}
  {552}\BibitemShut {NoStop}%
\bibitem [{\citenamefont {Kerschen}\ \emph {et~al.}(2009)\citenamefont
  {Kerschen}, \citenamefont {Peeters}, \citenamefont {Golinval},\ and\
  \citenamefont {Vakakis}}]{Kerschen2009}%
  \BibitemOpen
  \bibfield  {author} {\bibinfo {author} {\bibfnamefont {G.}~\bibnamefont
  {Kerschen}}, \bibinfo {author} {\bibfnamefont {M.}~\bibnamefont {Peeters}},
  \bibinfo {author} {\bibfnamefont {J.}~\bibnamefont {Golinval}},\ and\
  \bibinfo {author} {\bibfnamefont {A.}~\bibnamefont {Vakakis}},\ }\bibfield
  {title} {\bibinfo {title} {Nonlinear normal modes, part i: A useful framework
  for the structural dynamicist},\ }\href
  {https://doi.org/10.1016/j.ymssp.2008.04.002} {\bibfield  {journal} {\bibinfo
   {journal} {Mechanical Systems and Signal Processing}\ }\textbf {\bibinfo
  {volume} {23}},\ \bibinfo {pages} {170} (\bibinfo {year} {2009})}\BibitemShut
  {NoStop}%
\bibitem [{\citenamefont {Avramov}\ and\ \citenamefont
  {Mikhlin}(2013)}]{Avramov2013}%
  \BibitemOpen
  \bibfield  {author} {\bibinfo {author} {\bibfnamefont {K.~V.}\ \bibnamefont
  {Avramov}}\ and\ \bibinfo {author} {\bibfnamefont {Y.~V.}\ \bibnamefont
  {Mikhlin}},\ }\bibfield  {title} {\bibinfo {title} {Review of applications of
  nonlinear normal modes for vibrating mechanical systems},\ }\bibfield
  {journal} {\bibinfo  {journal} {Applied Mechanics Reviews}\ }\textbf
  {\bibinfo {volume} {65}},\ \href {https://doi.org/10.1115/1.4023533}
  {10.1115/1.4023533} (\bibinfo {year} {2013})\BibitemShut {NoStop}%
\bibitem [{\citenamefont {Peeters}\ \emph {et~al.}(2009)\citenamefont
  {Peeters}, \citenamefont {Vigui{\'{e}}}, \citenamefont {S{\'{e}}randour},
  \citenamefont {Kerschen},\ and\ \citenamefont {Golinval}}]{Peeters2009}%
  \BibitemOpen
  \bibfield  {author} {\bibinfo {author} {\bibfnamefont {M.}~\bibnamefont
  {Peeters}}, \bibinfo {author} {\bibfnamefont {R.}~\bibnamefont
  {Vigui{\'{e}}}}, \bibinfo {author} {\bibfnamefont {G.}~\bibnamefont
  {S{\'{e}}randour}}, \bibinfo {author} {\bibfnamefont {G.}~\bibnamefont
  {Kerschen}},\ and\ \bibinfo {author} {\bibfnamefont {J.-C.}\ \bibnamefont
  {Golinval}},\ }\bibfield  {title} {\bibinfo {title} {Nonlinear normal modes,
  part {II}: Toward a practical computation using numerical continuation
  techniques},\ }\href {https://doi.org/10.1016/j.ymssp.2008.04.003} {\bibfield
   {journal} {\bibinfo  {journal} {Mechanical Systems and Signal Processing}\
  }\textbf {\bibinfo {volume} {23}},\ \bibinfo {pages} {195} (\bibinfo {year}
  {2009})}\BibitemShut {NoStop}%
\bibitem [{\citenamefont {Tao}\ and\ \citenamefont {Gibert}(2019)}]{Tao2019}%
  \BibitemOpen
  \bibfield  {author} {\bibinfo {author} {\bibfnamefont {H.}~\bibnamefont
  {Tao}}\ and\ \bibinfo {author} {\bibfnamefont {J.}~\bibnamefont {Gibert}},\
  }\bibfield  {title} {\bibinfo {title} {Periodic orbits of a conservative
  2-{DOF} vibro-impact system by piecewise continuation: bifurcations and
  fractals},\ }\href {https://doi.org/10.1007/s11071-018-04734-4} {\bibfield
  {journal} {\bibinfo  {journal} {Nonlinear Dynamics}\ }\textbf {\bibinfo
  {volume} {95}},\ \bibinfo {pages} {2963} (\bibinfo {year}
  {2019})}\BibitemShut {NoStop}%
\bibitem [{\citenamefont {Moussi}\ \emph {et~al.}(2015)\citenamefont {Moussi},
  \citenamefont {Bellizzi}, \citenamefont {Cochelin},\ and\ \citenamefont
  {Nistor}}]{Moussi2015}%
  \BibitemOpen
  \bibfield  {author} {\bibinfo {author} {\bibfnamefont {E.}~\bibnamefont
  {Moussi}}, \bibinfo {author} {\bibfnamefont {S.}~\bibnamefont {Bellizzi}},
  \bibinfo {author} {\bibfnamefont {B.}~\bibnamefont {Cochelin}},\ and\
  \bibinfo {author} {\bibfnamefont {I.}~\bibnamefont {Nistor}},\ }\bibfield
  {title} {\bibinfo {title} {Nonlinear normal modes of a two degrees-of-freedom
  piecewise linear system},\ }\href
  {https://doi.org/10.1016/j.ymssp.2015.03.017} {\bibfield  {journal} {\bibinfo
   {journal} {Mechanical Systems and Signal Processing}\ }\textbf {\bibinfo
  {volume} {64-65}},\ \bibinfo {pages} {266} (\bibinfo {year}
  {2015})}\BibitemShut {NoStop}%
\end{thebibliography}%
\unappendix
\clearpage
\newpage
\pagebreak
\widetext
\begin{center}
	\textbf{\large Supplemental Materials: Wavenumber Scattering and Inter-band Targeted Energy Transfer in Phononic Lattices with Local Vibro-Impact Nonlinearities}\\
	Joshua R.~Tempelman, Alexander F.~Vakakis, Kathryn H.~Matlack\\
	\textit{Department of Mechanical Science and Engineering, University of Illinois at Urbana Champaign}
\end{center}
\setcounter{equation}{0}
\setcounter{figure}{0}
\setcounter{table}{0}
\setcounter{section}{0}
\setcounter{page}{1}
\makeatletter
\renewcommand{\theequation}{S\arabic{equation}}
\renewcommand{\thefigure}{S\arabic{figure}}
\renewcommand{\bibnumfmt}[1]{[S#1]}
\renewcommand{\citenumfont}[1]{S#1}


\section{Additional information for wavenumber scattering}

\begin{figure}[h!]
	\includegraphics[width=\linewidth]{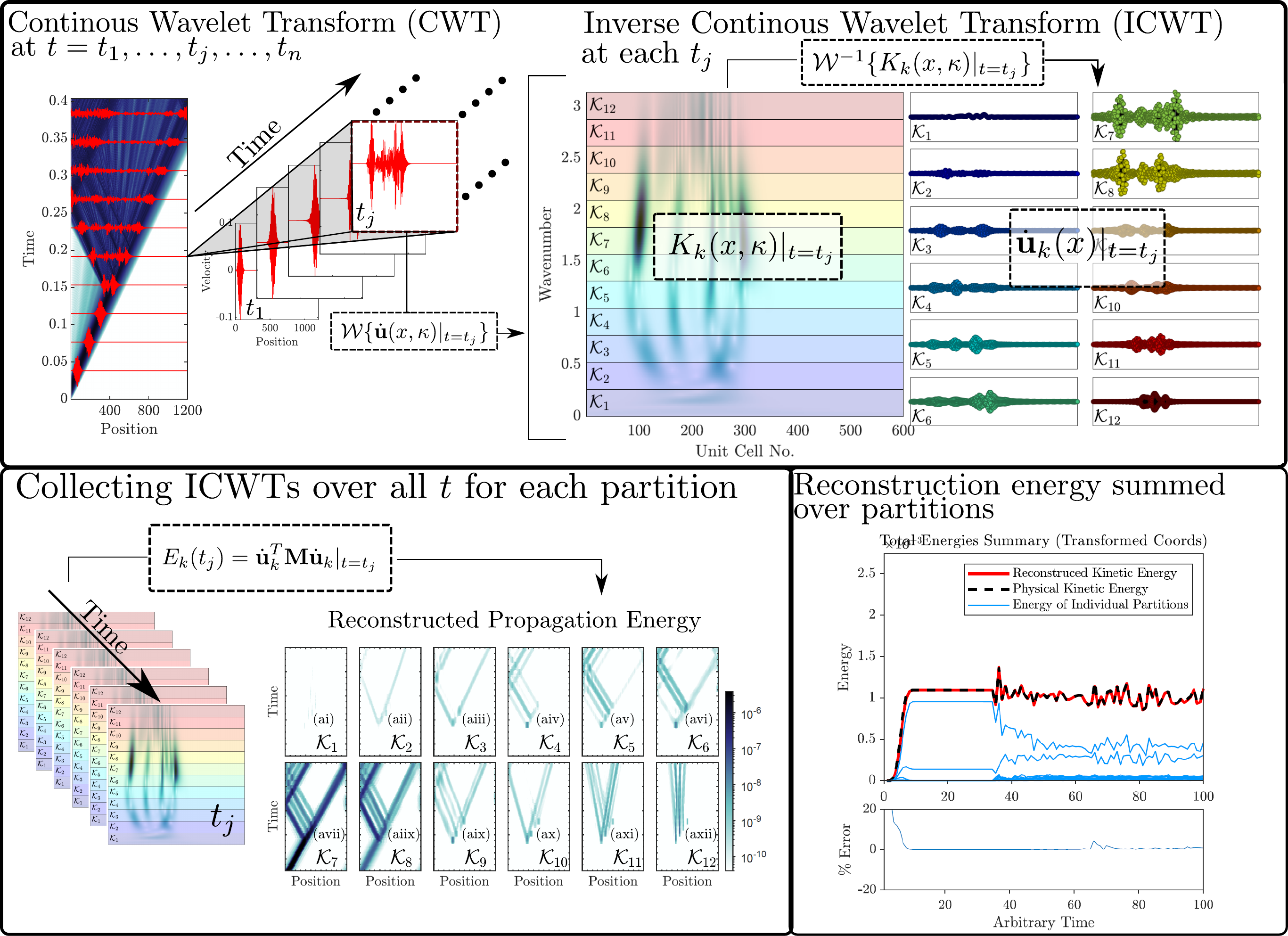}
	\caption{Graphical illustration of the wavelet-based wavenumber partitioning processes.}
\label{FIG:WnumDetailed}
\end{figure}

Fig.~\ref{FIG:WnumDetailed} provides a graphical illustration of the signal processing processes described in section~\ref{SEC:WavenumberScatter} and Appendix~\ref{SEC:apx_NNM}. Starting in the spatio-temporal domain, snap-shots of the wave velocity are taken successively and converted into the wavelet domain. This domain is partitioned into 12 bands (Fig.~\ref{FIG:WnumDetailed}(b)). The inverse transformation of the $k$-th band partition gives at a fixed point in time gives the velocity vector $\dot{\textbf{u}}_k(x)$. The instantaneous energy of the $k$-th band is then conveniently computed as $KE=\frac{1}{2}\dot{\textbf{u}}^T\textbf{M}\dot{\textbf{u}}$ or equivalently, $KE = \frac{1}{2}\sum_n \dot{u}_{n}^2m_n$. The energies are contacted over time to deliver the energy corresponding to wave propagation on the $k$-th band; note that minimal is shown for wave energy reconstruction when the sum of energy over all 12 partitions is compared to exact corresponding energy computed  by direct numerical integration of the governing equations of motion.

\clearpage
\newpage
\section{Extended results for wavenumber entropy}

\begin{figure}[h!]
			\includegraphics[width=\linewidth]{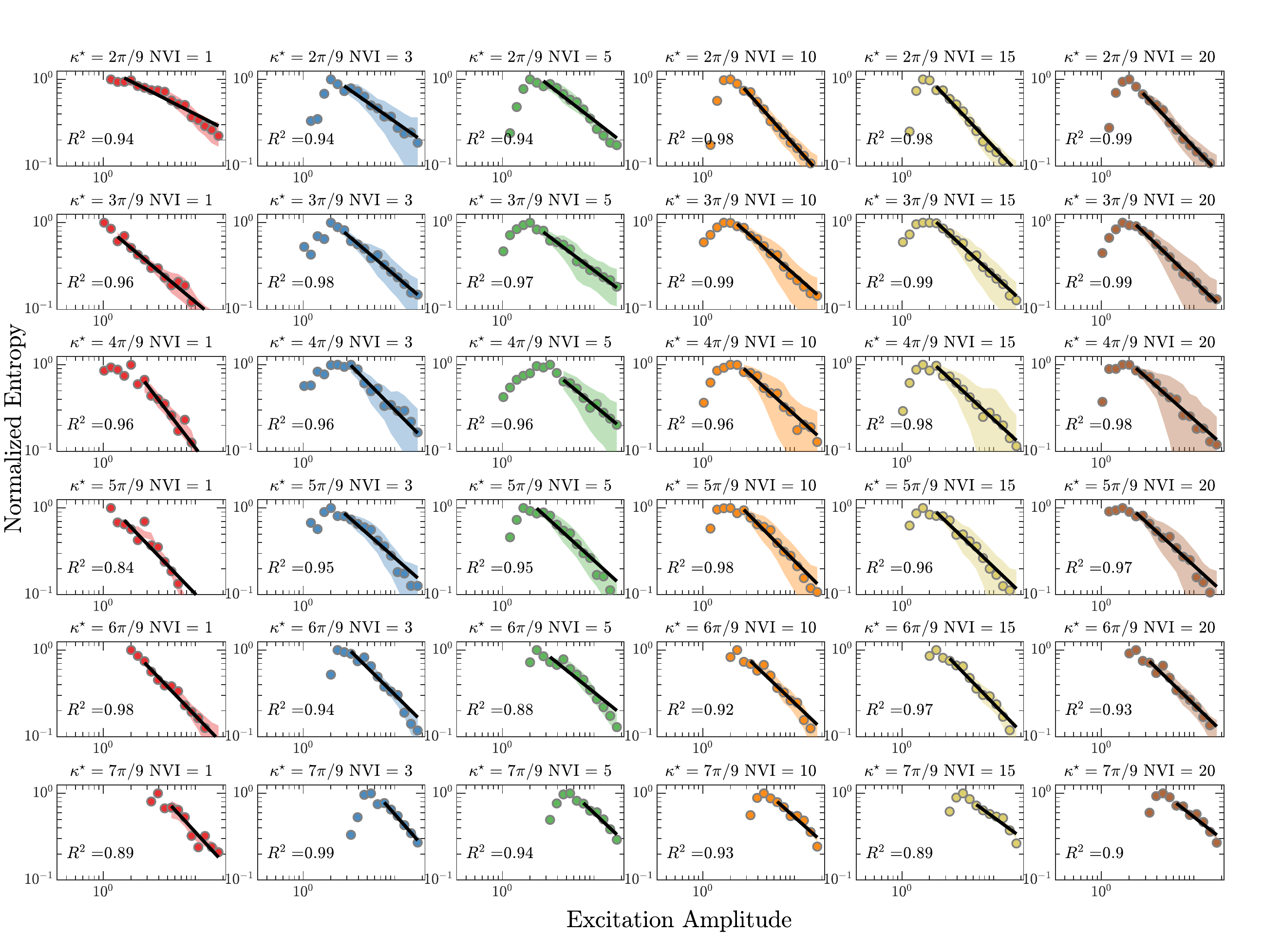}
			\caption{Wavenumber entropy versus excitation amplitude for all datasets generated for the diatomic lattice system of section~\ref{SEC:WavenumberScatter}.}
			\label{FIG:EntAll}
	\end{figure}

The results of Fig.~\ref{FIG:Entropy} were recovered for the entire ensemble of simulations conducted for the diatomic (2-band) lattice of section~\ref{SEC:WavenumberScatter}. The entire ensemble considered VI configurations depicted in Fig.~\ref{Fig:Diatomic_Cell} for excitation wavenumbers ranging from $2\pi/9$ to $7\pi/9$. The resulting normalized  wavenumber entropy trends with respect to input forcing are given in Fig.~\ref{FIG:EntAll} for all simulations, where it is seen that the trends presented in section~\ref{SEC:WavenumberScatter} are agnostic to the excitation wavenumber. Power law fits are superimposed onto each subplot, and the adjusted R-squared values of the fits range between 0.9 and 0.99 for nearly every simulation. 

\clearpage
\newpage
\section{Dispersion band selection for the 4-band lattice}
\begin{figure}[b!]\centering
	\includegraphics[width=\linewidth]{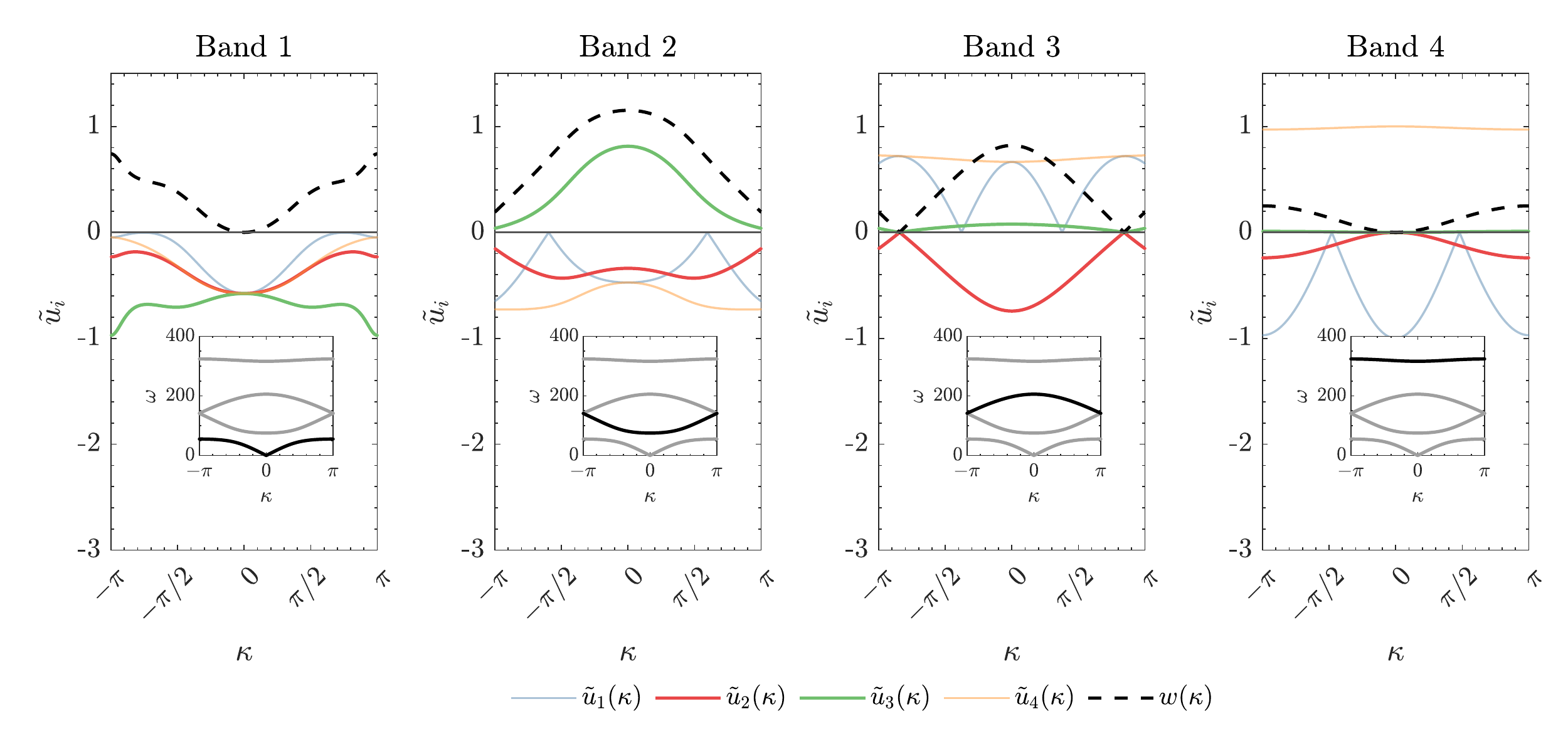}
	\includegraphics[width=\linewidth]{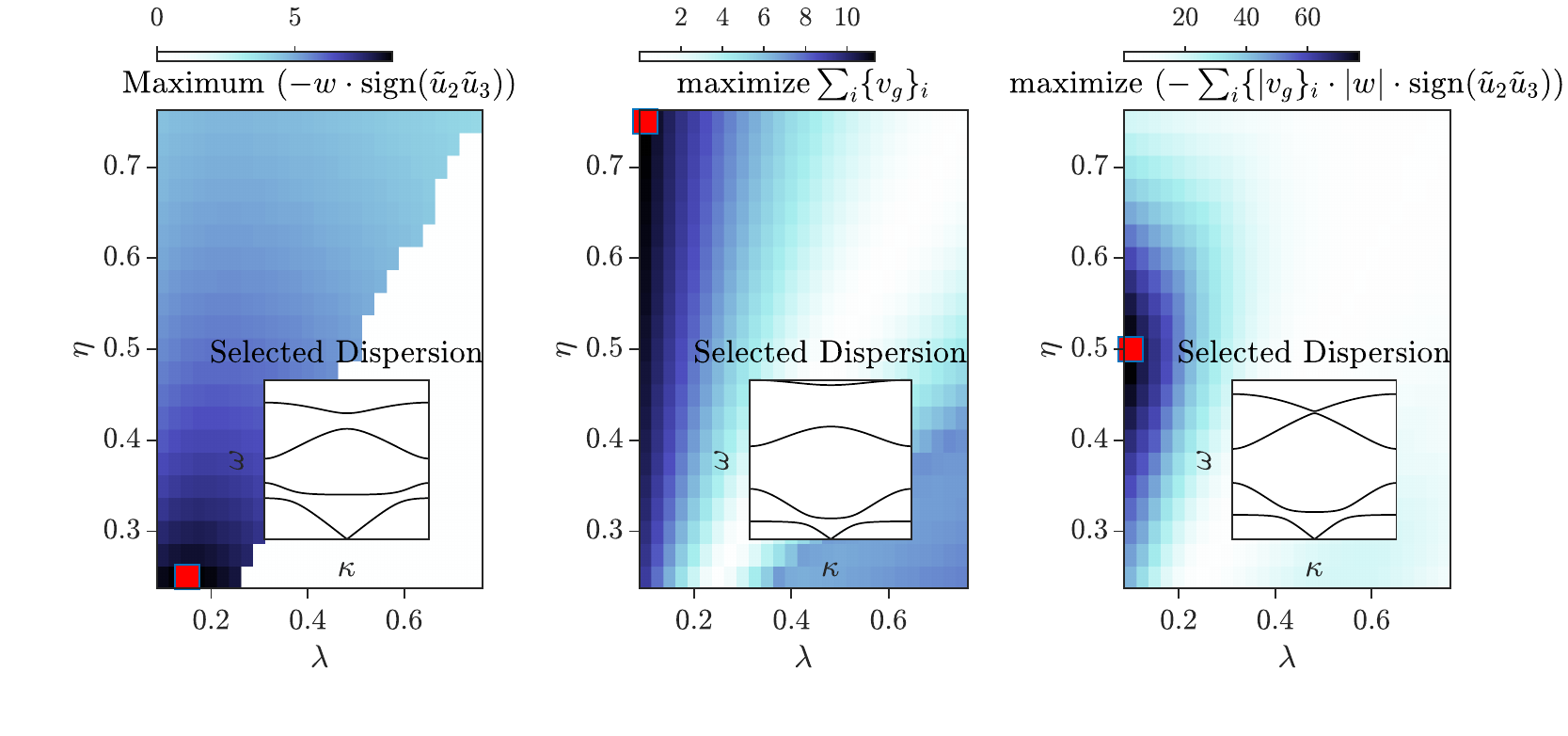}
	\caption{Top: The deflections of the Bloch-eigenmodes for oscillators $\tilde{u}_1$-$\tilde{u}_4$ of the 4-band lattice for each band, as well as $w=|\tilde{u}_3-\tilde{u}_2|$ depicting total deflection of the resonator; bottom: The \textit{cost-function} with respect to maximum deflection of the resonator on the second band ($w$) subject to out-of-phase motions, maximum group velocity, and a weighed measure considering both the deflection $w$ and the group velocity; the red squares the \textit{optimal} pairing of the parameters ($\lambda$,$\eta$), and the insets depict the resulting dispersion relations.}
	\label{FIG:dispersionselection}
\end{figure}

Details on the dispersion band selection for the 4-band lattice considered in section III are provided in Fig.~\ref{FIG:dispersionselection}. The deflections of the Bloch-eigenmodes of the lattice were computed by solving the Bloch-eigenproblem over a sweep of waveumbers in the Irreducible Brillouin Zone (IBZ). 
Within a unit cell, the deflection of the resonator is computed as, $w=|\tilde{u}_2-\tilde{u}_3|$, of the Bloch-eigenmode in terms if of $\lambda$ and $\eta$ as stated in the main text: $m_1=m_4= m = 0.005$ kg and $k_1=k_4=k = 2\times10^4$ N/m by $m_2 = m(1-\eta)$, $m_3 = m\eta$, and $k_3 = k\lambda$ while we fix $k_2 = 10^4$ N/m. Note that the notation $\tilde{u}$ indicates displacement defined over the Bloch-eigenmode, not to be confused with the notation $u$ which corresponds to coordinate displacements of the finite lattice in the main text.
The Bloch-eigenmodes thus satisfy the following eigenvalue problem:
\begin{equation}
	\left(
	m\omega^2\begin{bmatrix}
		1&0&0&0\\
		0&1-\eta&0&0\\
		0&0&\eta&0\\
		0&0&0&1
	\end{bmatrix}-
k
\begin{bmatrix}
	3/2 & 0 &0 &-1e^{-i\kappa}\\
	-1/2&1+\lambda	&-\lambda&-1/2\\
0&-\lambda&\lambda&0\\
-e^{i\kappa}&-1/2&0&3/2
\end{bmatrix}
\right)
\begin{pmatrix}
	\tilde{u}_1\\\tilde{u}_2\\\tilde{u}_3\\\tilde{u}_4
\end{pmatrix}=\textbf{0}.
\label{EQ:dispersion}
\end{equation}
This gives four Bloch-eigenmode solutions for $\bm{x}(\kappa)$ corresponding to the four bands of the lattice. The resonator of the 4 DoF model is described by $\tilde{u}_2(\kappa)$ and $\tilde{u}_3(\kappa)$. As explained in the main text, it is best that the second band corresponds to out-of-phase motion between these two coordinates, and that the deflection is maximized with respect to the system parameters.
To maximize deflection subject to only out-of-plane motion, the signs of $\tilde{u}_2$ and $\tilde{u}_3$ are to be different, and hence this is recovered by maximizing $|w|{\rm sign}(-\tilde{u}_2\tilde{u}_3)$. 
The group velocities over the bands is considered as well by finding the maximum in the IBZ,yielding the use of the weighted measure, $[\max{u}]_{\lambda,\eta} (v_g|w|{\rm sign}(-\tilde{u}_2\tilde{u}_3))$ where $\lambda$ and $\eta$ relate stiffnesses and masses in the unit cell. 

The \textit{cost-function} recovered for deflection, group velocity, and the weighted measure between the two are graphically shown in~\ref{FIG:dispersionselection}, together with the dispersion that is recovered by selecting the \textit{optimal} point in a parameter grid.  The  parameter pairing best suited for maximizing the previous weighted measure was taken as the \textit{ideal} parameter settings to achieve inter-band energy transfers from low-to-high bands (section~\ref{SEC:IBTET}).  The grid approach was selected because the eigensolutions of Eq~\eqref{EQ:dispersion} are too cumbersome to write-out analytically, and were not amenable for Newton-based straightforwardly. While a numerical scheme based on finite differences could resolve this, the search space was sufficiently confined and the problem was sufficiently small that direct grid search was not costly to perform. Moreover, the cost functions of Fig.~\ref{FIG:dispersionselection} show trivial minimum and maximum solutions. 


\clearpage
\newpage
\section{Additional results for IBTET}

\subsection{Recovered phase trajectories in the full lattice system}

The phase trajectories on branches of NNMs in the FEPs of the ROM  reported in the main text (Fig.~\ref{FIG:FEP}) revealed that the VI oscillator undergoes various dynamic regimes with varying energy, ranging from a low-energy linear system to a high energy smooth system governed by the elastic vibro-impact potential. The phase trajectories across regions I-IV of Fig.~\ref{FIG:FEP} can be compared to the corresponding phase plots of the full lattice in order to confirm that this physical mechanism is indeed seen in the lattice. To do this, simulations were considered whereby only one VI unit cell is embedded in the lattice with either Hertzian or bilinear contact law. The time series of the oscillators comprising the VI unit cell of the lattice  were then considered, and phase trajectories could be recovered in the $u_1$-$\dot{u}_1$ and $u_2$-$\dot{u}_2$ planes, where  $u_1$ corresponds to the outer mass of the unit cell and $u_2$ to the inner mass (the VI resonator). 

\begin{figure}[h!]
	\includegraphics[width=\linewidth]{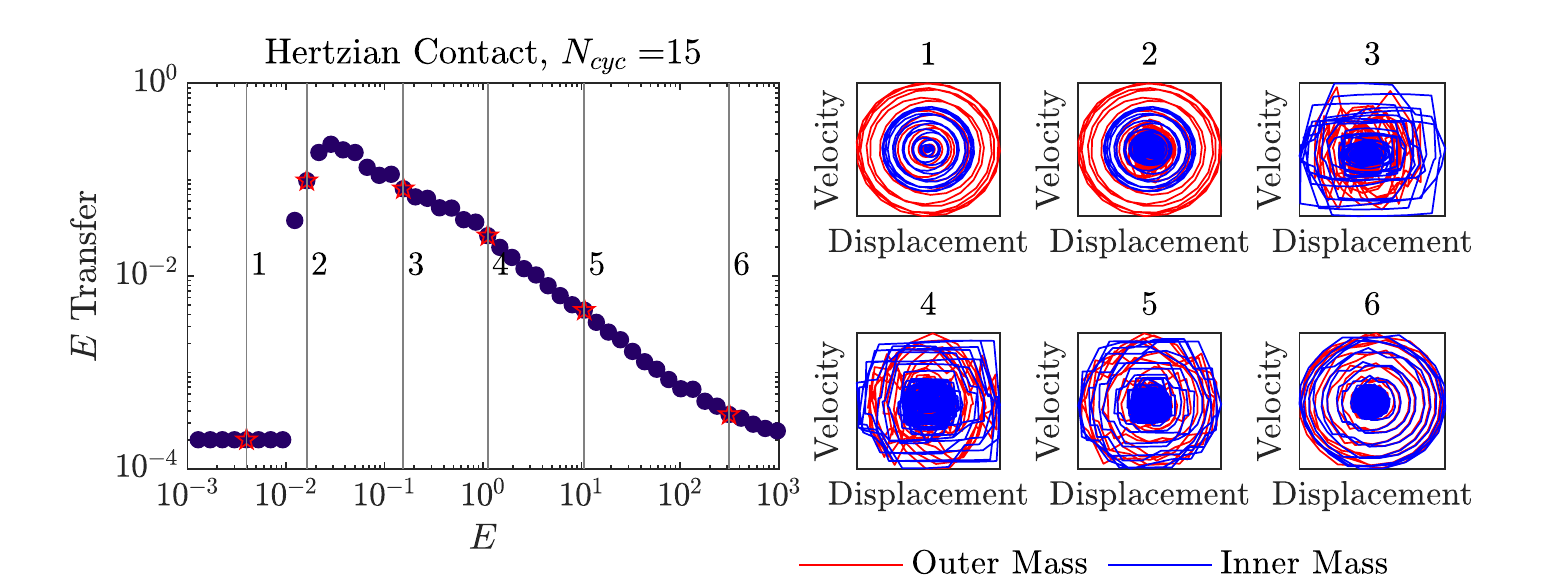}%
	\caption{The phase trajectories of the masses of a single VI unit cell obeying the Hertzian contact law embedded in a full lattice masses of a single VI unit cell obeying the Hertzian contact law, plotted for various energies (right panels), and the corresponding normalized IBTET with respect to input energy (left panel - red dots).}
	\label{FIG:phasefull_hertz}
\end{figure}

\begin{figure}[h!]
	\includegraphics[width=\linewidth]{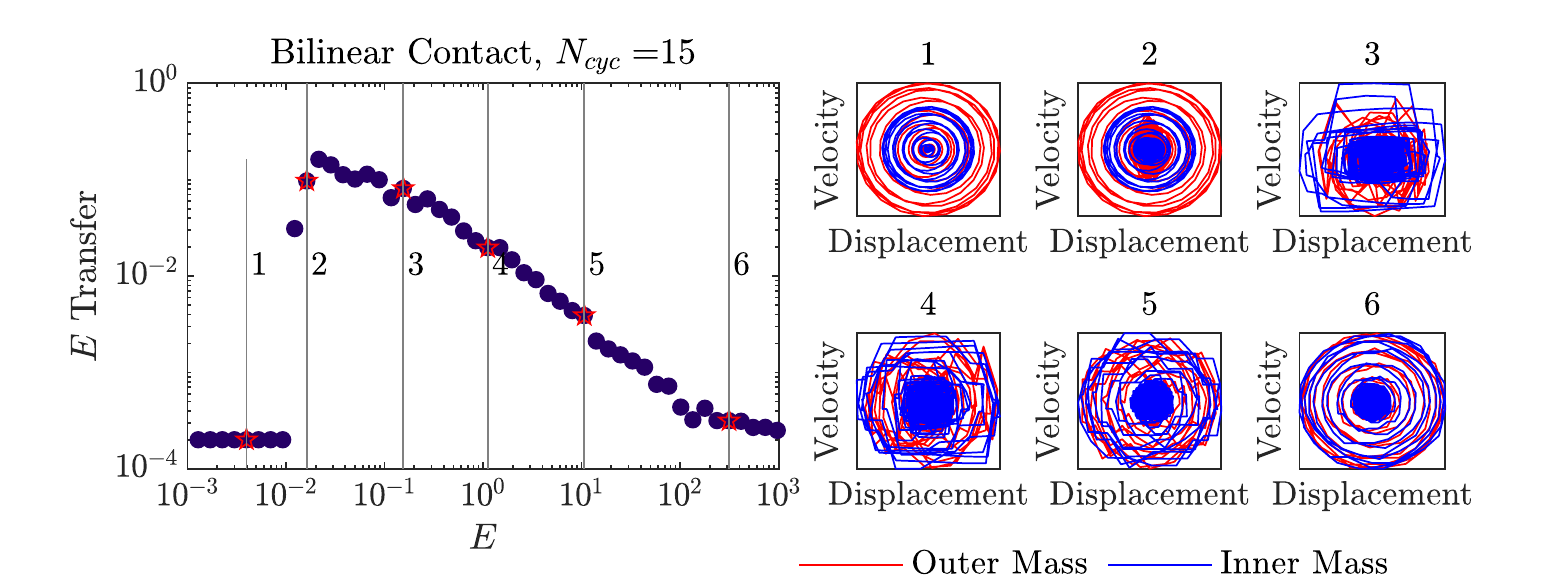}%
	\caption{The phase trajectories of the masses of a single VI unit cell obeying the bilinear contact law embedded in a full lattice masses of a single VI unit cell obeying the bilinear contact law, plotted for various energies (right panels), and the corresponding normalized IBTET with respect to input energy (left panel - red dots).}
\label{FIG:phasefull_bilin}
\end{figure}

Figs.~\ref{FIG:phasefull_hertz} and~\ref{FIG:phasefull_bilin} show the resulting phase portraits recovered for simulations of the full phononic lattice excited at various energies for both Hertzian and bilinear contact models, respectively. Low energy orbits are smooth and circular, indicating a linear response. Responses in the low-energy VI region (phase trajectory 2) are nearly the same, but with clear modulation and irregularity shown towards to origin of the host mass orbit (red), directly corresponding to the grazing region II of the FEP of the unit cell ROM. Higher-energy excitations (plots 3-4) in the fully VI energy regimes reveal non-smooth temporal dynamics, as predicted by region III of the unit cell FEP. Finally, high energy simulations result in phase trajectories that are nearly regular again, with motions of the host mass and resonator being in-phase and nearly completely overlaying each other indicating that the clearance now has nearly no effect, directly in correspondence of region IV of the unit cell FEP of the ROM. 

\clearpage
\newpage
\subsection{Detailed simulation response for bilinear system}

\begin{figure*}[h!]
	\includegraphics[width=\linewidth]{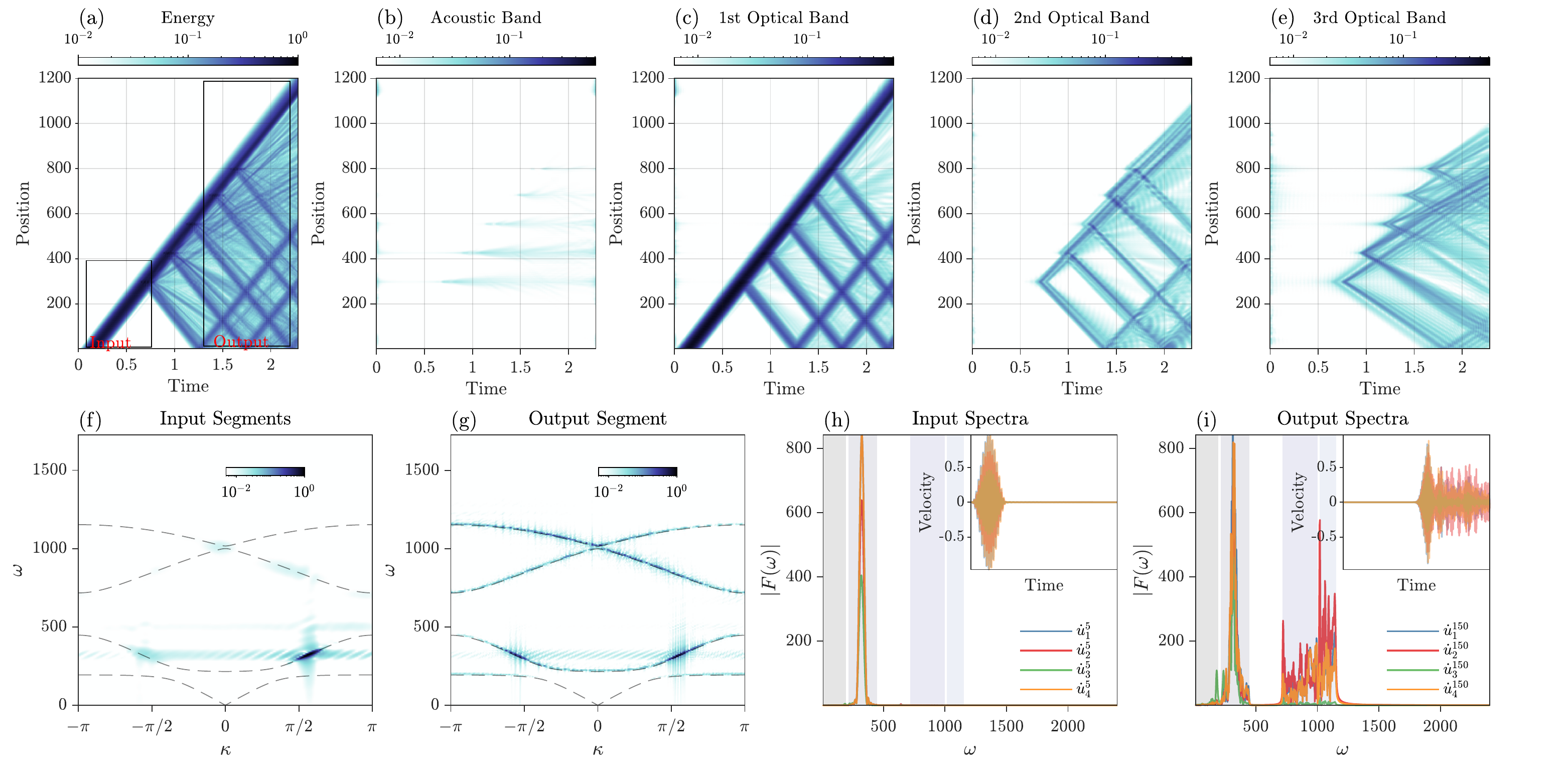}
	
	\caption{IBTET in the 4-band lattice with bilinear VI nonlinearity and 5 VI sites: (a) shows the evolution of the propagating wave energy; (b-e) propagation of the wave energy corresponding to each band of the lattice based on the numerically recovered dispersion of the full simulation; (f,g) dispersion of the input and output segments (labeled in (a)) demonstrating the targeted energy transfer to the higher bands; (h,i) Fourier spectra corresponding to the velocity of the four unit cell DoFs selected before (5-th unit cell) and after (150-th unit cell) VI engagement, with the four band-pass regions depicted with shading and insets depicting the corresponding velocity time histories. }
	\label{FIG:Bilin_Jump}
\end{figure*}

Fig.~\ref{FIG:Hertz_Jump} of the main text depicts a graphical summary of computational and post-processing results for the 4-band lattice with embedded Hertzian VI nonlinearities. For completeness, Fig.~\ref{FIG:Bilin_Jump} depicts the same computational summary computed for a system with embedded bilinear VI nonlinearity. The same remarks stated for Fig.~\ref{FIG:Hertz_Jump} in the main text apply to Fig.~\ref{FIG:Bilin_Jump} as well, further corroborating the similarities in behavior between Hertzian VIs and bilinear VIs with respect to IBTET.

\clearpage
\newpage
\subsection{Influence of input bandwidth and number of VI}

\begin{figure}[h!]			  		 
	\includegraphics[width=.3\linewidth]{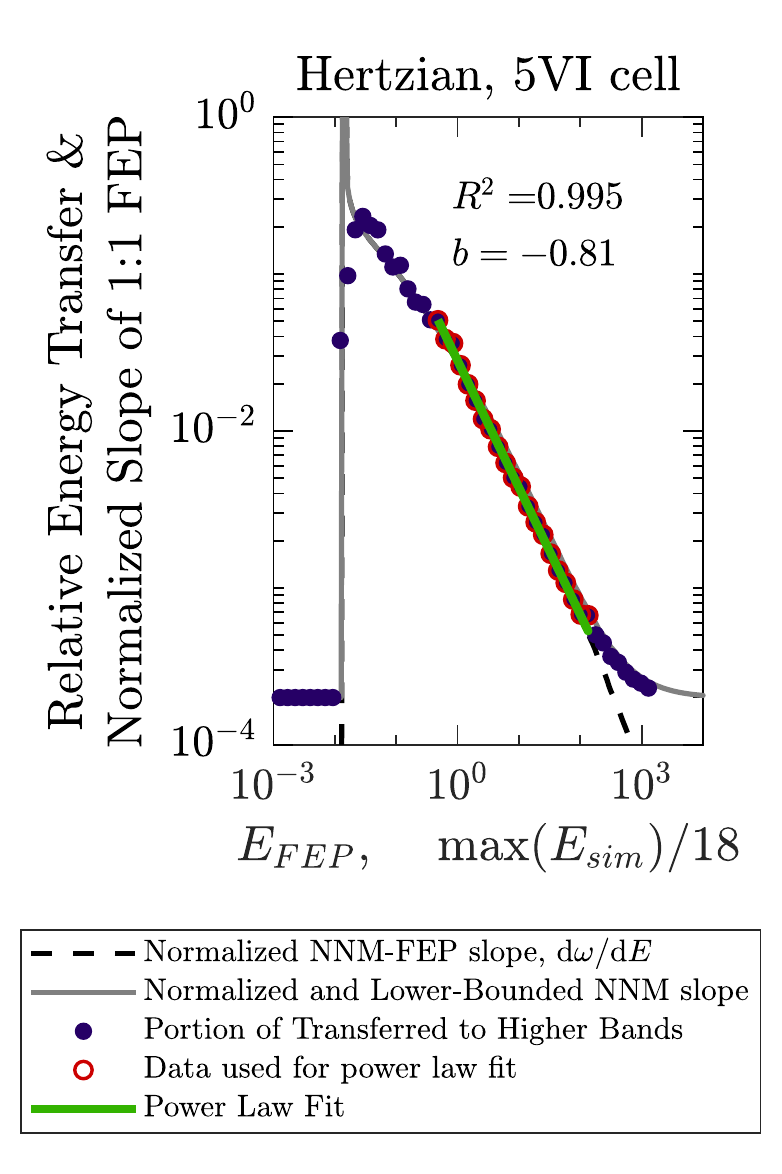}%
	\includegraphics[width=.3\linewidth]{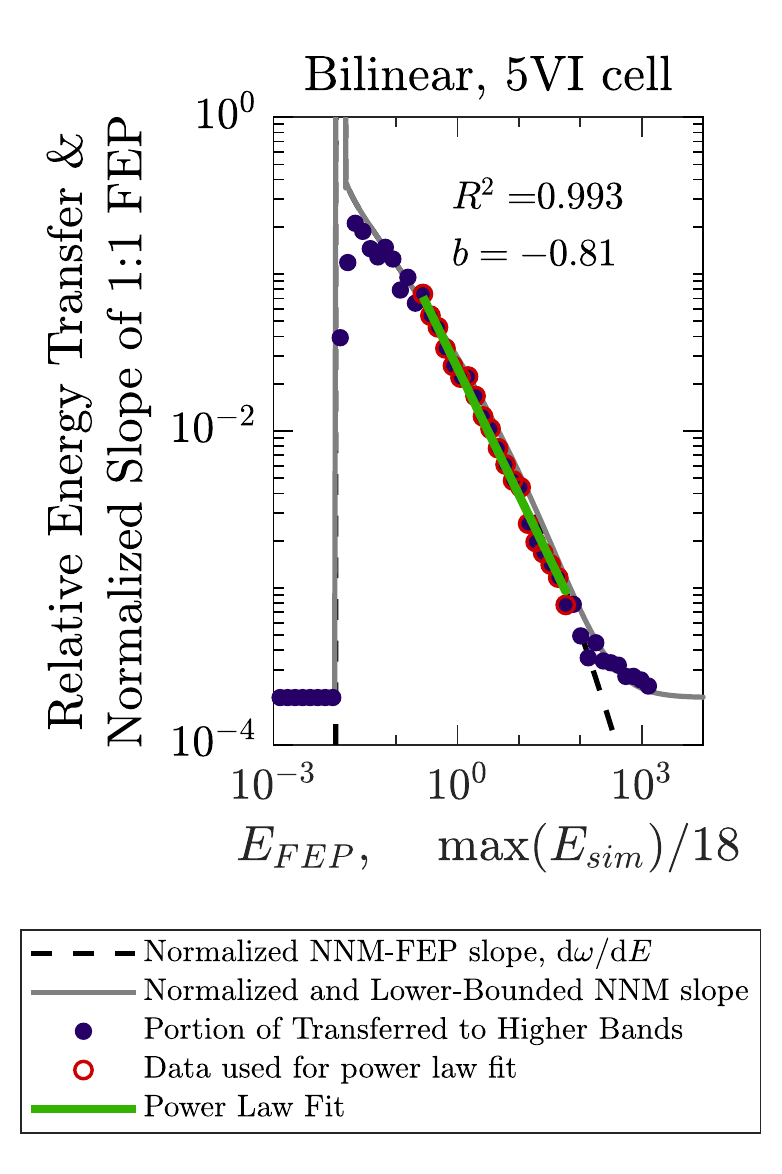}%
	\caption{The same as Fig.~\ref{FIG:Ejump_FEP}, but for 15 cycles of input excitation instead of 30. The relative energy inter-band energy transfer, with the normalized slope from the ROM-FEP superimposed for (a) Hertzian and (b) bilinear contact models; the dashed lines depict the normalized FEP slopes,  the gray lines depict the normalized FEP slopes lower-bounded by the initial (linear) energy of the higher bands, and green lines depict a power law fit to red dots, with the adjusted R-squared value shown with the inset.}
\label{FIG:ejump_5cyc}
\end{figure}

To understand the effect that the forcing profile has on the results presented in section~\ref{SEC:IBTET}, an additional set of simulations was performed subject to 15 cycles of input forcing instead of 30. The results are given in Fig.~\ref{FIG:ejump_5cyc} where very similar trends to Fig.~\ref{FIG:Ejump_FEP} are recovered. This indicates that the mechanisms for energy transfer are indeed non-resonant, as the duration of the oscillations that the VIs are subject to does not modify overall performance. 

\begin{figure}[h!]			  
		\includegraphics[width=.3\linewidth]{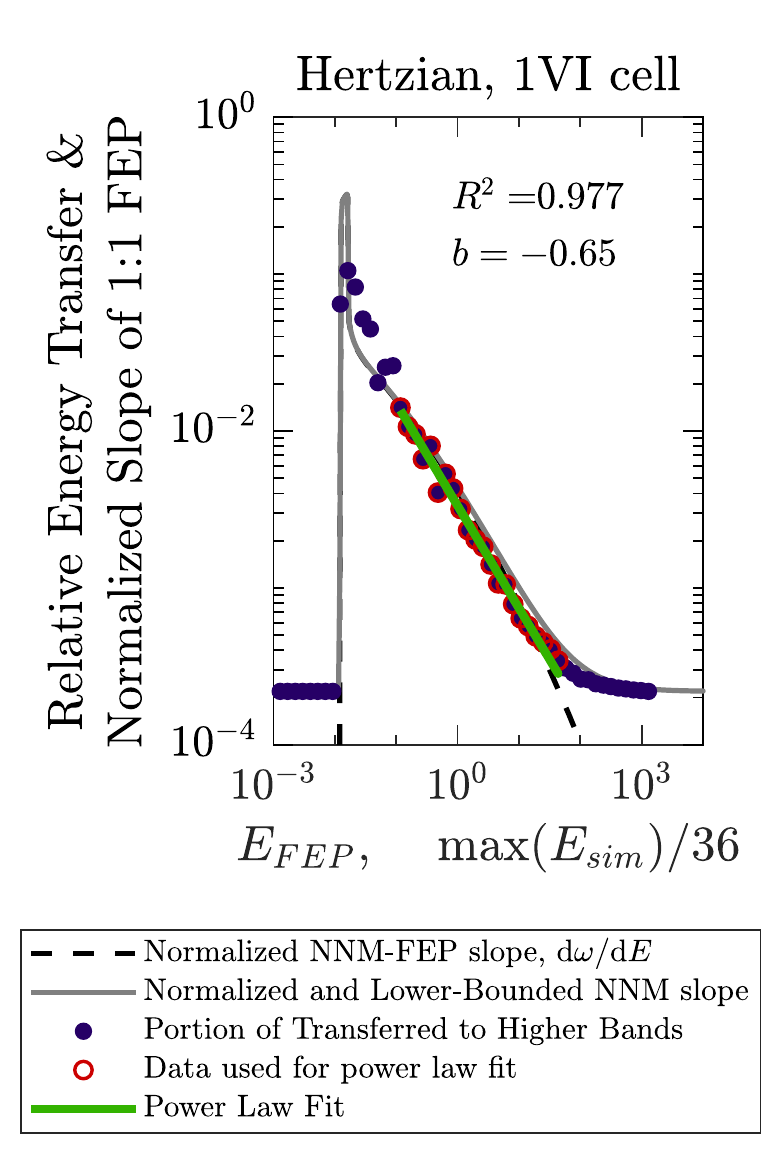}%
		\includegraphics[width=.3\linewidth]{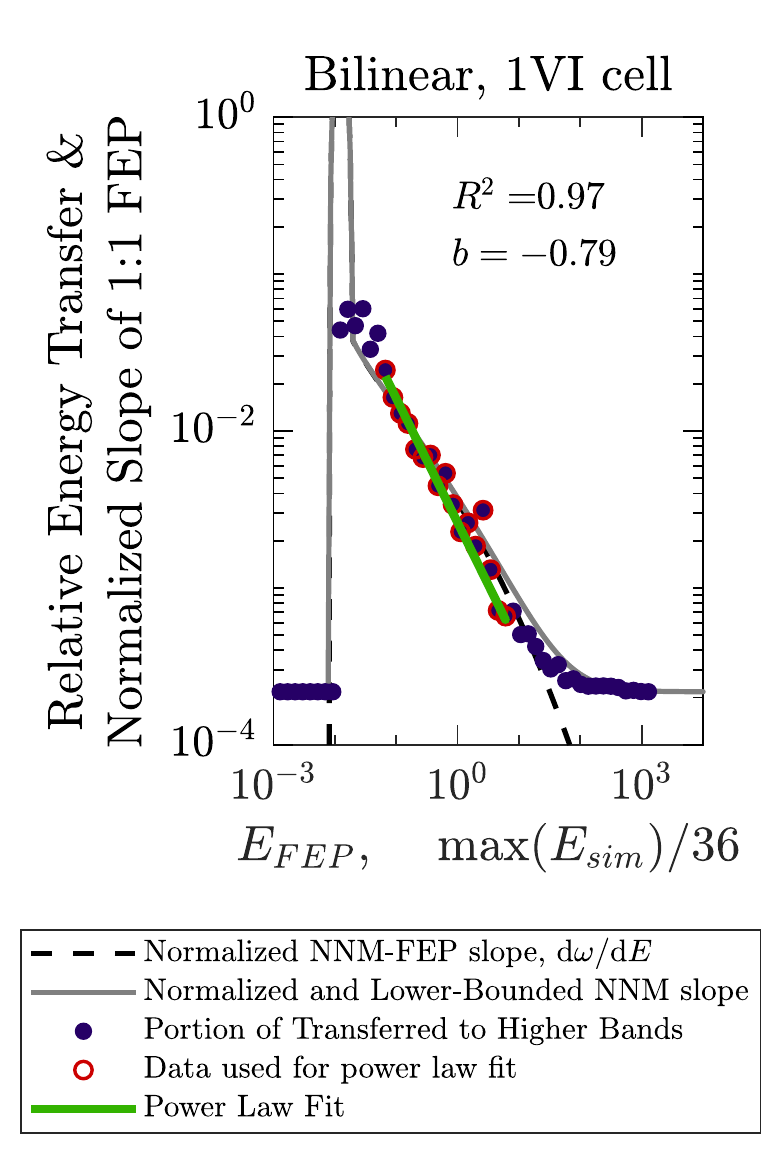}%
		\caption{The same as Fig.~\ref{FIG:Ejump_FEP}, but for 1 VI engaged instead of 5. The relative energy inter-band energy transfer, with the normalized slope from the ROM-FEP superimposed for (a) Hertzian and (b) bilinear contact models; the dashed lines depict the normalized FEP slopes,  the gray lines depict the normalized FEP slopes lower-bounded by the initial (linear) energy of the higher bands, and green lines depict a power law fit to red dots, with the adjusted R-squared value shown with the inset.}
	\label{FIG:ejump_1VI}
\end{figure}

Moreover, the effect of having only a single VI unit cell configuration is was considered as well. To this end, another set of simulations was performed subject to the 30 cycle excitation as the case for Fig.~\ref{FIG:Ejump_FEP} of the main text, but now for only 1 VI embedded within the finite lattice. 
The resulting IBTET are given in Fig.~\ref{FIG:ejump_1VI} with the normalized FEP slope 
superimposed. The same trends are recovered again, but with some minor differences. The total energy transfer achievable is unsurprisingly less (maxing out at approximate 10 percent). Hence, the normalization constants for the FEP slopes are slightly different, which is why the FEP slopes superimposed appear slightly different in Fig.~\ref{FIG:ejump_1VI}. Moreover, there are more pronounced perturbations  from the smooth decay trends as compared to the 5 VI case, and this is due to the volatility of the non-resonant VI dynamics which are \textit{smoothed-out} by incorporating more VIs. In other words, the energy transfer is dependent on the momentum transfer of incident waves. With additional VIs, this momentum transfer is better \textit{averaged} out across the system as compared to the single VI case. However, the agreement in the overall trends of Fig.~\ref{FIG:ejump_1VI} supports the arguments developed in section~\ref{SEC:ROM} for the evolution of the BTET  mechanism with respect to system energy.


\end{document}